\def\etal{{\em et al.$\;$}}
\def\H0{$H_0$ = 100 {\it h} km s$^{-1}$ Mpc$^{-1}$}
\documentstyle[12pt,psfig,amstex]{l-aa}     
\def\today{\ifcase\month\or
  January\or February\or March\or April\or May\or June\or
  July\or August\or September\or October\or November\or December\fi
  \space\number\day, \number\year}

\begin{document}
\topmargin=1.0cm
\title{The ESO-Sculptor Survey: Spectral classification of 
galaxies with $z$ $\le$ 0.5
	\thanks{Based on observations collected at the European Southern 
Observatory (ESO), La Silla, Chile.}}
\author{Gaspar Galaz 
	\and
	Val\'erie de Lapparent} 
\offprints{G. Galaz}
\institute{CNRS, Institut d'Astrophysique de Paris, 
	98 bis, Boulevard Arago, 75014 Paris, France.}
\date{Received: --- ; accepted: ---}

\maketitle

\begin{abstract}

Using the ESO-Sculptor galaxy redshift survey data (ESS), we have extensively
tested the Principal Components Analysis (PCA) method to perform 
the spectral classification of galaxies with $z \la$ 0.5. This method
allows us to classify all galaxies in an {\em ordered} and {\em 
continuous} spectral sequence, which is strongly correlated with the
morphological type. 
The PCA allows to quantify the systematic physical properties of the galaxies in
the sample, like the different stellar contributions to the
observed light as well as the stellar formation history. 
We also examine the
influence of the emission lines, and the signal-to-noise ratio of the data. 
This analysis shows that the emission lines play a significant role in the
spectral classification, by tracing the activity and abnormal spectral
features of the observed sample. The PCA also provides a powerful tool to
filter the noise which is carried by the ESS spectra.

By comparison of the ESS PCA spectral sequence with that for a selected
sample of Kennicutt galaxies (Kennicutt 1992a), we find that the 
ESS sample contains 26\% of E/S0, 71\% of Sabc 
and 3\% of Sm/Irr. The type fractions for the ESS show no significant changes
in the redshift interval $z \sim 0.1-0.5$, and 
are comparable to those found in other galaxy surveys at intermediate
redshift. The PCA can be used independently from any set of synthetic
templates, providing a completely 
objective and unsupervised method to classify spectra.
We compare the classification of the ESS
sample given by the PCA, with a $\chi^2$ test between the ESS
sample and galaxy templates from Kennicutt (Kennicutt 1992a), and obtain
results in good agreement. The PCA results are also in agreement with the
visual morphological classification carried out for the 35 brightest
galaxies in the survey.

\end{abstract}

\section{Introduction}

The classification of the galaxies in a 3-D galaxy map
provides invaluable information for studying the formation and
evolution of galaxies in relation to the large-scale structure.  With
these goals in mind, we have performed a spectral classification for
the ESO-Sculptor Faint Galaxy Redshift Survey (ESS,
hereafter; \cite{delapparent93}). The photometric catalogue of the ESS is
based on CCD 
imaging and provides the B, V and R(Johnson-Cousins)
photometry of $\sim13000$ galaxies (\cite{arnouts97}).  The
spectroscopic catalogue provides the flux-calibrated spectra of a
complete sub-sample of $\sim$700 galaxies with $R_c\le20.5$ obtained
by multi-slit spectroscopy (\cite{bellanger95a}).  The ESS allows for
the first time to map in detail the large-scale clustering at $z\la
0.5$ (\cite{bellanger95b}).

Morphological classification (\cite{hubble36}, 
\cite{devaucouleurs61}, \cite{sandage75}), is based on the recognition of
image patterns and it naturally started with the investigation of the nearest
galaxies. For non-local galaxies, producing an {\em objective} morphological
classification in the same classification system as for local galaxies is
extremely difficult and would require a very complex taxonomy
(\cite{ripley93}). The major limitation is the angular resolution given by
ground-based telescopes. Only a rough classification can be made, for example
by fitting de Vaucouleurs or exponential profiles or using the
relationship between the central concentration index and the mean surface
brightness (\cite{doi93}). Moreover, this method can only be applied up to modest
redshifts ($z \sim 0.2$). Recently, with the refurbished Hubble
Space Telescope (HST), can one see the detailed morphology of galaxies up to
$z \sim 0.7$ (or I $\la 25$) (\cite{abraham94}) and derive an acceptable
morphological 
description up to $z \sim 3.0$ (\cite{vandenbergh96}). Galaxies at these very
high redshifts present a wide variety of morphologies, when
compared to the nearby galaxies. However, when high redshift galaxies ($z
\ga 2.0$) are observed through  
visual photometric filters ({\it e.g.\/,} the HST filters), the 
morphology is delineated by the redshifted blue or the UV emission due to
young stars or 
by star-forming regions, making the objects appear of later morphological
types than they really are. This effect could partially explain the high rate
of distorted galaxies in the Hubble Deep Field (HDF)
(\cite{vandenbergh96}). In summary, the existing morphological
classifications are 
severely dependent on the image spatial resolution, on the photometric
filter, and as a result, on the redshift of the objects. 

In contrast to the qualitative approach of the morphological classification,
the principal physical characteristics of galaxies can be efficiently
quantified by their spectral energy distributions (SED). For a given galaxy,
the SED measures the relative contribution of the most representative stellar
populations and constrains the gas content and average metallicity. 
It is therefore sensible to classify galaxies along a spectral
sequence rather than a morphological sequence. Morgan \& Mayall (1957) have
shown 
that indeed there is a fundamental relationship between spectra of galaxies
and their morphologies: three different populations which in general
constitute every galaxy $-$the gas and the young and old stars 
(\cite{bershady93}, \cite{bershady95})$-$ contribute
both to delineate the main morphological features (bulge, spirals arms,
etc...), and the spectral features (the continuum shape, the emission lines
and absorption bands). The spectral classification has
several advantages over the morphological classification. The spectral range
covered by low resolution spectroscopy (R $\sim$ 500) is wider than the
standard filters, and thus allows to define a common interval for
objects describing a wide range of redshifts. Furthermore, spectra are easier
to handle than 2-D images when a large amount of data is processed. 

In this paper
we perform the spectral classification of the galaxies in the ESS, using
the Principal Component Analysis (PCA). The PCA technique has been applied to
many problems of variate 
nature, from social to biological sciences. In astronomy, it is frequently
used for compressing data to extract the variables which are truly
correlated (\cite{bija74}; \cite{faber73}; \cite{efst84}). 
The PCA has already been used to study inherent relationships between some
selected features or quantities calculated from the spectra of Seyfert
galaxies (equivalent widths and line ratios) and their photometric magnitudes
(\cite{dulzin96}), and on QSO spectra (\cite{francis92}). 
In a recent study, Connolly \etal (1995) have tested the PCA
using the spectral and morphological templates of Kinney \etal (1996), to
show how the spectral properties and the Hubble sequence are related. Using
Kennicutt spectra (Kennicutt 1992a), Folkes, Lahav \& Maddox (1996) and Sodre
\& Cuevas (1997),
show the correlation between spectral properties and morphological type for
normal galaxies.

Here we further test the PCA technique as a tool to achieve a reliable
spectral classification for a new sample of distant galaxies. 
The PCA method is shown to be a powerful tool for measuring 
both, the systematic {\em and} non-systematic spectral properties of a galaxy
sample. We
also study the behavior of the PCA with respect to the data noise level.

The paper is organized as follows. In \S 2 we describe the ESO-Sculptor
(ESS) spectroscopic data.
In \S 3 a brief overview of the PCA technique and its application to the
spectral 
classification are given, as well as the classification procedure
using the $\chi^2$ test. In \S 4 we apply the PCA to a sample of
normal Kennicutt galaxies, and illustrate some specific features of the 
method. In \S 5 the PCA is applied to the ESS. 
The analysis and the spectral classification are described in \S 6, along
with the visual classification for the brightest galaxies. In \S 7 we compare
our main results with those of other studies. The conclusions and prospects
are summarized in \S 8. 

\section{The Data}

Table \ref{info_survey} below lists the main parameters and characteristics
of the spectroscopic sample of the ESS. The redshifts are measured by
cross-correlation with galaxy templates which have been tested for
the reliability of the redshift scale which they provide
(\cite{bellanger95a}). For the $\sim 55\%$ of 
galaxies with emission lines, an ``emission redshift'' is also measured by
fitting the detected lines (mostly [OII], H$\beta$, and [OIII] at 4958 \AA$\;$
and 5007 \AA). When the absorption and emission redshift agree, a weighted
average is derived. The mean errors in the redshifts are given in Table
\ref{info_survey}. Detailed information on the
acquisition and redshift measurements are
described in Bellanger \etal (1995). 
The present spectral analysis is 
based on the subsample of 347 spectra having R(Cousins) $\le$ 20.5, S/N $\ge$
5, a reliable redshift measurement, and a 
spectro-photometric quality (see below). The remaining data
to bring the redshift survey to completion are in the course of
reduction. The only bias affecting the sub-sample used here is the tendency
to observe the brightest galaxies in the R filter (see Figure
\ref{compl}). There is no intended bias related to morphological type in the
observing procedure. The full ESS spectroscopic sample is defined by only one
criterion: R$_c \le 20.5$ (R$_c$ is an estimate of aperture magnitudes in
the R Cousins filter, using the Kron estimator [Arnouts \etal 1996]). 

\begin{table}
\caption[]{Characteristics of the ESO-Sculptor
spectroscopic survey.}
\label{info_survey}
\begin{tabular}{l|l}
\hline \hline
Center  		&  $\alpha(J2000) \sim 0^{h} 20^{m}$ \\
			&  $\delta(J2000) \sim -30^{\circ}$   \\
Sky coverage		&  ($\alpha$) 1.3$^{\circ}$ $\times$ 
($\delta$) 0.27$^{\circ}$  \\
b$^{II}$ 		&  $\sim -83^{\circ}$ \\
Magnitude limit		&  R$_{c} =$ 20.5 \\
Effective depth		&  $z$ $\sim$ 0.5 \\
Telescopes used		& ESO 3.6m and 3.5m NTT \\
Instruments		& EFOSC (3.6m) and EMMI (NTT) \\
			& with multi-object spectroscopy \\
Wavelength coverage	& $\sim$ 4300-7000 \AA$\;$ (3.6m) \\
			& $\sim$ 3500-9000 \AA$\;$ (NTT) \\
Total \# galaxies		&  669 \\
Spectral resolution	& $\sim$ 20 \AA$\;$(3.6m), $\sim$ 10
\AA$\;$(NTT) \\
Slit width		& 1.3 $-$ 1.8 arcsec \\
Redshift error		& $\sim$ 100$-$150 km sec$^{-1}$ \\
\hline
\end{tabular}
\end{table}

\begin{figure}
\centerline{\vbox{\psfig{figure=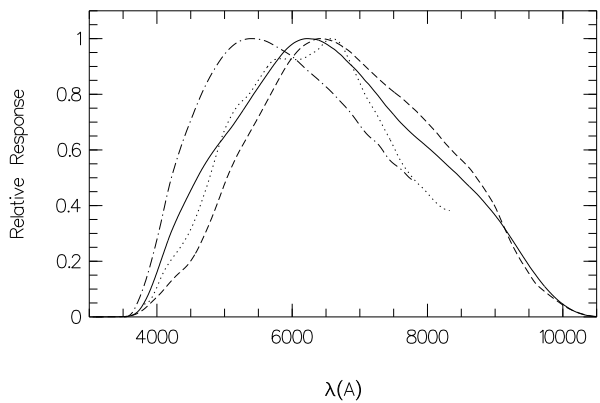,angle=-90,height=6.0cm}}}
\caption[]{Different ``calibrating curves'' (CC), corresponding to different
instrumental set-ups. Solid and dashed curves represent typical CCs for the
NTT telescope, and dotted and dot-dashed curves represent CCs for the 3.6m
telescope.}
\label{cal_curve}
\end{figure}

Because spectral classification techniques are sensitive to the continuum
shape 
of the spectra, the flux-calibration is a crucial step which we now
describe.  This stage amounts to the calculation of the instrumental
response curve, which depends on the telescope, instrument, and CCD
combination, and is 
modulated by the variations in the transparency conditions at the moment of
observation. We denote ``calibrating curve'' the product of these two
independent functions. The calibrating curves for some of the different
instrumental set-ups used for the observations of the ESS are shown in
Figure \ref{cal_curve}. 
The instrumental response for each instrument is
calculated from the observation of spectro-photometric standard stars and is
the average ratio between the {\em observed} spectrum of the star and the
{\em reference} spectrum, in good spectro-photometric conditions. For the ESS
sample we used the
standard stars LTT 377, LTT 7987 and Feige 21 (see \cite{hamuy92}). 
Several standards (2-3) were observed each night or one standard
was observed several times per night (2 to 3 times). 
The resulting r.m.s. variations in the calibrating curves during
a night reported as photometric by the observer, and from one such
night to another, are $\la 10$\%. We therefore select from the available ESS
spectroscopic sample all spectra obtained during these ``stable''
nights. We then correct each spectrum by the mean calibrating curve derived
for the corresponding observing run. The resulting flux
calibrations are only relative. An absolute calibration could be obtained
using the photometric 
magnitudes (cf. \cite{arnouts97}), but this is not necessary for the present
analysis. The final subsample, which contains 347 spectra, represents 52\%
of the total of 669 galaxies with R$_c$ $\le$ 20.5 (see \S 6.5 for
completeness correction). Before the spectral analysis,
the atmospheric O$_2$ absorption bands of the spectra, near
6900 \AA$\;$ and 7600 \AA$\;$, are eliminated by linear interpolation from
the surrounding continuum.

\begin{figure*}
\centerline{\hbox{\psfig{figure=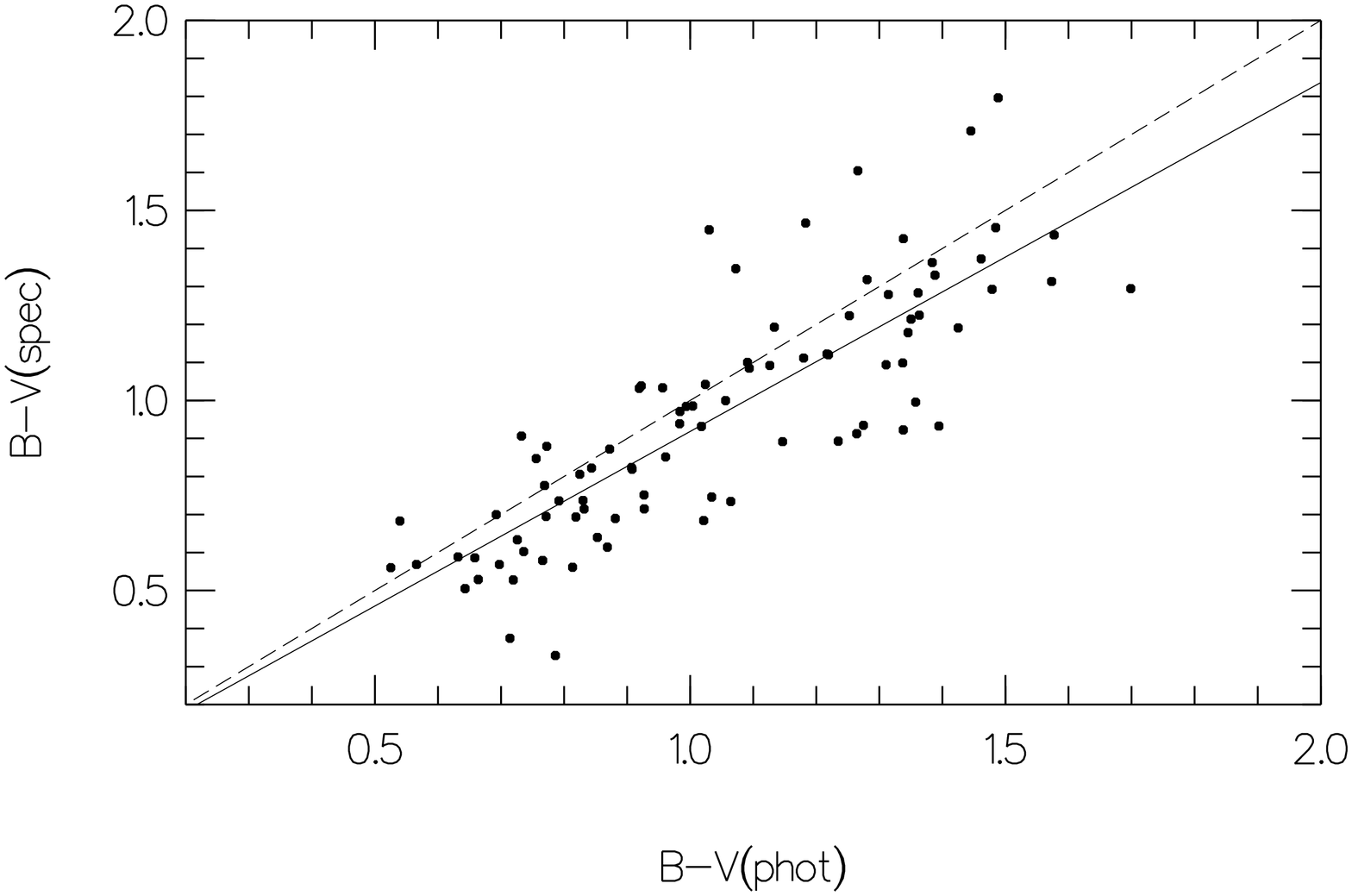,angle=-90,height=7.0cm}\psfig{figure=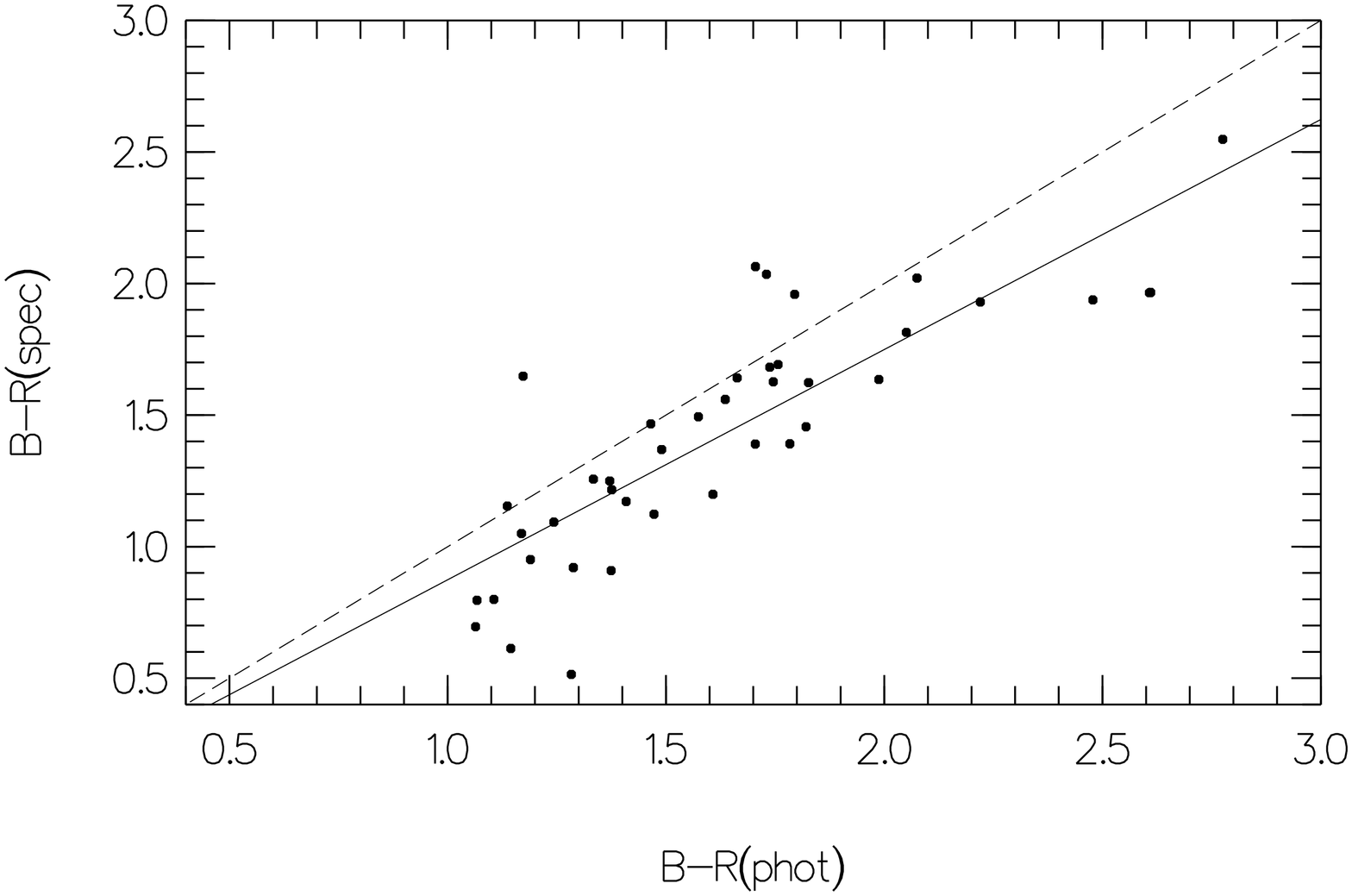,angle=-90,height=7.0cm}}}
\caption[]{Relationship between the photometric and spectro-photometric
colors B$-$V and B$-$R. The parameters for the best linear fits (solid lines)
are given in the text. The dashed lines indicate the
locus for an hypothetical perfect correspondence (B--V)$_{spec}$ =
(B--V)$_{phot}$, and (B--R)$_{spec}$ = (B--R)$_{phot}$.}
\label{comp_colors}
\end{figure*}

To assess quantitatively the spectro-photometric quality of the selected
sample of 347 spectro-photometric calibrated spectra, two tests are performed:
(1) the comparison of 
the spectra of the same galaxy, observed twice or more, and (2), the
comparison of the photometric colors with the synthetic spectro-photometric
colors. First, we found that 40 galaxies from the available spectral sample 
have 2 measured spectra. The r.m.s. variations in the ratios of the
spectra for each pair are $\sim$ 7-10\% when both are obtained in
spectro-photometric conditions, and $\ga$ 10\% when at least one spectrum of
the pair is taken during a non-spectro-photometric night. This confirms that the
spectro-photometric stability indicated by multiple observations of standard
stars during each night is a reliable indicator of the spectro-photometric
quality of the resulting calibrated spectra. Second, we calculate synthetic
colors from the calibrated spectra, and compare the results with the standard
colors obtained from the CCD photometric catalogue (see
\cite{arnouts97}). The photometric magnitude system is B(Johnson),
V(Johnson), and R(Cousins). We compare colors rather than magnitudes in order
to cancel out the unknown absolute flux calibration. We then fit a polynomial
of degree 1 to the spectro-photometric versus photometric colors, for B$-$V
and B$-$R. The slope is 0.952 $\pm$ 0.07 and 0.905 $\pm$ 0.08 for B$-$V and
B$-$R, respectively (see Figure \ref{comp_colors}). 
For a perfect correspondence, the slope should be
1.0. The dispersion around the fit are 
$\sigma$[(B--V)$_{spec}$--(B--V)$_{phot}$] = 0.17,
and $\sigma$[(B--R)$_{spec}$--(B--R)$_{phot}$] = 0.19.
These values are consistent with
the dispersion resulting from the intrinsic photometric and spectrophotometric
errors, which is $\sim \sqrt{2(0.04)^2 + 2(0.10)^2}$ $\sim 0.15$, where 0.04
and 0.10 are the intrinsic errors of the photometric and spectro-photometric
data, respectively.
Therefore, there is a good agreement between the
spectro-photometric and photometric B$-$V and B$-$R colors for spectra taken
during photometric nights, and our estimate of $\sim$ 10\% for the external
uncertainty in our relative flux calibrations appears valid.

We examine one last effect which could bias our flux-calibrated spectra. The
1-D spectra are obtained from the 2-D spectra using the optimal extraction
weight method (\cite{robertson86}). This method weights differently the wings
and the central parts of the light distribution in such a way that the
noisier parts of 
the spectrum (the outer regions of the galaxy) have a smaller weight than the
high S/N part of the spectrum (the core of the object). 
Typically, the weight in the wings is 12 to 20\% of
the weight in the center. Due to well known color gradients in the surface
photometry of individual galaxies, the extracted 
spectra can be affected
differently for different wavelengths, in such a way that 
the extracted spectra are dominated by the stellar content of the center of
the light distribution. However, comparison of the spectra
obtained using the weighted and the un-weighted extraction for 27 objects,
shows that the optimal extraction method does not  
change the {\em shape} of the spectra by more than 3\% (for spectra with S/N
$\ga$ 12). This is well inside the 10\% spectro-photometric uncertainty in
our flux-calibrated spectra. 
\section{PCA and $\chi^2$ test: the formalism}

A detailed description of the PCA technique can be found in
Murtagh \& Heck (1987), and in \cite{kendall80}. Here we summarize its main
characteristics. 
The PCA is applied to a data set of $N$ vectors with 
$M$ coordinates. In this $M$-dimensional space,
each object is a point and the sample forms a cloud 
of points. The central problem which is solved by the
PCA is the description of the cloud of points 
by a set of P vectors of a new orthonormal base,
with P $\ll Min \{N$,$M\}$ and with 
a minimal Euclidean distance from {\em each} point to 
the axes defined by the new base.
The eigenvectors of this new base are
called principal components (PC). Minimizing the sum of 
distances between spectra and axes is equivalent to maximizing
the sum of squared projections onto axes, {\it i.e.\/,} maximizing 
the variance of the spectra when projected onto these new axes.

The input for the PCA is a matrix of $N$ spectra $\times$ $M$ variables, 
which in our case are spectral elements
with 2 to 10 \AA/pix, depending on the resolution
of the data. Each spectrum ${\bf S}$ is normalized by its norm
(the square root of its scalar product with itself), yielding the $N$
normalized spectra ${\bf S}^{norm}$ which serve as input to the PCA:
\begin{equation}
\label{norm}
{\bf S}^{norm} = \frac{{\bf S}}{\sqrt{\sum_{j=1}^{Nbins} {\bf S}_{j}^{2}}}.
\end{equation}
Other normalizations can be used (for example a flux normalization), but it
was shown by 
Connolly \etal (1995) that the details of the normalization
applied to the input $N \times M$ matrix do not
have a strong influence onto the PCA results. However, the interpretation 
of the principal components does depend on the technique used to reduce 
the input matrix.
Because our input vectors are normalized 
by their norm, we can apply the PCA onto the sum of squares
and cross product (intermediate) matrix (SSCP method), which does not 
rescale the data nor center the data cloud. 
The normalized spectra then lie on the surface of a $M$-dimensional
hyper-sphere of radius 1, and 
the first PC has the same direction as the
average spectrum, but with norm equal to 1.
Two other procedures 
are based on the variance-covariance matrix (VC method) and the correlation
matrix (C method), respectively.
The VC method places the new origin onto the centroid of the sample
and the C method also re-scales the data in such a way that 
the distance between variables is directly proportional to
the correlation between them. For the VC method the average
spectrum has to be used in order to reconstruct individual
spectra. We emphasize
that neither the PC's nor the projections given by the SSCP method, used in this
paper, are the same as those given by the 
VC method. However, if the normalized cloud of points is concentrated in a small
portion of the hyper-sphere, then the first PC of the VC method will have
almost the same direction as the second PC given by the SSCP method
(see \cite{francis92}, \cite{folkes96}).
Although these different methods give different PC's, if we take into account
the underlying transformations explained above,
the physical interpretation of the PC's and the projections does not change,
and the final result always satisfies the 
maximization conditions and the orthonormality among the different principal
components. 

After application of the PCA using the SSCP method, we can write
each spectrum ${\bf S}^{norm}$ as 
\begin{equation}
\label{Sapprox}
{\bf S}_{approx} = \sum_{k=1}^{N_{pc}} \alpha_{k} {\bf E}_{k},
\end{equation}
where ${\bf S}_{approx}$ is the reconstructed spectrum of ${\bf S}^{norm}$, 
$\alpha_{k}$ is the projection of spectrum ${\bf S}^{norm}$ onto the
eigenspectrum ${\bf E}_{k}$ and 
$N_{pc}$ is the number of PC's taken into account for
the reconstruction. In equation (2), the PC's are in decreasing order of
their contribution to the total variance.

We show in \S 5 below, that if the
S/N is high enough ({\it i.e.\/,} $\ga$ 8), then we 
can take $N_{pc}$ = 3 or 4 to reconstruct $\sim$ 97 to 98\% of the signal,
respectively.
If the S/N $\la$ 8, it requires a 
higher number of PC's to reproduce the initial
spectrum to such high accuracy because of the noise pattern. Therefore, 
the first 2 or 3 components 
carry most of the signal in each spectrum, which leads us to
use $\alpha_1$, $\alpha_2$ and
$\alpha_3$ to describe the spectral sequence. We choose to reduce
these 3 parameters to the radius $r$ and the angles $\delta$ and $\theta$
defined by the spectrum 
$\alpha_1{\bf E}_1 + \alpha_2{\bf E}_2 + \alpha_3{\bf E}_3$ (as in
\cite{connolly95}) in spherical coordinates ($\delta$ 
the azimuth and $\theta$ the polar angle taken from the equator),
\begin{gather}
\label{sphere}
\alpha_{1} =  r \cos \theta \cos \delta \tag{\ref{sphere}$a$} \\
\alpha_{2} =  r \cos \theta \sin \delta \tag{\ref{sphere}$b$} \\
\alpha_{3} =  r \sin \theta. \tag{\ref{sphere}$c$} 
\end{gather}
\addtocounter{equation}{1}
We express the values of $\delta$ and $\theta$ {\em independently} of the
value of $r$:
\begin{gather}
\label{delta_theta1}
\delta = \arctan \left(\frac{\alpha_{2}}{\alpha_{1}}\right)\tag{\ref{delta_theta1}$a$} \\ 
\theta = \arctan \left\{\left(\frac{\alpha_{3}}{\alpha_{2}}\right) \sin \left[\arctan \left(\frac{\alpha_2}{\alpha_1}\right)\right]\right\}. \tag{\ref{delta_theta1}$b$} 
\end{gather}
\addtocounter{equation}{1}
Note that we prefer the use of $\delta$ and $\theta$ (rather than the ratios
$\alpha_2$/$\alpha_1$ and $\alpha_3$/$\alpha_2$) for defining the spectral
sequence because they have a geometrical meaning. 
In the next section, we show that the physical meaning of $\delta$ is the
relative contribution of the red (or early) and the blue (or
late) stellar populations within a  galaxy. Note that if $r \sim 1$,
then from equation \ref{sphere}c, equation 4b approximates to $\theta
\approx$ arcsin$\alpha_3$. 

For comparison with the PCA, we have implemented a simple
$\chi^{2}$ test between the galaxies of the ESS sample and a set of templates
derived from the Kennicutt sample (Kennicutt 1992a, see \S 5 and \S6).
In contrast to the PCA, the $\chi^{2}$ test is dependent on the set of templates
used and can only provide a constrained classification procedure.
The $\chi^{2}$ between an observed spectrum and a template can 
be written as
\begin{equation}
\label{chi2}
\chi^2 = \sum_{j=1}^{N_{bins}} \frac{({\bf S}_{j} - 
{\bf T}_{j})^{2}}{{\bf S}_{j} + {\bf T}_{j}},
\end{equation}
where ${\bf S}_{j}$ and ${\bf T}_{j}$ are the values in the spectral element
or bin $j$ of the flux-calibrated spectrum and the template, respectively. 
$N_{bins}$ is the total number
of wavelength bins for both the spectrum and the template (we take the largest
wavelength interval common to the spectrum and template, and rebin both
to a common wavelength step of 5 \AA/pix). 
The denominator measures the variance of the spectrum and the template,
assuming that the noise is Poissonian. 
Because for a given observed spectrum $N_{bins}$ is the same for all the 
comparison templates, the $\chi^{2}$ value does not need to be normalized.
Therefore, if we have a
set of $P$ templates, then the closest template $k$ to the spectrum
{\bf S} is the one which satisfies
\begin{equation}
\label{closest1}
\chi^2_{{\bf T}_k} = Min \left(\chi^{2}|{\bf T}_1,
{\bf T}_2,...,{\bf T}_P\right).
\end{equation}
Note that in the PCA treatment, the wavelength interval of all input spectra
must be identical. For the  
$\chi^2$ test, the wavelength interval can be larger than the one used for
the PCA and varies from spectrum to spectrum. This difference will allow us
to check the dependence of the PCA 
classification on the wavelength interval (cf. \S 6).

\section{PCA and spectral sequence: test on Kennicutt galaxies}

Connolly \etal (1995) have shown using the spectra from \cite{kinney96}, that the
first 2 projections of the PCA define a sequence tightly correlated with the
morphological type. Folkes, Lahav \& Maddox (1996) and Sodre \& Cuevas (1997)
have demonstrated this 
property using a larger sample of spectra of local galaxies, namely the
sample of \cite{kennicutt92}. Here we use again the Kennicutt sample to
complement the previous studies and to serve as comparison sample for the ESS
sample. We have selected 27 normal Kennicutt
galaxies from Hubble types E0 to Im, by discarding peculiar morphological
types, and excluding spectra of galaxies with a particular spatial sampling
(strong HII regions or high  extinction zones). Table \ref{kenni_gal} 
lists the ID, the names and morphological types of the selected galaxies. We
apply the PCA to these spectra restricted  
to the spectral range 3700 to 6800 \AA,
with a pixel size of 5 \AA, which is the highest possible resolution 
for that sample (see Kennicutt 1992a). 

\begin{table}
\caption[]{Kennicutt galaxies selected for PCA.}
\label{kenni_gal}
\begin{tabular}{lll|lll}
\hline \hline
\# &  Galaxy 	   &  	Type 	&  \# 	 &  Galaxy &  Type  \\ \hline
1 	 & NGC3379 &	E0 	& 15	 & NGC3627 &	Sb \\      
2 	 & NGC4472 &	E1/S0   & 16	 & NGC2276 &	Sc \\    
3 	 & NGC4648 &	E3      & 17	 & NGC4775 &	Sc \\
4 	 & NGC4889 &	E4   	& 18	 & NGC5248 &	Sbc  \\         
5 	 & NGC3245 &	S0  	& 19	 & NGC6217 &	SBbc \\ 
6 	 & NGC3941 &	SB0/a	& 20	 & NGC2903 &	Sc \\
7 	 & NGC4262 &	SB0  	& 21	 & NGC4631 &	Sc \\
8 	 & NGC5866 &	S0   	& 22	 & NGC6181 &	Sc \\
9 	 & NGC1357 &	Sa   	& 23	 & NGC6643 &	Sc \\
10 	 & NGC2775 &	Sa	& 24	 & NGC4449 &	Sm/Im \\
11 	 & NGC3368 &	Sab	& 25	 & NGC4485 &	Sm/Im \\
12 	 & NGC3623 &	Sa 	& 26	 & NGC3227 &	Sb \\
13 	 & NGC1832 &	SBb	& 27 	 & MK270   &	S0 \\
14 	 & NGC3147 &	Sb &  &  &   \\ \hline  
\end{tabular}
\end{table}

\begin{figure*}
\centerline{\hbox{\psfig{figure=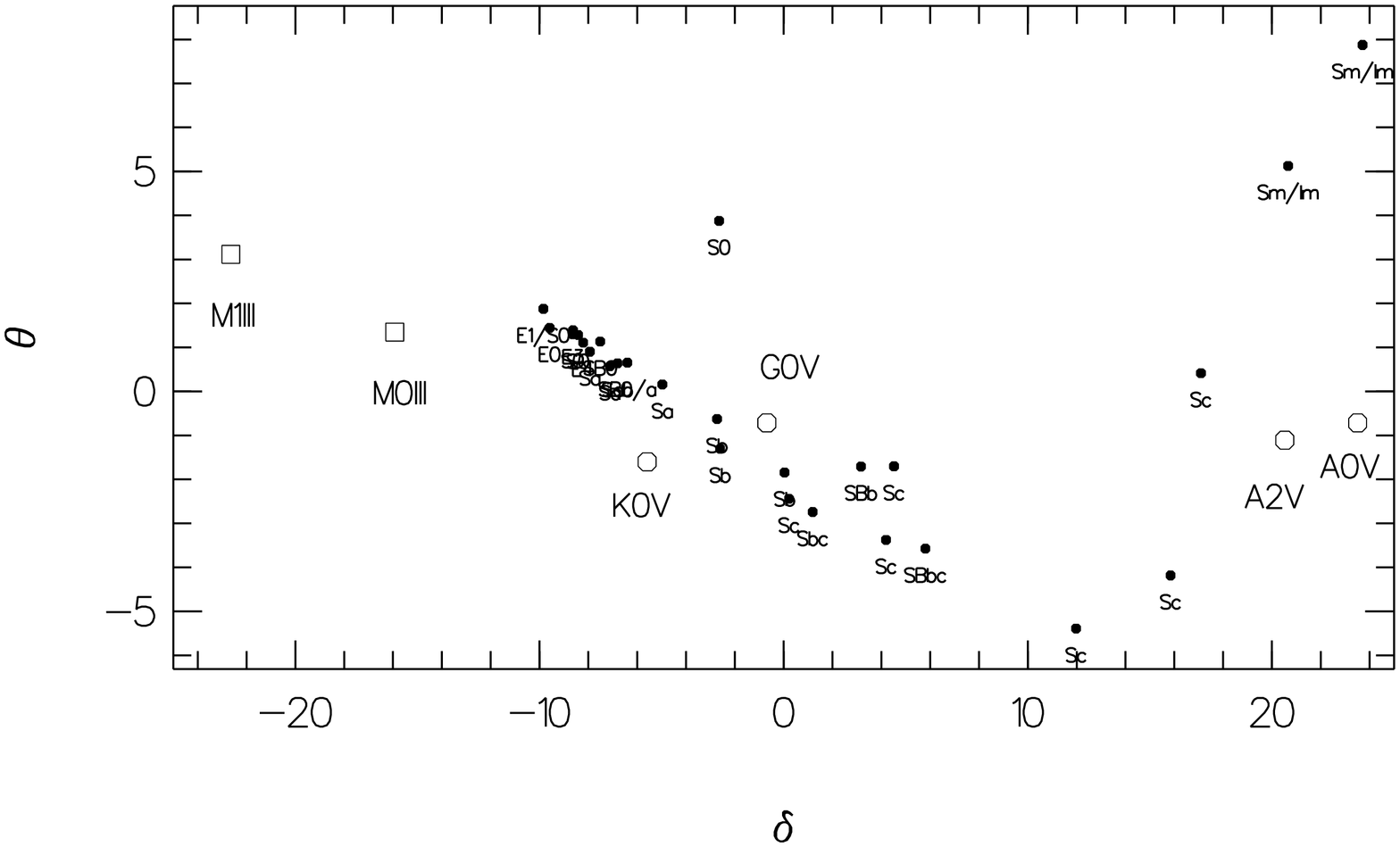,angle=-90,height=6.5cm}\psfig{figure=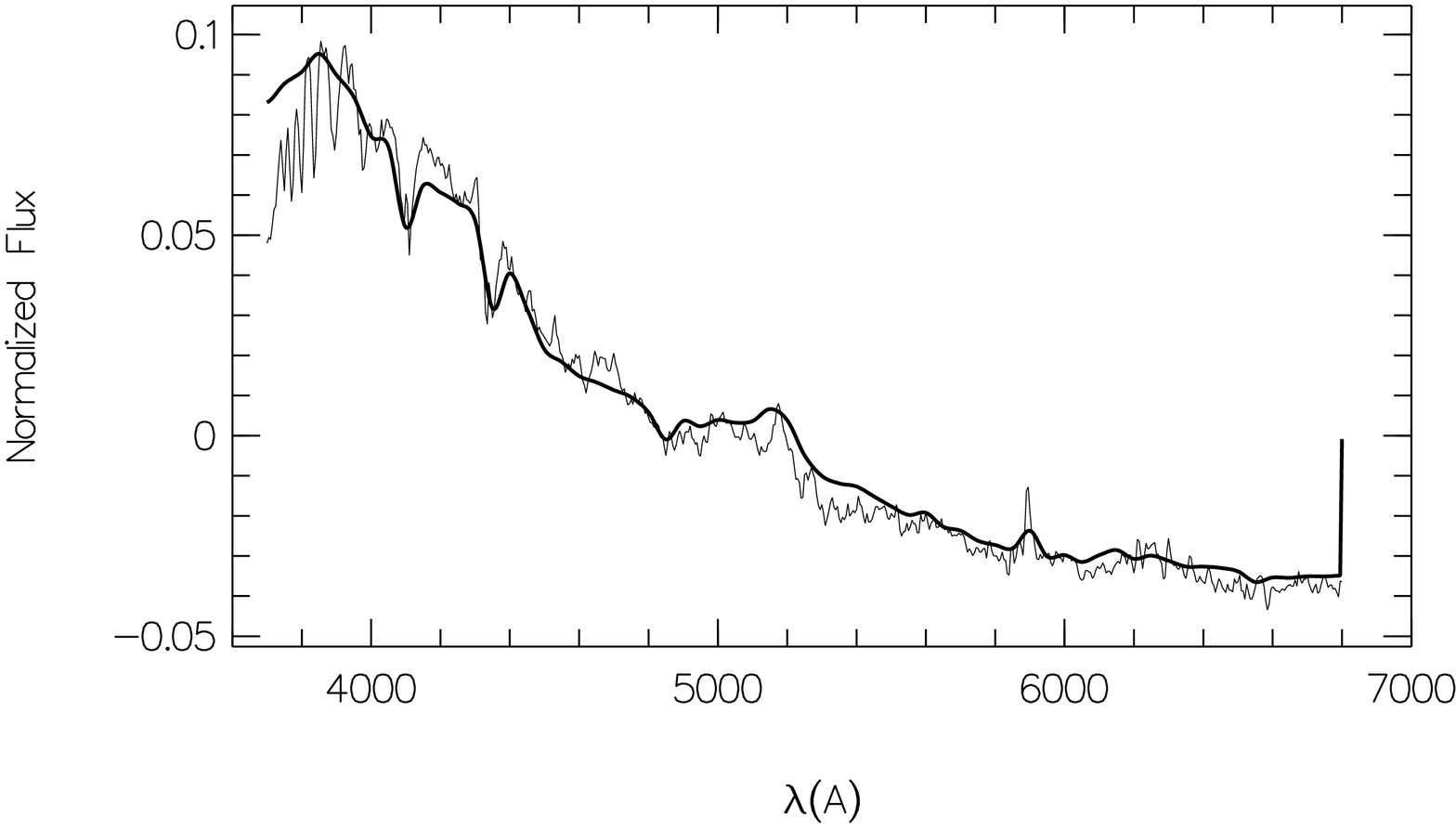,angle=-90,height=6.5cm}}}
\caption[]{The Kennicutt spectra of normal Hubble types (left Figure, dots),
onto the classification plane. Red or early type galaxies are
to the left with $\delta$ $\la$ 0 and blue or late types are to the right with
$\delta$ $\ga$ 0. The deviation in the $\theta$ parameter
is mainly related to the emission lines. The circles and squares indicate
the position of the spectra of main sequence stars and giant stars,
respectively. The right panel shows a comparison between the second PC from
the sample of Kennicutt normal galaxies (thin line) and the PC from the
sample of stars appearing in the left panel (thick line).}
\label{delta_theta_type}
\end{figure*}

Left panel of Figure \ref{delta_theta_type} shows the angles $\delta$ and
$\theta$ (see equations [4a] and [4b]) for the 27 chosen Kennicutt
spectra (see Table \ref{kenni_gal}), showing
the tight sequence strongly correlated with the morphological type, already
shown by Sodre \& Cuevas (1994), Connolly \etal (1995) and Folkes, Lahav \&
Maddox (1996), using different coordinates. 

With only the first 3 PC's, we can reconstruct,
on average, 98\% of the signal of each Kennicutt spectrum in Table
\ref{kenni_gal}. PC's of superior order do not contribute more than 2\% to
the signal. This was already demonstrated by
Connolly \etal (1995), using the observed spectra of \cite{kinney96}. The
physical reason for this striking feature is closely related to the fact that
the fundamental spectral features of normal galaxies can be described by
a reduced number of stellar spectra, namely spectral types AV and M0III. This
was first suggested by Aaronson (1978), using UVK color-color diagrams
(see also \cite{bershady93}, and \cite{bershady95}). 
To probe this effect using the PCA approach, we 
project stellar spectra (from \cite{sviderkien88}) of stars with types A0,
A2, G0, and K0 of the main 
sequence, and two spectra corresponding to giants M0 and M1, onto the first 3
PC's from the Kennicutt sample and derive the values of $\delta$ and
$\theta$. Symbols other than points in the left panel 
of Figure \ref{delta_theta_type} show that the A stars and the M giants mark
the extreme regions (or the extrapolation) of the Hubble
sequence, whereas the G and K stars are located inside the sequence. In
addition, the right panel of Figure 
\ref{delta_theta_type} shows the surprising similarity between the second PC
of the Kennicutt sample (with the emission lines eliminated) and the second
PC from the stellar spectral sample. This extends and further demonstrates the
results of Aaronson (1978)
and allows us to conclude that the spectra of nearby galaxies with normal
Hubble types, may be described with a reduced number of stellar spectra (2
types), at least in the spectral range which is considered here. 
Because the position of the observed galaxies along the $\delta$ axis 
accounts for the relative contributions of the red and blue stellar
populations in the observed galaxies, we adopt the $\delta$ parameter to
describe the spectral sequence.

\begin{figure}
\centerline{\psfig{figure=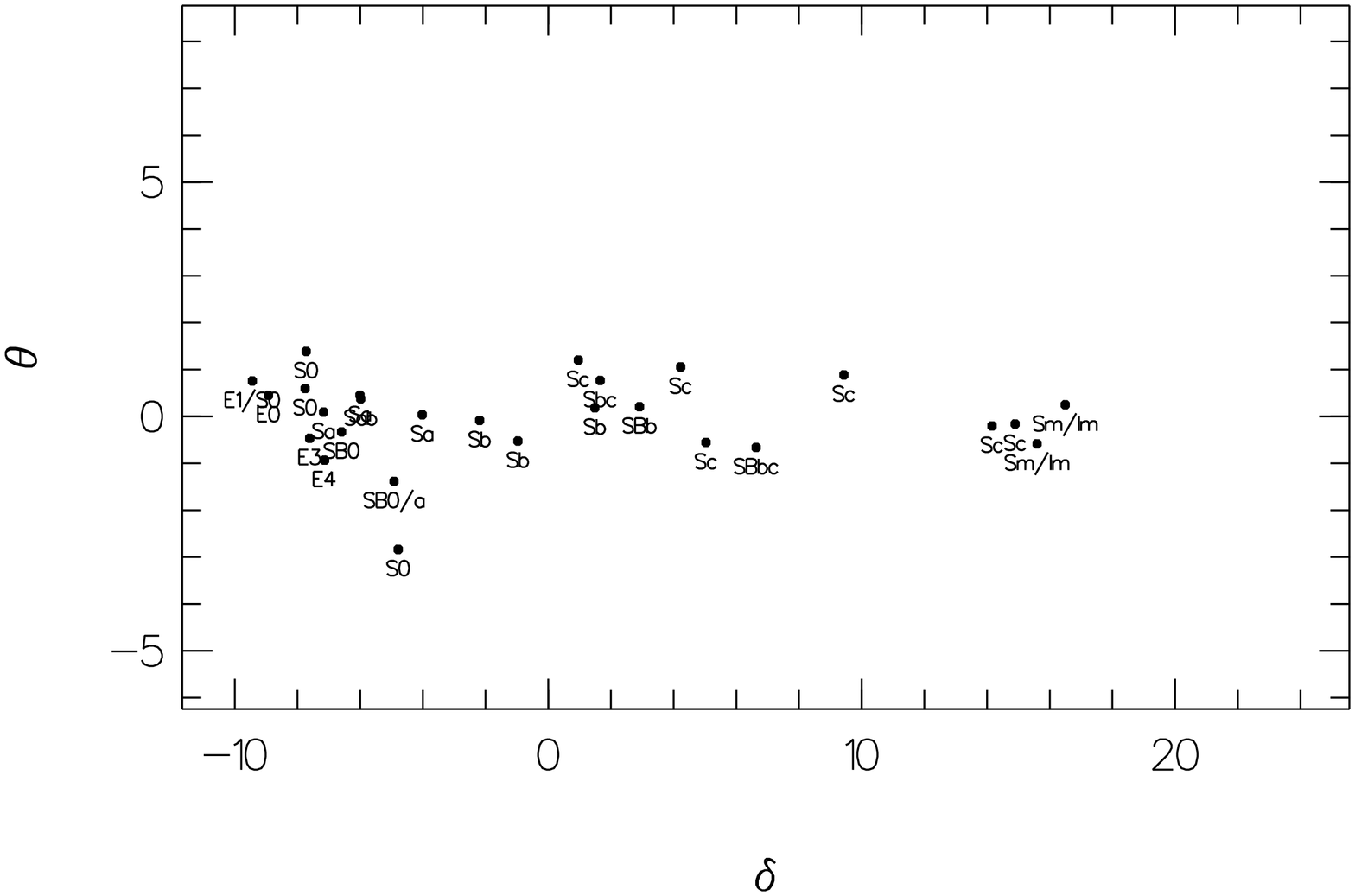,angle=-90,height=6.5cm}}
\caption[]{Figure showing the Kennicutt templates with the emission lines
removed, projected onto the
spherical space ($\delta$,$\theta$), with the same scale as in Figure 
\ref{delta_theta_type}.}
\label{class_lines_nolines}
\end{figure}

As a complement, the parameter $\theta$ conveniently characterizes the
presence of emission lines. 
The emission lines play an important role in the spectral classification of
galaxies. They serve to characterize the strength of star formation, the
nuclear activity and abundances, using for example the ratio between
the strength of different emission lines. Francis \etal (1992), apply
successfully the PCA technique to understand the systematic properties of
QSO's. The role of $\theta$ is demonstrated by truncating the emission lines
from all the Kennicutt spectra. This is done by fitting a polynomial of
degree one to the adjacent continuum for each line. The resulting values of
$\delta$ and $\theta$ are shown in Figure \ref{class_lines_nolines}. 
The ordering of the Hubble sequence along the
$\delta$ axis remains the same as for the sample with emission
lines. However, all spectra now have smaller $|\theta|$ values.
Figure \ref{class_lines_nolines} therefore shows that the emission lines
increase the dispersion in the ($\delta$,$\theta$) plane, 
placing galaxies with strong emission lines far from the equator
defined by $\theta$ = 0. The known correlation between star formation and/or
activity and the Hubble type for morphologically normal galaxies explains the
observed correlation between $\theta$ and $\delta$ in Figure
\ref{delta_theta_type}. 

Folkes, Lahav \& Maddox (1996) have studied in detail the reconstruction
error as a function 
of the S/N in the input spectra, using simulated spectra constructed from the
Kennicutt sample. They also demonstrated the greatly improved capability of
the PCA for filtering the noise over other standard techniques. 
Here we also find that the noise in the input spectra has no effect onto the
classification space ($\delta$,$\theta$) and that the observed $\delta$
sequence remains unchanged when adding arbitrarily high noise onto the
Kennicutt spectra of Table \ref{kenni_gal}: decreasing the S/N of the spectra
down to 10\% of their original value yields a change of
$\frac{|\Delta\delta|}{|\delta|} \la 7\%$ on the average ($\theta$ changes by
$\Delta \theta \la 0.1$ for galaxies without emission lines, that is types E0
to Sa, with $|\theta| \la 4^\circ$; $\Delta \theta \sim 3$ for types Sc and
Sm/Im, with $|\theta| \ga 4^\circ$). 

\section{PCA applied to the ESS sample}

In this section we apply the PCA to the ESS sample of 347
flux-calibrated spectra described is \S 2. 
As the input spectra for the PCA must have
identical wavelength intervals and number of bins, each spectrum is rebinned
to rest frame wavelength with a step of 5 \AA/pix
which is small enough for not destroying spectral features 
and large enough for not introducing non-existent patterns in the signal. This
step is 
slightly larger than the typical steps obtained with the 3.6m and the NTT,
which are $\sim$ 3.4 and 2.3 \AA/pix, respectively (see \S 2). Because the
spectra  
were obtained with multi-object spectroscopy, the wavelength 
coverage is not the same for all the spectra in the catalogue and is a
function of the position of the slit along the dispersion direction on the
aperture mask. In order to maximize the number of spectra to be analyzed, one
must carefully select the wavelength domain. Sample 1
defined in Table \ref{lambmin_lambmax} provides a good compromise between a
wide wavelength interval and a large number of spectra (80\% of the total
number of spectra analyzed). The 
wavelength interval contains major emission and absorption features
usually present in galaxy spectra: [OII] (3727 \AA), the
H \& K CaII lines (3933 and 3968 \AA), CaI line (4227 \AA),
H$\delta$ 
(4101 \AA), the G band (4304 \AA), H$\beta$ (4863
\AA), [OIII] (4958 \AA$\;$ and 5007 \AA), and
MgI (5175 \AA). Samples 2 and 3, contain
the ``blue'' and ``red'' spectra caused by extreme positions on the aperture
masks (near the edges) sometimes combined accidentally with high or low
redshift. The PCA is applied separately to samples 1, 2 and 3. For the
10 objects which do not belong to any of samples 1, 2 or 3, we only apply 
the $\chi^2$ method described in \S 3.
For application of the PCA to the 3 samples defined in Table
\ref{lambmin_lambmax}, we normalize each spectrum by its norm as defined in
equation \ref{norm} and we use the sum of squares and cross product matrix 
method for the PCA, described in \S 3. 

\begin{center}
\begin{table}
\caption[]{Minimum and maximum rest wavelengths defining samples
1, 2 and 3. Also shown is the number of galaxies in each sample.}
\label{lambmin_lambmax}
\begin{tabular}{llll}
\hline \hline
Sample	&  $\lambda_{min}$ (\AA)  &  $\lambda_{max}$ (\AA)  & N \\ \hline
1	&  3700			  &	5250		    & 277 \\
2	&  3700			  &	4500		    & 33  \\
3	&  4500			  &	6000		    & 27  \\ \hline
other   &  			  &   			    & 10  \\ 
\hline \hline
\end{tabular}
\end{table}
\end{center}

\begin{figure}
\centerline{\vbox{\psfig{figure=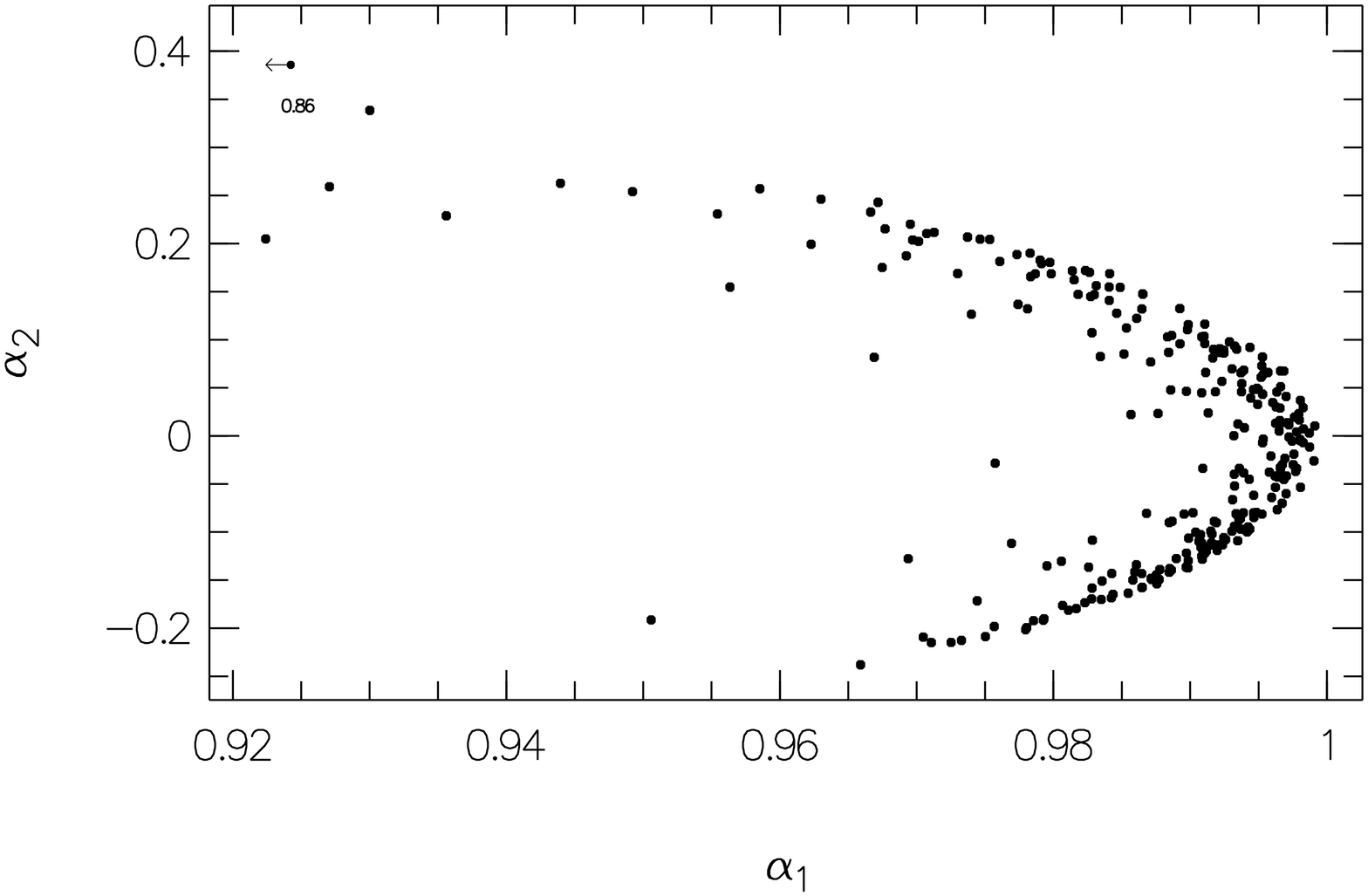,angle=-90,height=6.5cm}\psfig{figure=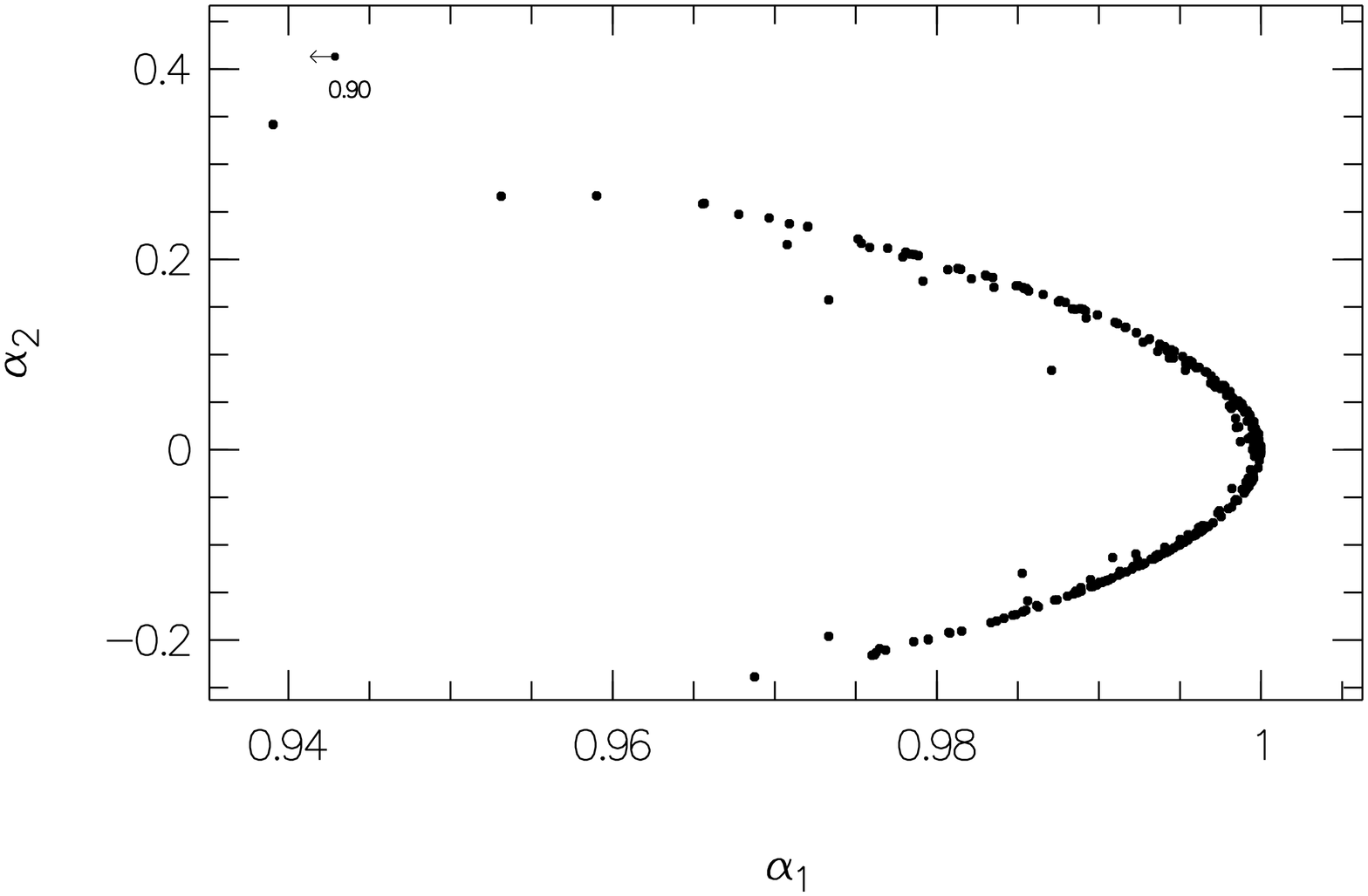,angle=-90,height=6.5cm}\psfig{figure=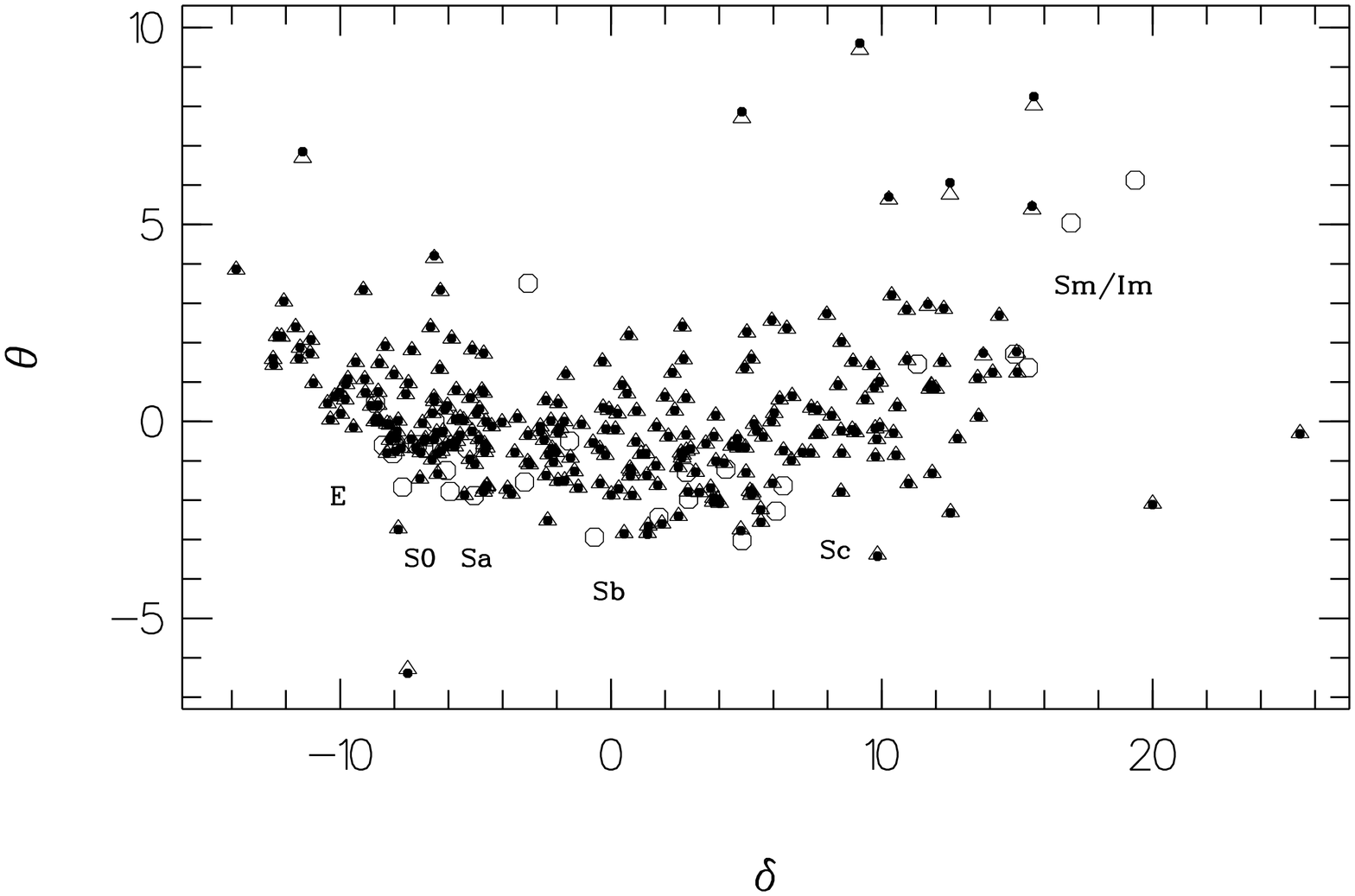,angle=-90,height=6.5cm}}}
\caption[]{Uppermost two panels: (a), projections onto the first 
2 eigenvectors for the spectra of sample 1 of the ESS (see text and 
Table \ref{lambmin_lambmax}). (b), same as upper panel,
with normalization to the first 3 projections:
$\sqrt{\alpha_1^2 + \alpha_2^2 + \alpha_3^2} = 1$. The arrow indicates an
extremely blue galaxy.
Bottom panel (c): the 277 galaxies of sample 1 in the 
classification space ($\delta$,$\theta$) 
with normalization to the first 3 projections (open triangles) and without
normalization(filled circles). Open circles map the projections of the 27
Kennicutt templates of Table \ref{kenni_gal} onto the first 3 PC's obtained
from Sample 1 and with the $\sqrt{\alpha_1^2 + \alpha_2^2 + \alpha_3^2} = 1$
normalization.} 
\label{pca_reg1}
\end{figure}

The main PCA analysis of the ESS data is performed on sample 1. 
The redshift distribution for this sample has the same shape as
for the full sample of 347 objects (a Kolmogorov-Smirnov test 
shows that the two distributions have a 78.2\% probability to originate from
the same parent distribution, with a confidence level of 87.7\%).
We are therefore not introducing a redshift bias when using sample 1. 
Figure \ref{pca_reg1}a shows the projections of the 277 spectra of sample 1
onto the first 2 PC's derived from that sample. The galaxy marked with an
arrow is an extremely blue galaxy. Figure \ref{pca_reg1}b shows the projections
after normalizing to the first 3 projections ($\sqrt{\alpha_1^2 + \alpha_2^2
+ \alpha_3^2} = 1$). Although the normalization to 
the first 3 PC's
artificially decreases the scatter in the ($\alpha_1$,$\alpha_2$) sequence, 
this normalization changes the position of the points in the
($\delta$,$\theta$) plane by a very small amount: $\frac{\Delta
\delta}{\delta} \sim 0.0003$ and $\frac{\Delta \theta}{\theta} \sim 0.002$
(see Figure \ref{pca_reg1}c). 

\begin{figure}
\centerline{\hbox{\psfig{figure=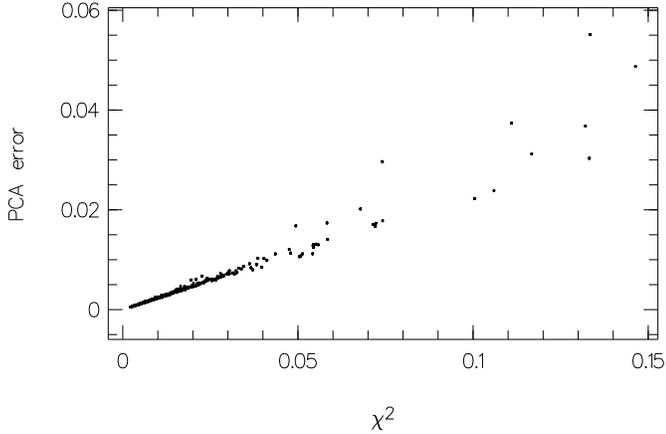,angle=-90,height=6.5cm}}}
\caption[]{The relationship between the error in the
PCA (defined by equation \ref{pca_error2}) and the $\chi^2$ value between the
input spectrum and its reconstruction using the first 3 PC's.}
\label{pca_error_sample1}
\end{figure}

Justification for adopting the $\sqrt{\alpha_1^2 + \alpha_2^2 + \alpha_3^2} =
1$ normalization comes from the high
reconstruction level reached using the first 3 PC's, 
with $<\sqrt{\alpha_1^2 + \alpha_2^2 + \alpha_3^2}> \sim$ 98\% for sample 1.
The distribution of errors we make with this approximation for sample
1 is shown in
Figure \ref{pca_error_sample1}, where we plot the $\chi^2$ value
between the original and reconstructed spectra (using 3 PCs) and the error in
the reconstruction defined as 
\begin{equation}
\label{pca_error2}
PCA_{error} = 1 \; - \; \sqrt{\alpha_1^2 + \alpha_2^2 + \alpha_3^2}.
\end{equation}

It can be shown analytically that the following relation exists:
\begin{equation}
\label{pca_error3}
PCA_{error}(n) \simeq \; B \; + \; A \;\chi^2_n, 
\end{equation}
with $A$, $B$ constants $\ge 0$. The label $n$ denotes the spectrum $n$.
The linear
relation between $PCA_{error}$ and $\chi^2$ is clearly visible in Figure
6. This result also confirms the reliability of the PCA reconstruction and
associated error $PCA_{error}$. 

Figures \ref{pca_reg1}a,b show that the spectral sequence for the ESS galaxies
is similar to the sequence found by Connolly \etal (1995) for the 10 Kinney
nearby galaxies. The spectral sequence of Figure \ref{pca_reg1}c is also
similar to that shown in Figure \ref{delta_theta_type} (left), for the
Kennicutt spectra (see Table \ref{kenni_gal}). Note that we verified (as in
\S 4) that the inclusion or 
rejection of the emission lines mainly affects the $\theta$ parameter, and
therefore the presence of emission lines does not significantly affect the
results of the spectral classification. 

To compare the ($\delta$,$\theta$) sequence for sample 1 with that for the Kennicutt 
templates, we project the spectra of Table \ref{kenni_gal} onto the PC's
obtained by the application of the  
PCA to sample 1 of the ESS (open circles in Figure \ref{pca_reg1}c). The
normal Kennicutt galaxies lie along the sequence for sample 1, which shows that 
the ESS data describes the full Hubble sequence.
We also notice in the ESS sample the
existence of galaxies {\em redder} and {\em bluer} than the
earliest and latest normal Kennicutt Hubble types, respectively. These and other
cases will be discussed in \S 6. Application of the PCA to samples 2 and 3
yields a fairly well  
defined sequence in the ($\delta$,$\theta$) plane, similar to that for sample
1. However, 
the absence of the 4000 \AA$\;$ break within the wavelength interval for
sample 3 yields a larger 
scatter in ($\delta$,$\theta$). 

\begin{figure*}
\centerline{\hbox{\psfig{figure=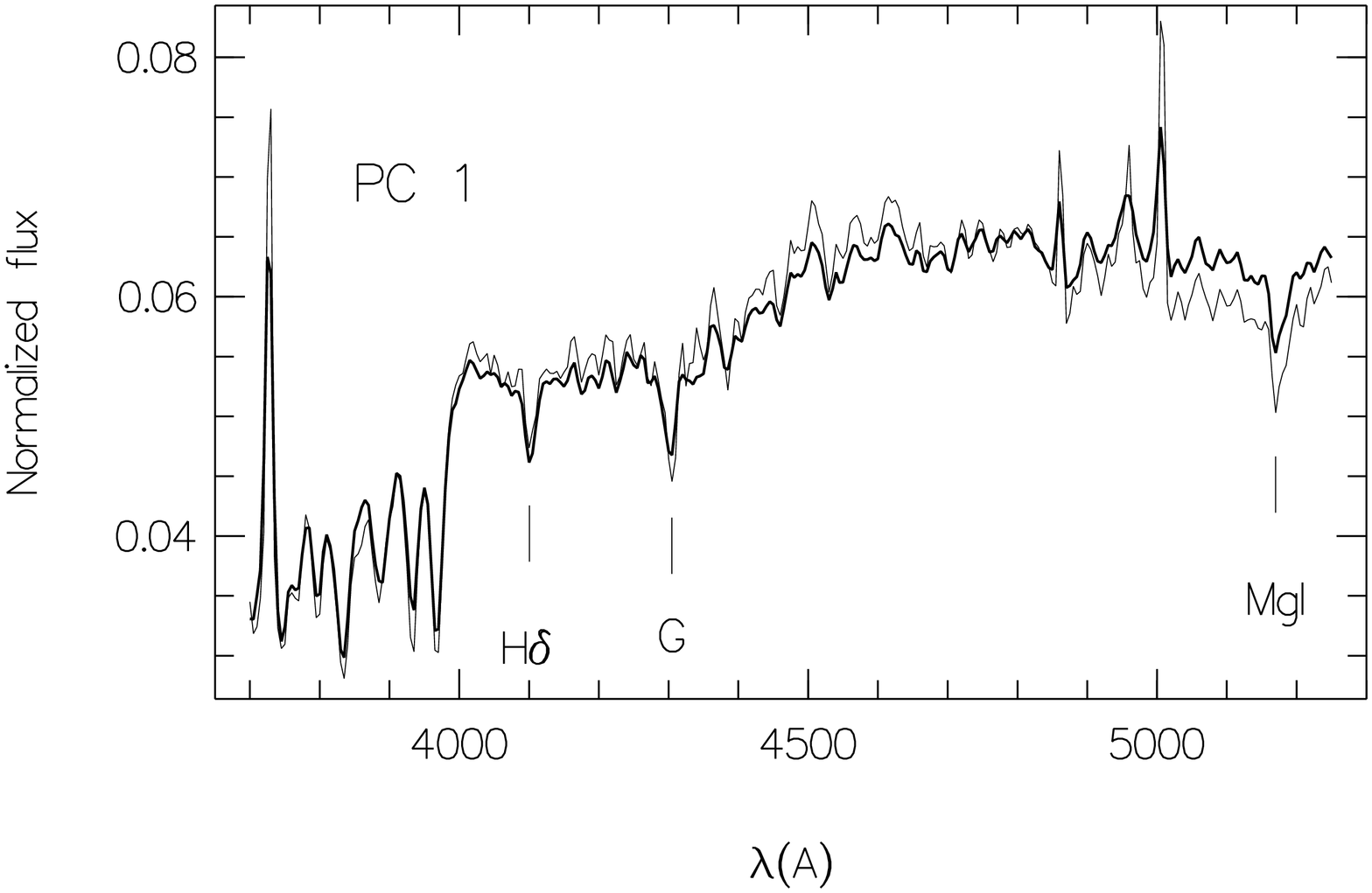,angle=-90,height=7.0cm}\psfig{figure=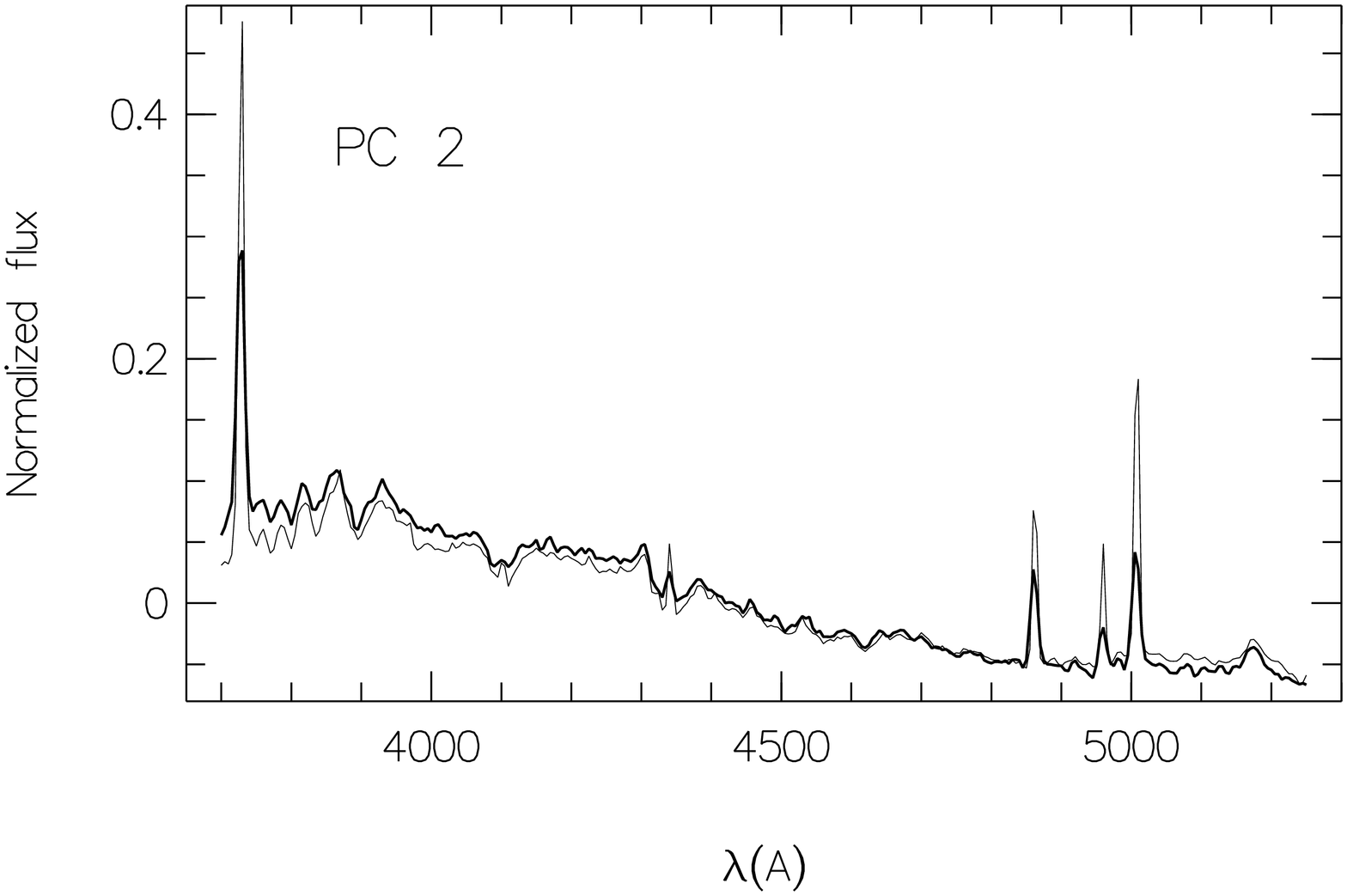,angle=-90,height=7.0cm}}}
\centerline{\hbox{\psfig{figure=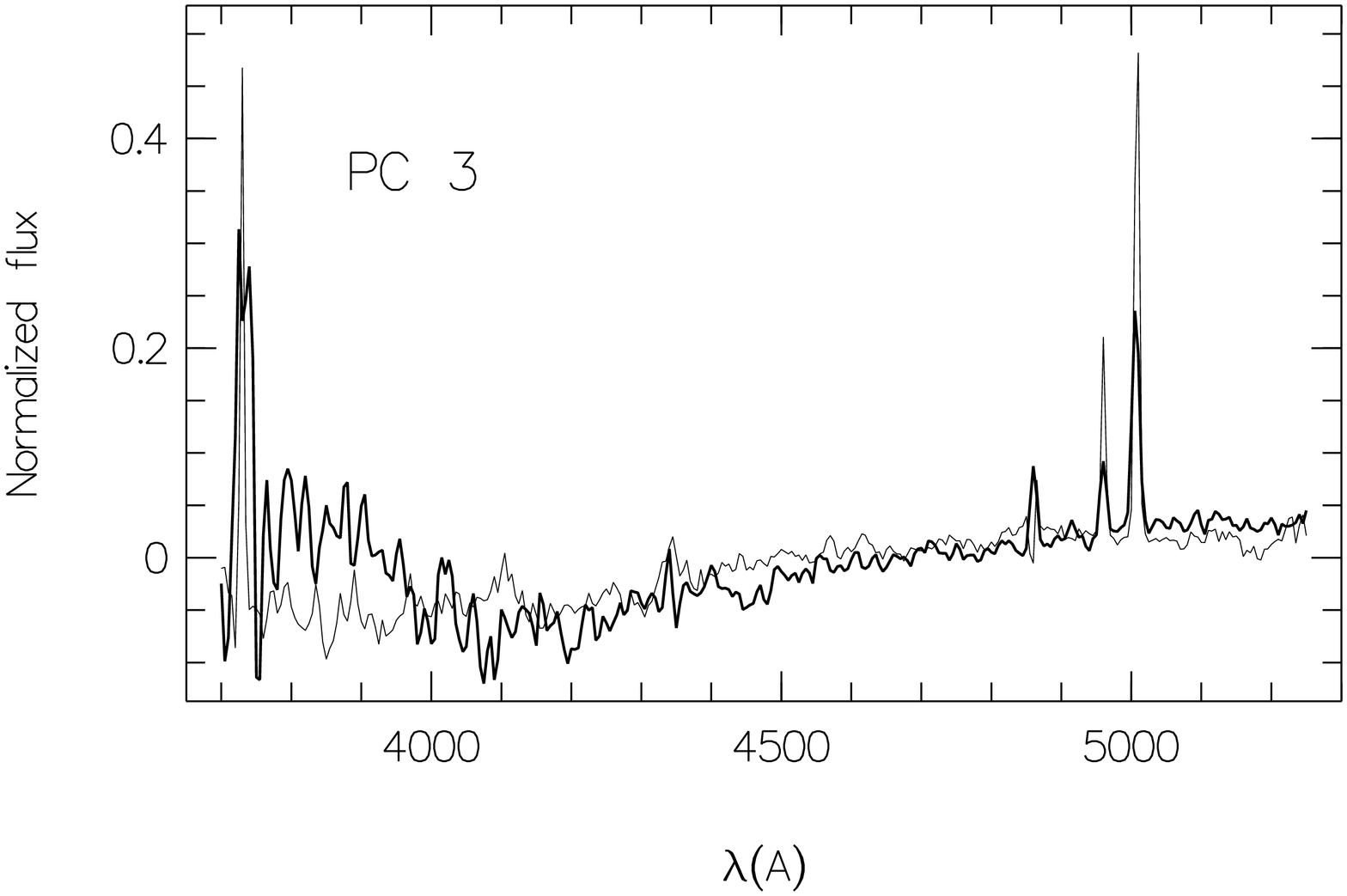,angle=-90,height=7.0cm}\psfig{figure=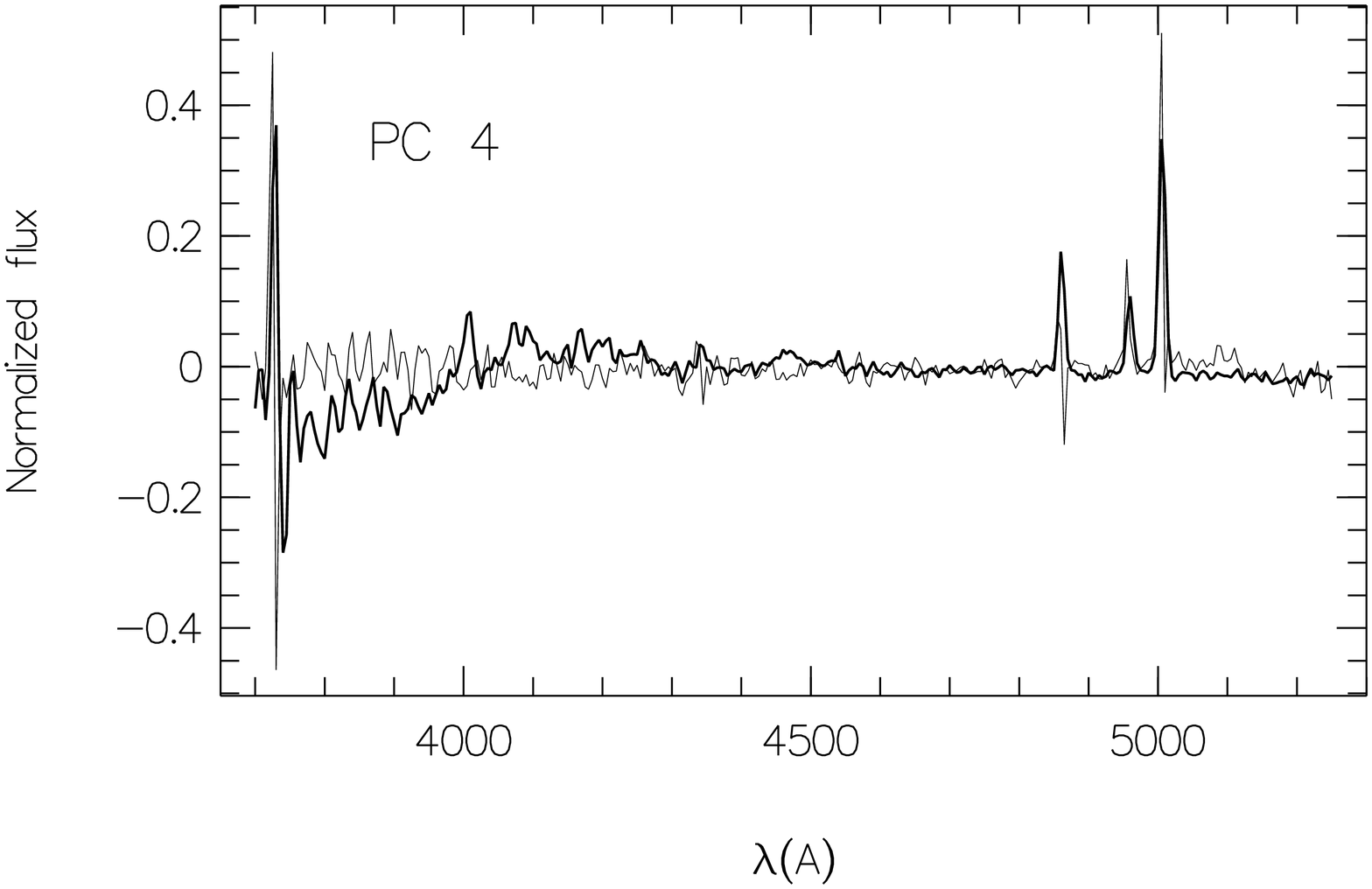,angle=-90,height=7.0cm}}}
\caption[]{First four eigenvectors for the PCA applied to sample 1 of the ESS
(bold lines) and for the selected galaxy sample from Kennicutt (thin
lines). The spectral range for both samples, when applying the PCA, is $3700
\le \lambda \le 5250$ \AA.}
\label{pca_reg1_vec}
\end{figure*}

Figure \ref{pca_reg1_vec} shows the first 4 PC's obtained for sample 1 (bold
lines). The first PC of sample 1 is
characterized by 
the CaII K and H absorption lines near 4000 \AA, a pronounced continuum break and
by the other 
absorption features typical of early-type galaxies. The first 4 PC's also
contain [OII], H$\beta$ and both [OIII] emission 
lines. The projections onto the first 4 PC's for sample 1 satisfy 
$<|\alpha_1|> = 0.97$, $<|\alpha_2|> = 0.10$, 
$<|\alpha_3|> = 0.023$, $<|\alpha_4|>
= 0.021$. In addition, $<|\alpha_5|> = 0.016$ and PC's of higher 
order contribute less than 1\% of the total flux.
Figure 7 also shows the PC's derived by application of the PCA onto the Kennicutt 
spectra (thin lines), using a wavelength interval restricted to the
same spectral range as for sample 1 (from 3700 $\le \lambda \le$ 5250 \AA).
The resemblance of the first 4 PC's for both
samples is striking, which shows that the galaxy
population in sample 1 has similar spectral properties to the
sample of normal, local galaxies selected by Kennicutt. One must however be careful
in comparing both samples, because the number of objects 
differ by a factor of $\sim 10$ 
and properties such as the blue and red continuum and the strength of emission
lines in the first PC's depend on the frequency of the particular Hubble
types. In this respect, although the selected Kennicutt 
spectra are representative of the spectral features observed in normal
galaxies, the population fractions are {\em not} representative of the local
universe. This could be responsible for the slight differences in the PC's 
between sample 1 and the Kennicutt sample: the 1$^{st}$ and the 2$^{nd}$
PC of the sample 1 are redder and bluer than the 1$^{st}$ and the 2$^{nd}$
PC's of the Kennicutt sample, respectively; and the strengths of the emission
lines for the 4 PC's differ.  

In conclusion, 
according to the PCA technique, the ESS galaxies with $z = 0.1-0.5$ have
similar spectral properties in the range $\lambda = 3700-5250$ \AA$\;$ to the
Kennicutt sample of galaxies with $z \la 0.025$, which supports a close
resemblance of the fractions of stellar populations between the two samples.

\section{Analysis}

\subsection{Classifying the galaxies}

In this paragraph we perform the spectral classification of the ESS galaxies,
using the $\delta$ sequence. Our goal is not an indirect morphological
classification using the spectral classification. The objective is to
establish a link between the
spectral classification and the known Hubble sequence in
order (1) to test the reliability of the PCA classification by comparison with
the $\chi^2$ technique, and (2) to compare the ESS classification with that for
other redshift surveys. 
The major assumption is that the spectral
trends for the observed galaxies have the
same nature as the spectral trends followed by the Kennicutt sequence, in
which we know {\em a priori} the morphological type. In
order to assign discrete types for comparison with the Hubble sequence, we
define a type-$\delta$ relationship using
the Kennicutt templates. 

\begin{figure}
\centerline{\hbox{\psfig{figure=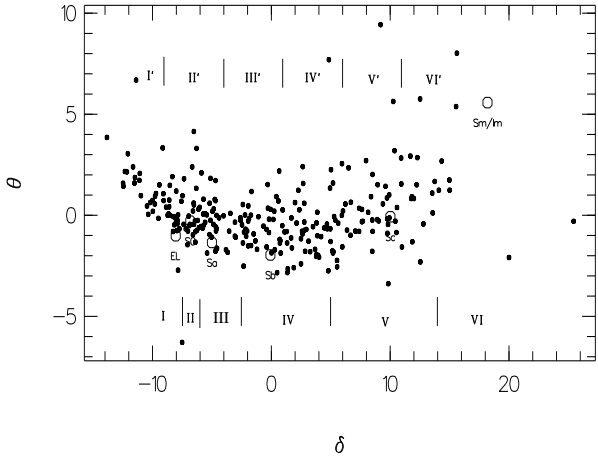,angle=-90,height=7.0cm}}}
\caption[]{Position of galaxies from sample 1 in the 
($\delta$,$\theta$) plane ($\bullet$), and of the 6
averaged Kennicutt templates ($\circ$) projected onto the PC's of
Sample 1. The spectral sequence is binned in two different ways: a variable
binning 
in $\delta$ (marked as I, II, etc...) which follows the position of the
Kennicutt templates, and a uniform binning in $\delta$ (marked as I', II',
etc...). Vertical  lines indicate the boundaries of the $\delta$ classes.}
\label{reg123_bin}
\end{figure}

Figure \ref{reg123_bin} shows the galaxies of sample 1 in the
($\delta$,$\theta$) plane (dots), and 6 galaxies of known Hubble type, which
are the average of several of the Kennicutt templates from Table
\ref{kenni_gal} with the same type (open circles). The EL type is the average
of galaxies \# 1, 2, 3 and 4 of Table \ref{kenni_gal}. The other types are S0 
(average over objects \# 5, 7 and
8), Sa (\# 9, 10, and 12), Sb (\# 14, 15 and 
26), Sc (\# 16, 17, 20, 21, 22, and 23) and Sm/Im (\# 24 and
25). These average spectra are then projected onto the PC's derived from the
ESS sample 1. The different averaged Kennicutt spectra 
in Figure \ref{reg123_bin}
are not equally separated in the ($\delta$,$\theta$) plane. It was already
known that 
spiral galaxies have larger differences in their spectra than ellipticals
(\cite{morgan57}, \cite{wyse93}). Figure \ref{reg123_bin} provides 
a quantitative demonstration of this effect: the change by one morphological
type for early-type galaxies in the averaged Kennicutt galaxies, 
corresponds to a small variation in $\delta$, compared to the late-types. In
Figure 8, the density of objects $\rho(\delta)$ is 
significantly higher for 
low values of $\delta$ than for high values of $\delta$: $\sim 15$ gal/deg for 
$\delta \sim [-10,-2.5]$ and $\sim 8$ gal/deg for $\delta \sim [-2.5,8.0]$. 
The large $\delta$ distances between the Sb and Sc and between 
the between the Sc and the Sm/Im leave space for the intermediate types
Sbc, Scd, etc..., and for the Sd type, respectively, for which there are no
templates in the Kennicutt's sample.

The observed position 
of the average Kennicutt templates along the $\delta$ axis, provides 
a natural binning for correlating the spectral sequence and the Hubble type,
which we define by regions I, II, III, IV, V and VI, separated by vertical
lines in Figure \ref{reg123_bin}. The boundaries of these regions are the
averaged $\delta$ value between 2 adjacent average Kennicutt
templates. This variable binning accounts for the {\em varying} density
$\rho(\delta)$ which, as mentioned above, is inherent
to the frequency of the spectral properties of the ESS galaxies.
We also define a uniform binning in $\delta$ denoted I', II', III',
IV', V', and VI'. 
The length for
each bin in this case is the total span in $\delta$ divided by the number of
morphological/spectral types.
Table \ref{results_pca} shows the bin values in $\delta$ for the uniform and
non-uniform binning. Also shown in Table \ref{results_pca} are the number of
galaxies per bin and the corresponding fractions from the total of 310 galaxies
for the combination of Samples 1 and 2. 

A bootstrap test shows that the r.m.s. deviation
in the number of galaxies within each class in Table \ref{results_pca} is
$\sim 2$, that is 0.7\% in the fractional number per type given in Table
\ref{results_pca}. The largest source of error comes from the uncertainties
in the flux calibrations of the spectra. The 7\% external errors in the
calibrations curves induce a 5\% error in the number of galaxies per spectral
class. Using bootstrap experiments we also derive an r.m.s. deviation in
$\delta$ of 0.22$^\circ$, and an r.m.s. uncertainty in $\delta$ varying from
0.2$^\circ$ at $\delta = -10^\circ$, to 2.3$^\circ$ at $\delta = 20^\circ$.

\begin{figure}
\centerline{\hbox{\psfig{figure=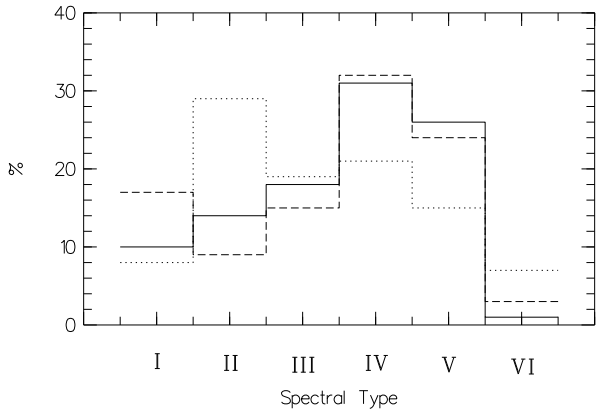,angle=-90,height=6.0cm}}}
\caption[]{Histogram showing the distribution of spectral types derived by the
$\chi^2$ method (solid line, over Samples 1, 2 and 3) and the PCA (Samples 1
and 2 included). Dashed and dotted lines
indicate the results using a non-uniform and uniform binning in
$\delta$, respectively (see Tables \ref{results_pca} and \ref{results_chi2}).}
\label{hist_12_types}
\end{figure}

Table \ref{results_chi2} shows the results of the $\chi^2$ (see \S 3) 
applied to the galaxies of Samples 1, 2 and 3
(347 in total) with the 6 averaged Kennicutt galaxies as
templates. Considering the discrete nature of 
the $\chi^2$ method, it is important to establish an indicator of the error
which we make in the association of a type. We define it as the fraction of
total galaxies for each type for 
which the second closest template has a $\chi^2$ value differing by less than 
20\% from the $\chi^2$ with the closest template (column 3 of Table
\ref{results_chi2}). We choose 20\% as a conservative
threshold: given the 7\% uncertainty in
the flux calibration, we consider that cumulative squared
differences less 
than 20\% between a spectrum and 2 templates is not significant.
\begin{table*}
\caption[]{PCA classification for samples 1 and 2. N(PCA) indicates the
numbers and fractions of galaxies for the defined spectral types, using the
position of the 
spectra along the $\delta$ axis (see Figure \ref{reg123_bin}). We show 
the results for uniform and variable bins in $\delta$. The variable
binning is obtained from the projected position of the averaged Kennicutt
templates in Figure 8 (see text).}
\label{results_pca}
\begin{tabular}{lllclll}
\hline \hline
\multicolumn{3}{c}{Uniform Binning$^{(a)}$} & 
\multicolumn{4}{c}{Variable Binning$^{(a)}$} \\ \hline
Spectral Type & $\delta$ range & N(PCA) & Spectral Type & Kenn. Temp$^{(b)}$ &
$\delta$ range & N(PCA) \\ \hline
I'   &  	]$-$14,$-$9] & 27(8\%)  & I & E		  & ]$-\infty$,$-$7.5] &
52(17\%) \\
II'  &	]$-$9,$-$4]  & 91(29\%) & II & S0		  & ]$-$7.5,$-$6]      &
29(9\%) \\
III' &	]$-$4,1]     & 59(19\%)	& III & Sa		  & ]$-$6,$-$2.5]      &
48(15\%) \\
IV'  &	]1,6]	     & 64(21\%)	& IV & Sb		  & ]$-$2.5,5]	       &
97(32\%) \\
V'   &	]6,11]       & 46(15\%)	& V & Sc		  & ]5,14]	       &
75(24\%) \\
VI'  &	]11,16]	     & 23(7\%)	& VI & Sm/Im		  & ]14,$\infty$[      &
9(3\%)
\\ \hline
I'-II' & 	]$-$14,$-$4] & 118(38\%) & I-II & E/S0		  & 
]$-\infty$,$-$6]   & 81(26\%) \\
III'-V' &	]$-$4,1]     & 170(55\%) & III-V & Sa/Sb/Sc		  & ]$-$6,14]
& 220(71\%) \\ \hline
\end{tabular}
\smallskip
\\
\footnotesize
\underline{Notes:} \\
$^{(a)}$ See Figure \ref{reg123_bin} and explanations in the text. \\
$^{(b)}$ Kennicutt averaged templates used in the definition of the variable 
binning (see text).
\end{table*}
\begin{table}
\caption[]{Results of the $\chi^2$ spectral classification method over
Samples 1, 2 and 3, using averaged Kennicutt templates described in the
text.} 
\label{results_chi2}
\begin{tabular}{lll}
\hline \hline
Kenn. Temp. & N($\chi^2$)$^{(a)}$ & $\Delta_{\chi^{2}}^{(b)}$ \\
\hline
E		   &	33(10\%)	&	5\% \\
S0		   &	50(14\%)	&	8\% \\
Sa		   &	64(18\%)	&	6\% \\
Sb		   &	107(31\%)	&	0.8\% \\
Sc		   &	89(26\%)	&	0\% \\
Sm/Im		   &	4(1\%) 		& 	0\% \\ \hline
E/S0		   &	83(24\%)	&	14\% \\
Sa/Sb/Sc	   &	260(75\%)	&	7\% \\ \hline
\end{tabular}
\smallskip
\\
\footnotesize
\underline{Notes:} \\
$^{(a)}$ Number of galaxies per type, and percentage of the total of 347
galaxies. \\
$^{(b)}$ Fraction of galaxies, for each Hubble type, for
which the $\chi^{2}$ value between the first closest template and 
the second one is less than 20\%.
\end{table}
The histogram of Figure \ref{hist_12_types} shows the corresponding
distribution of galaxy types for the sum of samples 1, 2 and 3 using the
$\chi^2$ method (solid lines). The dotted and dashed lines represent the
distributions of types derived from the PCA using the uniform and non-uniform
binnings in 
$\delta$, respectively, as defined in Table \ref{results_pca}. The reader
should recall that the $\chi^2$ test is performed over the {\em whole} spectral
range covered by each ESS spectrum, and therefore the comparison between the
PCA and $\chi^2$ method also provides a test of the influence of the spectral
range on the spectral classification. For a uniform spanning of 
types in $\delta$ (dotted line) there are  
large differences with respect to the $\chi^2$ method (solid line), for
almost all types. On the other hand, the non-uniform binning based on the
averaged Kennicutt templates (dashed line) gives a good agreement between the
PCA and the $\chi^2$ results, especially for late types, thus further
demonstrating the reliability of the PCA technique in classifying galaxy
spectral types. 
Note that the $\sqrt{\alpha_1^2 + \alpha_2^2 +
\alpha_3^2} = 1$ normalization 
changes the type fractions given in Tables 4 and 5 by less than 0.1\%. 
In \S 7.2 we compare the type fractions 
of Tables 4 and 5 with the results of other major surveys.

The fact that the $\chi^2$ test over the whole  
sample of 347 galaxies, using the largest possible spectral range for each
spectrum (see Table \ref{kenni_gal}) produces nearly the same\footnote{A
Kolmogorov-Smirnov test shows that the probability that the values of column
7 of Table 4 and column 2 of Table 5 originate from the same parent
distribution is 73.2\% with a 85\% of confidence level.} fractions of spectral
types (see Figure 18) than the PCA over sample 1 restricted to the spectral
range (3700-5250  \AA), confirms that this spectral interval is wide enough
for application of both techniques.

\subsection{Filtering effect of the reconstruction and type dependence}

\begin{figure}
\centerline{\hbox{\psfig{figure=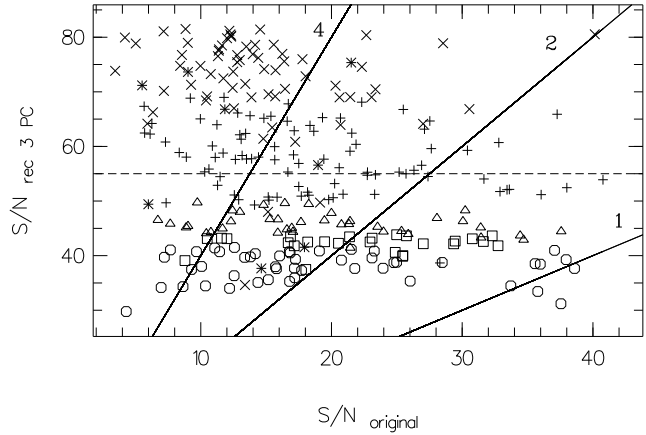,angle=-90,height=6.5cm}}}
\caption[]{S/N of the input spectra and their reconstructions using the first
3 PC's. The different symbols indicate spectral type: I($\circ$), II($\Box$),
III($\triangle$), IV(+), V($\times$), VI($\ast$). Lines indicate the boundary
of a gain in S/N equal to factors of 4, 2 and no gain. The dashed line indicate the S/N
of the first PC.} 
\label{gain_sample1}
\end{figure}

We now illustrate the filtering capability of the PCA using the ESS sample. 
Figure \ref{gain_sample1} shows the S/N of the reconstructed
spectra of sample 1 using 3 PC's, as a function of the original S/N, for 
the different spectral types. The filtering effect of reconstructing the
spectra with 3 PC's is striking. Whereas the S/N of the input spectra range
from 6 to 40, the reconstructed spectra have S/N between 35 and 80. The gain
in S/N 
is strongly dependent on the spectral type. For late types, the increase in
S/N has the largest values, which reaches more than a factor of 4 for
70\% of the types V and VI. This is due to the low S/N ratio in the 
weak continuum of the original spectra, which allows a relatively large
improvement in S/N.  
For most of the early types (I to II), the
gain in S/N is larger than 1.5 times, and can be as high as
a factor 5. The range of variation for the S/N of the reconstructed spectra is
related to the S/N of the PC's. The PC's have an intrinsic level of noise,
and there is a minimum and maximum S/N achieved with the permitted linear
combinations of the first 3 PC's for the defined spectral sequence. In Figure
\ref{gain_sample1}, the dashed line indicates the S/N of the first PC
(55). 

We conclude that the noise carried by the original spectra can be reduced to an
interval of well known S/N values, if one uses the reconstructed spectra. 
The S/N in the original spectra is a function of the apparent magnitude of the
objects and the observing conditions, whereas the S/N in the reconstructed 
spectra depends only on the systematic and statistically significant spectral
characteristics of the objects. Of course, in the limiting case when the
noise is so high as to hide all the spectral features, the PCA error
(equation 7) is large.
With data of sufficient S/N ratio, the possibility of
reconstruction allows to transform the original sample of spectra into a
sample with a reduced noise level. The filtered spectra can then be used for
follow-up study of 
the various galaxy populations, and for comparison of 
the spectral features with models of spectro-photometric evolution. 
Note however that the details of the line properties, like equivalent
width or line strength, are not linear and cannot be described in 
detail by the linear PCA approach.
 
\subsection{The emission line galaxies}

\begin{figure}
\centerline{\hbox{\psfig{figure=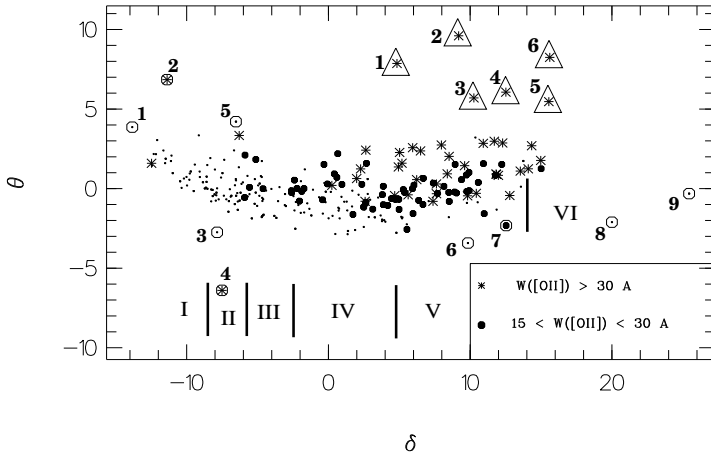,angle=-90,height=7.0cm}}}
\caption[]{Galaxies of sample 1 (277 galaxies) with W([OII]) $<$ 15  \AA$\;$ (dots), 
with 15 $\leq$W([OII])$\leq$ 30  \AA$\;$ (filled circles), and with
W([OII])$\geq$30  \AA$\;$ (stars). Open circles indicate the peculiar galaxies
discussed in \S 6.6, ordered following Table 8. Triangles denote emission
line galaxies discussed in \S 6.3.}
\label{reg12_Oxg}
\end{figure}

We now examine the properties of the [OII] emission line, present in 
$\sim 55\%$ of the ESS spectra, in
relation to the PCA classification. Figure \ref{reg12_Oxg} shows 
the ($\delta$,$\theta$) values for the galaxies of
sample 1, and indicates the points with measured equivalent widths (W,
hereafter) of [OII] 
(3727  \AA) satisfying W([OII]) $\geq$ 30  \AA $\;$ (asterisks), and 15  \AA $\le$
W([OII]) $<$ 30  \AA$\;$ (filled dots). 

\begin{table*}
\caption[]{Information on the emission-line galaxies$^{(a)}$.} 
\label{emilines}
\begin{center}
\begin{tabular}{lrrrrrr}
\hline \hline
Spectral type$^{(b)}$	& N   & Fraction & $\overline{z}$ & $\sigma_{z}$ &
$\overline{\theta}$ &  $\sigma_{\theta}$ \\ \hline
\multicolumn{7}{c}{15 $\leq$ W([OII]) $\leq$ 30  \AA} \\
\hline
I(E)		& 0   & 0\%	   &       &      &       &  \\
II(S0)		& 0   & 0\%	   &       &      &       &   \\
III(Sa)		& 7   & 18\%	   & 0.33  & 0.13 &  0.44 & 1.0\\
IV(Sb)		& 27  & 31\%     & 0.32    & 0.10 &  $-$0.17 & 0.97 \\
V-VI (Sc/Sm/Im)	& 29  & 38\%     & 0.28    & 0.10 &  $-$0.044 & 1.0 \\
\hline
\hline
\multicolumn{7}{c}{W([OII]) $\ge$ 30  \AA} \\ \hline
I(E)		& 3   & 6\%	 & 0.34    & 0.13 & 0.3 & 0.1 \\
II(S0)		& 1   & 4\%	 & 0.31    & 0.0  & 3.3 & 0.0   \\
III(Sa)		& 0   & 0\%	 &         &      &     &    \\
IV(Sb)		& 8   & 9\%      & 0.34    & 0.14 & 1.6 & 2.7 \\
V-VI (Sc/Sm/Im)	& 31  & 41\%     & 0.23    & 0.10 & 2.2 & 2.5 \\ \hline
\end{tabular}
\smallskip
\\ 
\end{center}
\footnotesize
\underline{Notes:} \\
$^{(a)}$ For sample 1, 277 galaxies, 89\% of the total number of galaxies analyzed. \\
$^{(b)}$ Using the non-uniform binning (see Table \ref{results_pca}).
\end{table*}

\begin{figure}
\centerline{\hbox{\psfig{figure=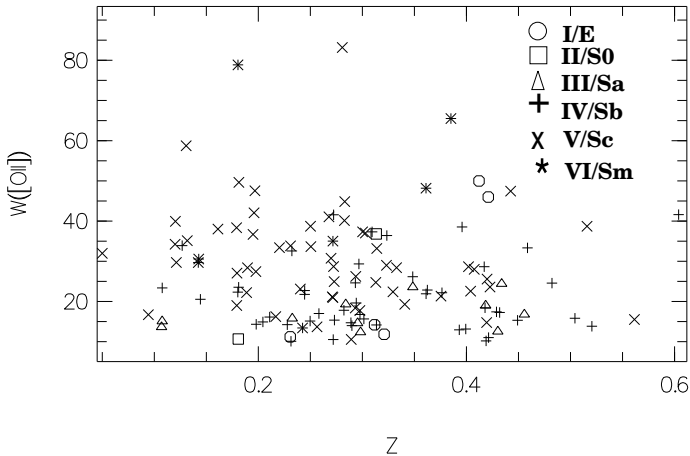,angle=-90,height=7.0cm}}}
\caption[]{Equivalent widths of [OII] for galaxies with W([OII])
$\ga$ 10 \AA, as a function of redshift and spectral type.}
\label{emi_types}
\end{figure}

In Figure \ref{reg12_Oxg}, most of the galaxies with $\delta \ga 5^\circ$ (types
V/Sc and later), have W([OII]) 
$\ge$ 15 \AA. There are only 4 galaxies with W([OII]) $\ge$ 30 \AA$\;$ and types
I/E or II/S0. Table \ref{emilines} shows the number and fraction of galaxies
with 15 $\leq$ W([OII]) $\leq$ 30  \AA$\;$ and W([OII]) $\ge 30$ \AA$\;$ for
different  
spectral types. We give the mean redshift and the mean value
of $\theta$ for each sub-sample and the corresponding standard deviations.
Table \ref{emilines} first displays the well known trend between
the spectral type and the frequency of strong emission lines: the later the
type, the larger the fraction with strong emission lines, and the stronger
the emission lines. For a given spectral class (IV, V, VI), 
$\theta$ is systematically larger for galaxies with larger W([OII]) (see Table
\ref{emilines}). This confirms the relationship between $\theta$ ({\it
i.e.\/,} the power of the third eigenvector for each spectrum), and the
strength of emission lines.
Figure \ref{emi_types} shows that within each subsample of Table \ref{emilines},
the objects span most of the redshift range for the ESS ($0.1 \la z \la
0.5$).

The equivalent width of [OII] allows us to examine the possibility of galaxy
evolution using the average value of W([OII]), as a 
function of redshift: W([OII]) is a direct measure of the degree
of star formation in a galaxy (see for example \cite{osterbrock89} and
references therein),  
because this radiation is produced by the interstellar medium (ISM) which is
excited by the ultraviolet (UV) 
radiation from hot stars. Several scenarios of galaxy evolution predict
an increase in the star formation rate with increasing look back
times. Observational evidence is provided by the
excess of blue galaxies in deep number counts (\cite{couch87},
\cite{colless93}, \cite{metcalfe95} and  
references therein) as well as the increasing
density of emission line galaxies in deep redshift surveys
(\cite{broadhurst88}). The Butcher-Oemler effect shows signs of recent
evolution in clusters of galaxies (at $z \sim 0.2$) which can be partly
interpreted in terms of increased 
star formation (\cite{butcher78}, \cite{dressler83}, \cite{lavery88}). One of
the current issues is whether 
analogous effects occur in the field and at which redshift. 
\cite{hammer97} show that the fraction of bright emission-line galaxies in
the field gradually increases with redshift, from 34\% to 75\% in the
redshift range $0.45 - 0.85$. Their [OII] luminosity
density of field galaxies 
increases only weakly from $z = 0$ to $z = 0.4$ (by a
factor 1.6), and by a large factor (8.4) between $z = 0.4$ and $z = 0.85$. 
Similar results are found from other data samples, using different selection
criteria and/or different spectroscopic techniques (\cite{heyl96}). 
In the ESS sample analyzed here, we do not detect any significant  
evidence for an increase of W([OII]) with redshift up to $z \sim 0.5$ (see
Figure 15). This result agrees with those given by \cite{hammer97} and by
Heyl \etal (1996).

We also note the presence of [OII] in several early type galaxies of the ESS
sample. In particular, there are 4 galaxies with types I-II/E-S0 which have
W([OII]) $> 30$ \AA$\;$ (with $\delta < 0$ and marked with an asterisk in 
Figure 11). The nature of this emission is not fully clear. 
However, some agreement exists (see for example \cite{dorman97}) to point out 
that very hot post-AGB stars present in such galaxies could be the source 
responsible for the [OII] emission, probably also related to the so-called 
Ultraviolet Upturn Phenomenon (``UVX''); NGC 1399 a well-known example
of this effect (\cite{dorman97}). It also was suggested that the environment
could play an important role in the ``UVX'' phenomenon (\cite{ellis93}). Note
that 2 of 4 early-types galaxies with emission lines in the ESS sample
belong to the same group of galaxies (at $z = 0.41$) and the other 2 have
$z = 0.19$ and 0.31.  

\begin{figure*}
\centerline{\hbox{\psfig{figure=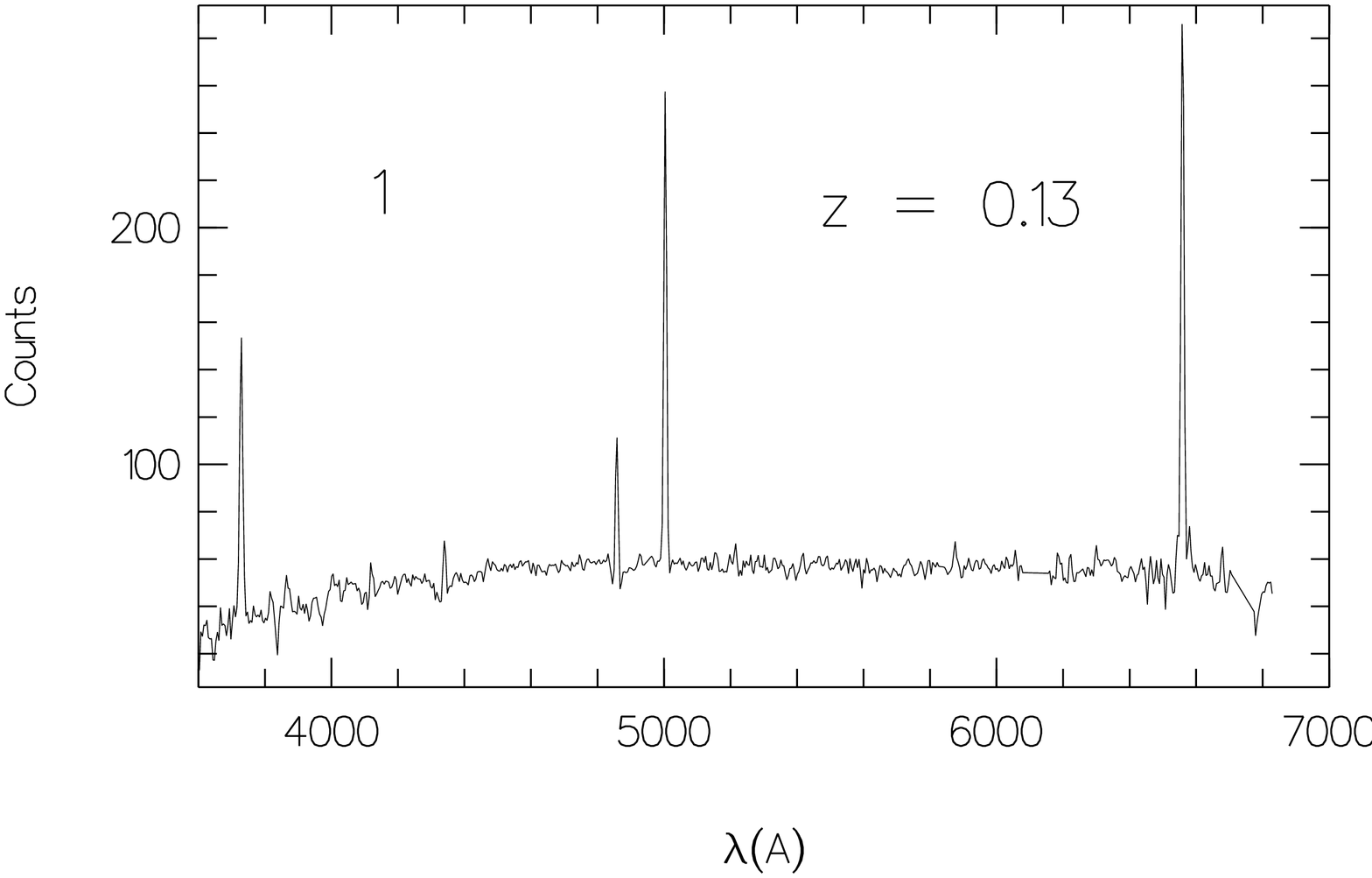,angle=-90,height=4.5cm}\psfig{figure=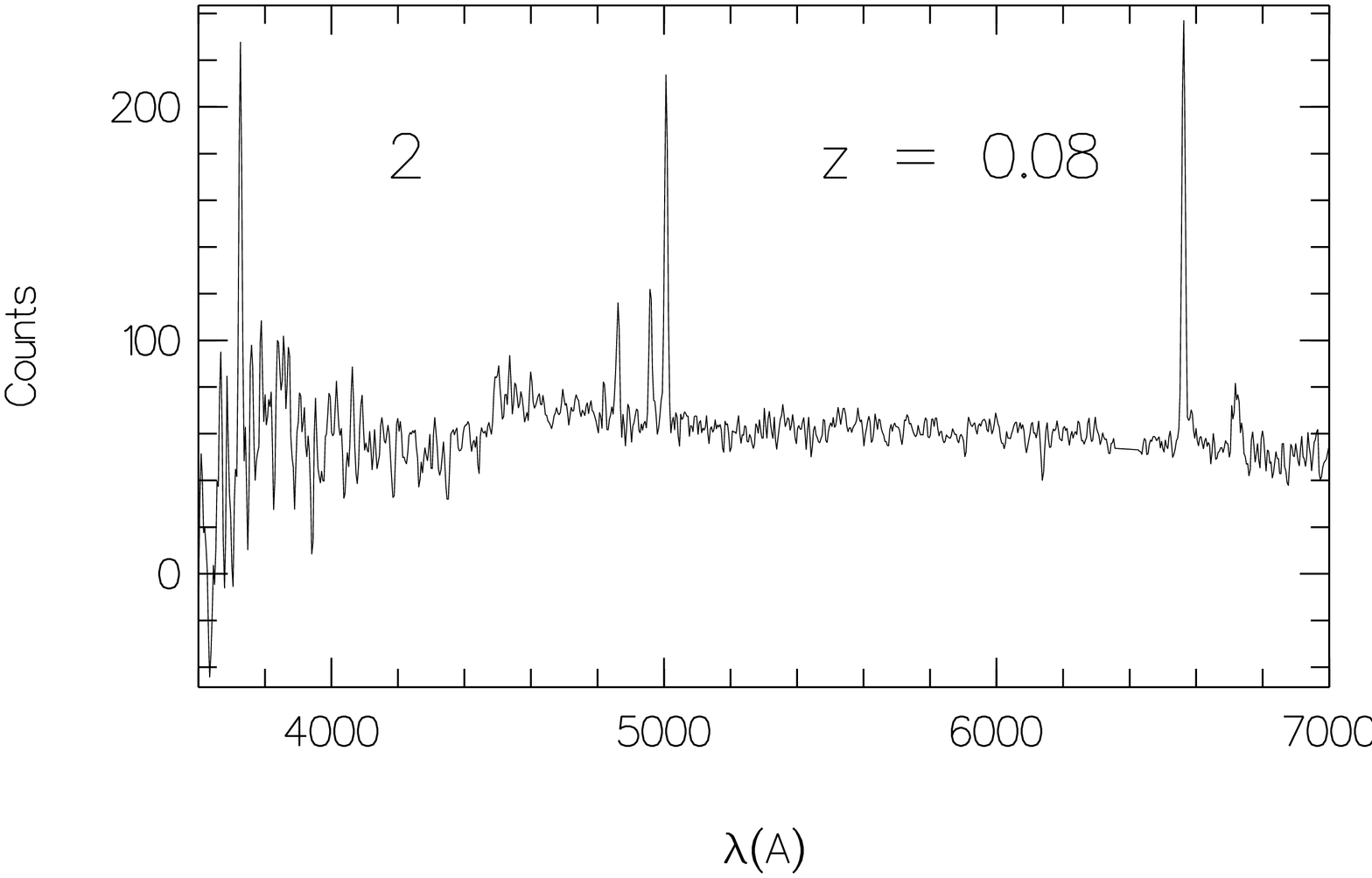,angle=-90,height=4.5cm}\psfig{figure=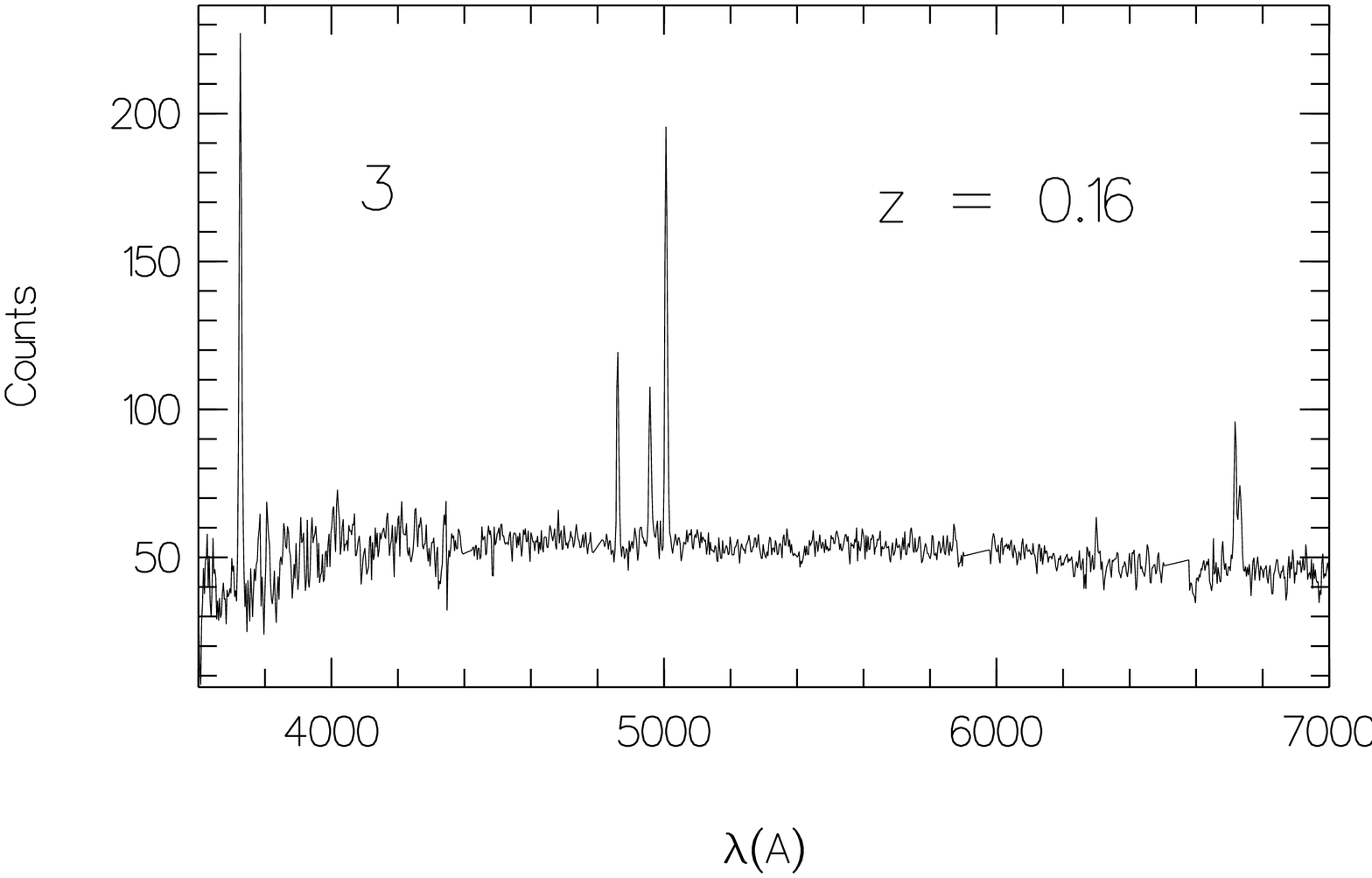,angle=-90,height=4.5cm}}}
\centerline{\hbox{\psfig{figure=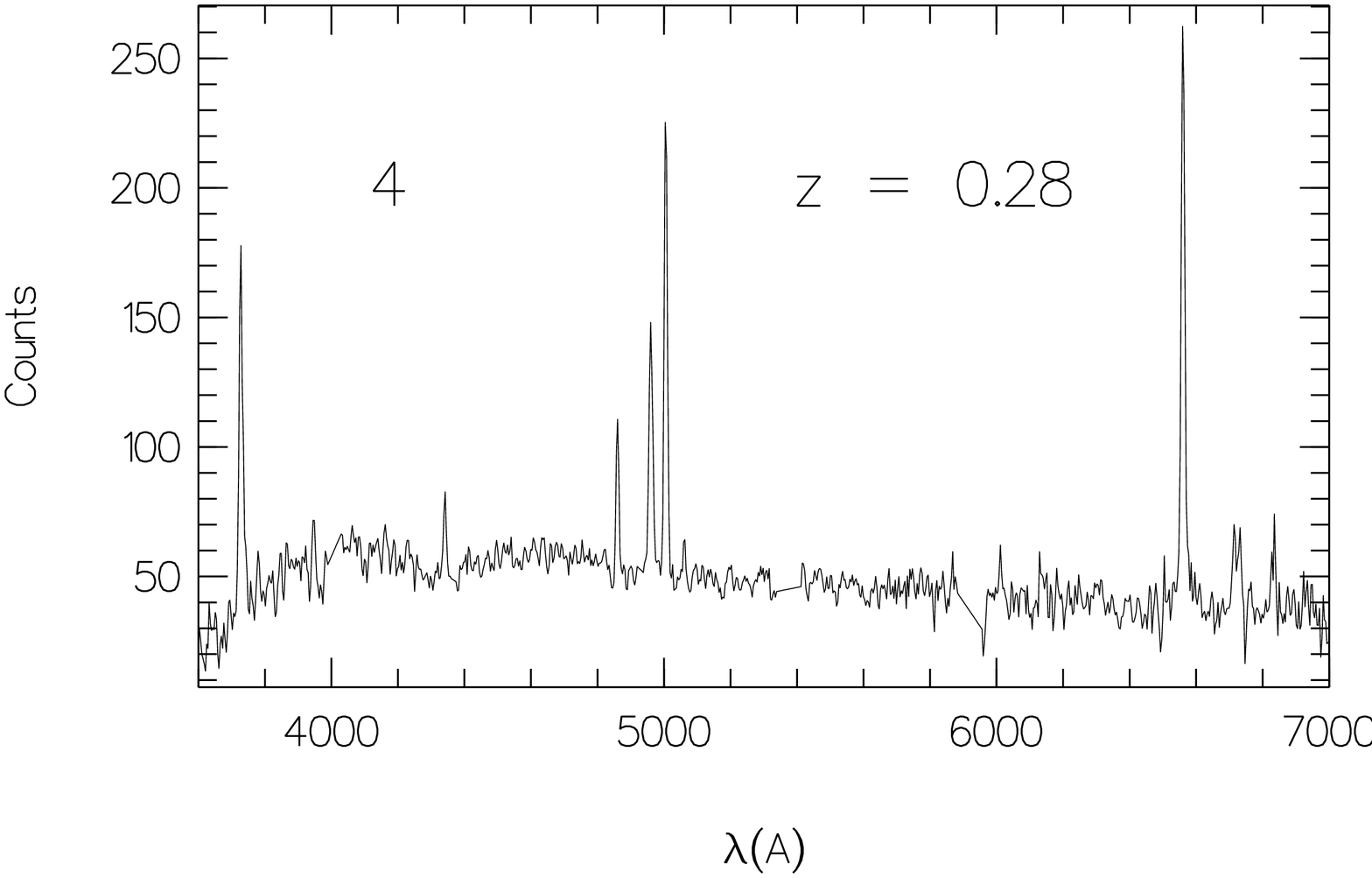,angle=-90,height=4.5cm}\psfig{figure=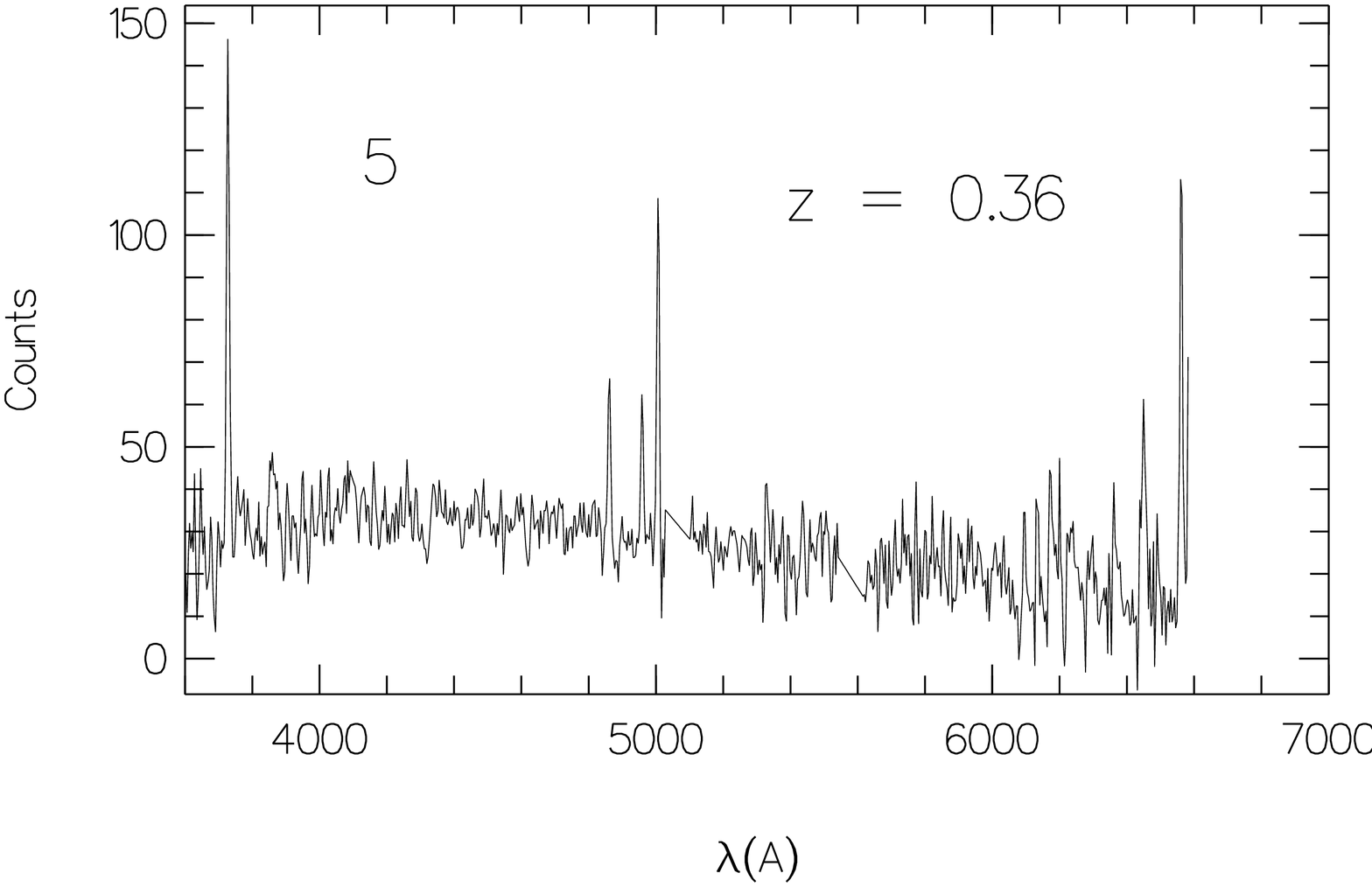,angle=-90,height=4.5cm}\psfig{figure=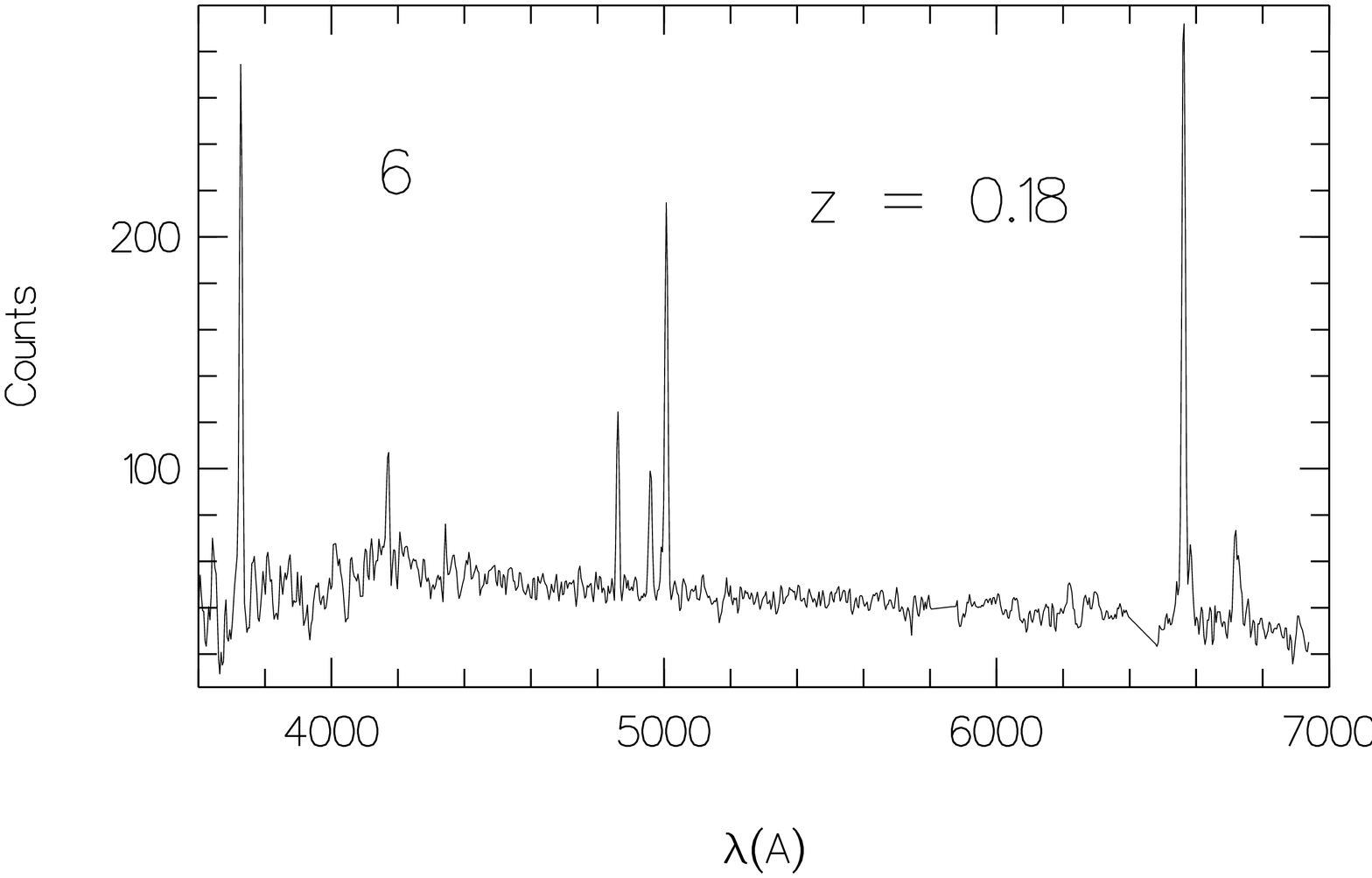,angle=-90,height=4.5cm}}}
\caption[]{Spectra of the galaxies in the ESS Sample 1 (at rest-wavelength) with
$\delta \ga 4^\circ$ and $\theta \ge 5^\circ$. The galaxies are marked as
open triangles in Figure 11 and have W([OII]) $\ge 30$ \AA. For each object,
the spectral type and W([OII]) in \AA$\;$are: \#1, IV/Sb, 52; \#2, V/Sc, 44;
\#3, V/Sc, 38; \#4, V/Sc, 50; \#5, VI/Sm-Im, 46; \#6, VI/Sm-Im, 51.}
\label{reg12_emi}
\end{figure*}

Figure \ref{reg12_emi} shows the spectra of the six galaxies with
$\delta \ge -4$ and $\theta \ge$ 5.0 (open triangles in Figure 11). They all
have W([OII]) $> 30$ \AA, and, 
except one, have spectral types later than IV/Sb. The spectra show clear signatures
of strong star formation or activity. If we place these 6 emission-line
galaxies onto diagnostic diagrams of log([OIII]$\lambda$5007/H$\beta$) versus
log([NII]$\lambda$6584/H$\alpha$) or log([OIII]$\lambda$5007/H$\beta$) versus
log([SII]$\lambda$6716+$\lambda$6731/H$\alpha$)
(\cite{villeux87}), we find that galaxies \# 1, 2, 5 and 6 are most
likely HII galaxies ({\em i.e.} have a high current stellar formation
rate). Only galaxy  
\# 3 lies clearly inside the Seyfert 2 region. Galaxy \# 4 was impossible to
classify due 
the absence of H$\alpha$ from the spectrum. Note that the ESS fraction of AGN
($\sim 2\%$) is in marked disagreement with the large fraction ($\sim 17\%$)
found by \cite{tresse96} at $z \la 0.3$ in the Canada France Redshift Survey
(CFRS).  

\subsection{Redshift distribution and completeness}

\begin{figure*}
\centerline{\hbox{\psfig{figure=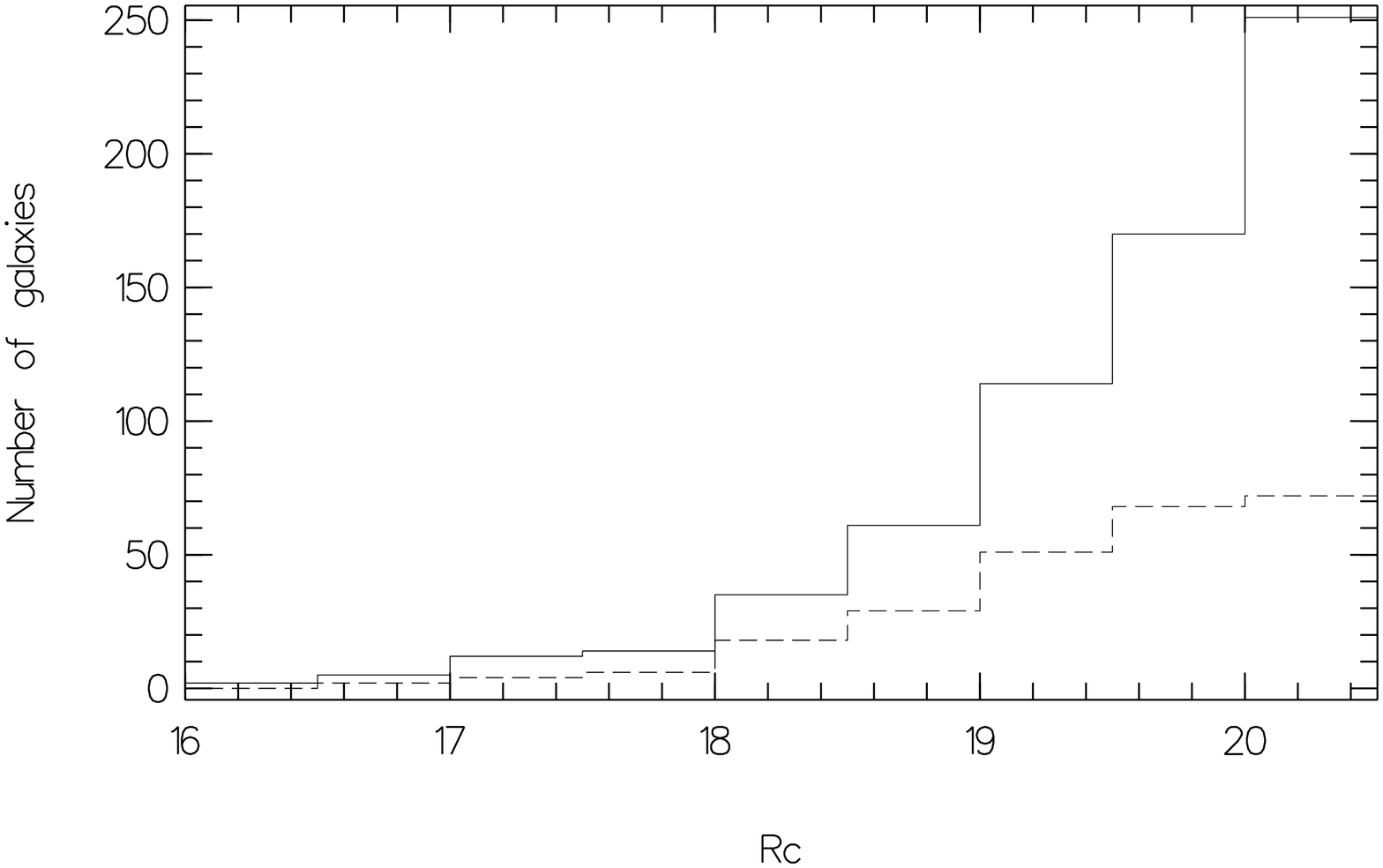,angle=-90,height=7.0cm}\psfig{figure=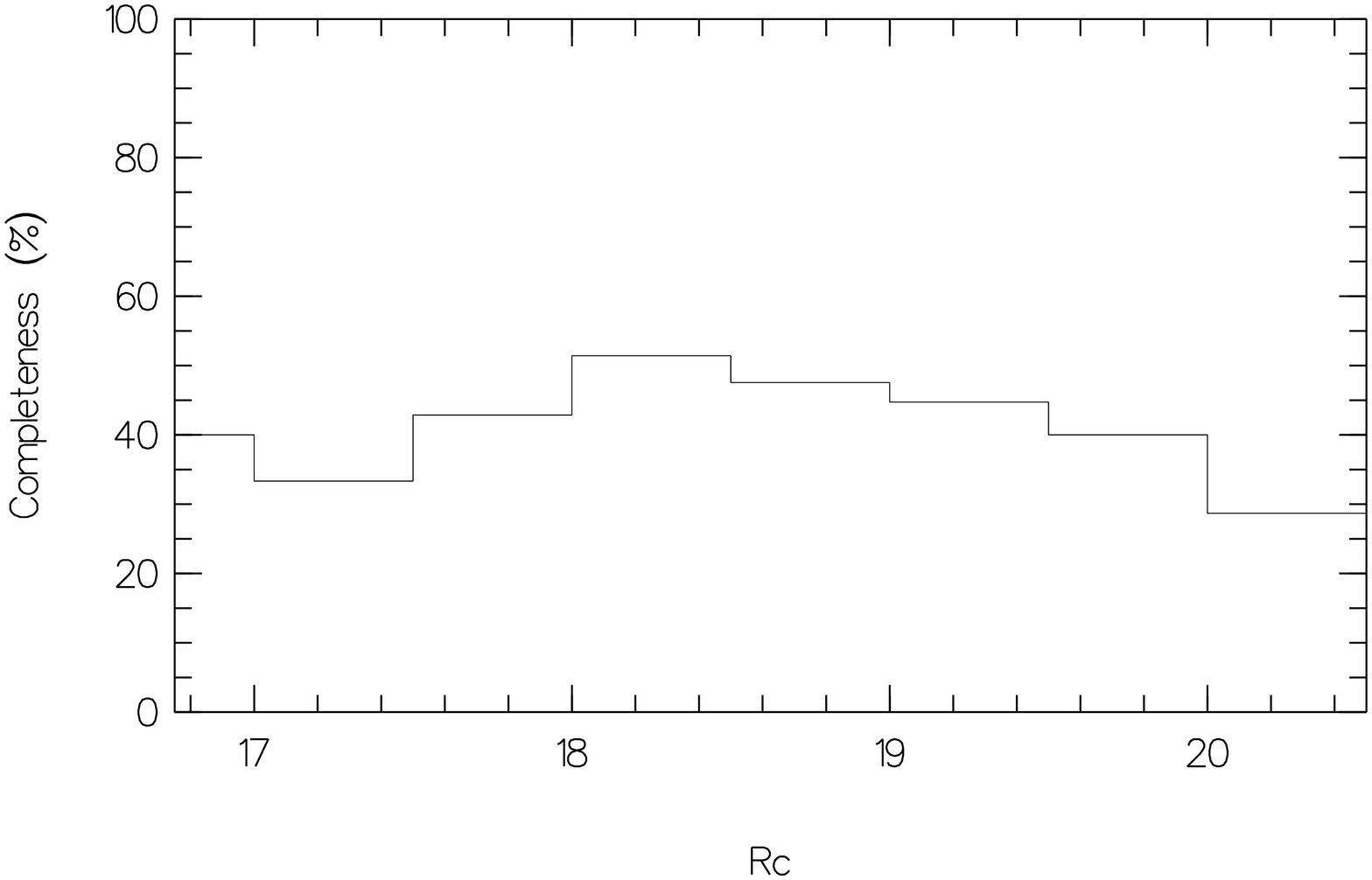,angle=-90,height=7.0cm}}}
\caption[]{Left panel: histograms showing the total number of galaxies per
bin of 0.5 magnitudes and with R$_c \le 20.5$ (total 669, solid), and the
number of galaxies used for the present analysis 
(the 277 galaxies of sample 1, dashed). Right panel: the fraction of the
total number of galaxies per 0.5 magnitudes in R$_c$ in sample 1.}
\label{compl}
\end{figure*}

To the limiting magnitude R$_c$ = 20.5, the spectroscopic sample used for the
spectral classification in this
paper (sample 1) represents 41\% of the complete magnitude-limited sample.
For a given 
magnitude bin, the inverse fraction of galaxies having a measured redshift
gives the completeness correction to apply for that bin. Left panel of
Figure \ref{compl} shows the histogram of galaxies per 0.5 magnitude bin.
The solid line represents the total number of galaxies in the ESS
spectroscopic sample with R$_c \le 20.5$ (669 galaxies). The dashed
line represents the histogram of the 277 galaxies of sample 1 used in most of
the analysis. The right panel of Figure \ref{compl} shows the completeness as
a function of R$_c$ magnitude for sample 1 (in 0.5 magnitude 
bins). We then correct the number of galaxies {\em per type} which are
obtained in \S6.1 and Table \ref{results_pca} by using the inverse of the
completeness curve. The resulting type fractions are nearly identical for all
spectral types, with absolute changes $\la$ 1\% in the type fractions. The
small variations result mainly from the homogeneous spread of different types
as function of apparent magnitude.

\begin{figure}
\centerline{\hbox{\psfig{figure=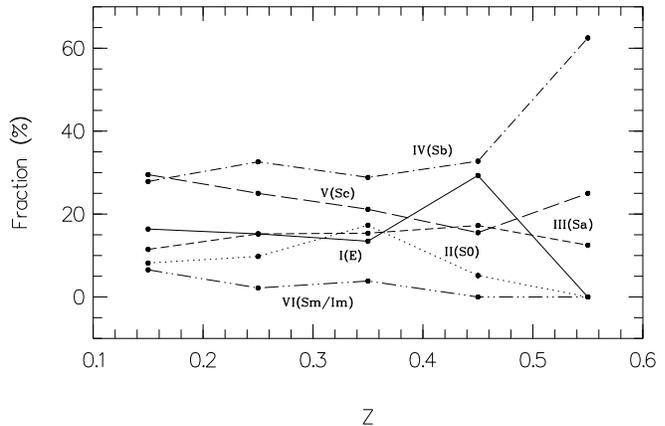,angle=-90,height=6.5cm}}}
\caption[]{Fraction of galaxies of each spectral type, per redshift
interval. The spectral type is provided by the PCA over sample 1 (277
galaxies). The bin size is $\Delta z =$ 0.1. The absolute 1$\sigma$ errors
in the type fractions are $\sim 5\%$ in $0.1 \la z \la 0.5$ and $\sim15\%$ for
$0.5 \la z \la 0.6$, this last value due to the reduced number of galaxies.}
\label{dist_types}
\end{figure}

Figure \ref{dist_types} shows the distribution of types in redshift
space (for 0.1 $\leq z \leq 0.6$), using the PCA spectral classification. The
type population is stable as a function of redshift for $z \la 0.4$, with
spectra of type IV(Sb) as the dominant type, followed by type V(Sc) with no
clear indication of evolution in the type populations with $z$. Recall that
the absolute errors in the population fractions are $\sim 5\%$ (note that the
last bin 
has a small number of objects, so the errors in the population fractions are
larger). Figure 
\ref{dist_types} indicates however a significant excess in 
the fraction of early types at
$z = 0.4-0.5$: the local density of galaxies with types I-II/E-S0 is
3.1$\sigma$ above the average value for $z = 0.1-0.6$
(using $\sigma_{types} = 5\%$). This effect could be caused by the presence
of an elliptical-rich group of galaxies (\cite{ramella97}). The complete
redshift sample is necessary for further investigation of this feature. 

\subsection{Morphology-spectral relationship for the ESS sample}

\begin{figure*}
\centerline{\hbox{\psfig{figure=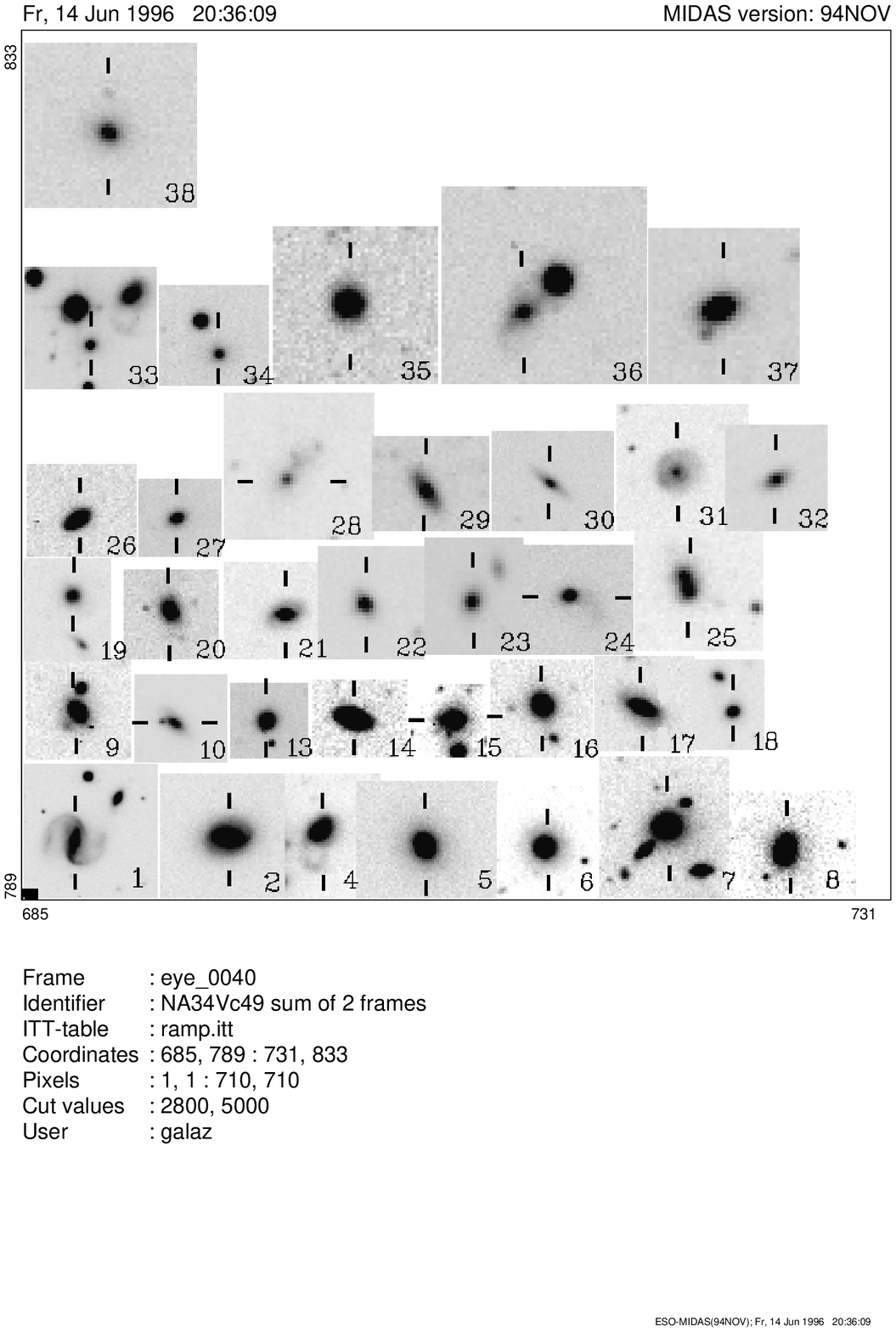,angle=0,height=10cm,clip=on}}}
\caption[]{The 35 brightest galaxies (filter R$_c$) is the ESS sample listed
in Table \ref{eye_types}. The spatial extension of galaxy \# 1 is
$\sim18^{\prime\prime}$ and it is $9^{\prime\prime}-12^{\prime\prime}$ for
galaxies labeled \# 2 to 38. ID number correspond to those in Table 7, where
some galaxies have more than one spectroscopic measure (see also Figure 17).}
\label{eye_gal}
\end{figure*}

We now examine the morphology-spectral relationship for the ESS sample by
testing whether our spectral classification procedure is consistent with the
morphology of some of the objects. We have performed a
visual morphological classification of the 35 brightest galaxies in Samples 1
and 2. CCD images of these objects in the R filter (\cite{arnouts97}) are
given in Figure 
\ref{eye_gal}, in decreasing order of brightness along with the orientation
of the slit used to obtain the 
spectra. The redshift, R$_c$ magnitude, morphological and PCA spectral
types are listed in Table \ref{eye_types}. These galaxies span the magnitude
range R$_c$ = 15.82-18.58 and the redshift range $z \sim$ 0.10-0.25, with
one galaxy having $z =$ 0.42 (\# 28). The morphological classification is 
inspired from that for the Revised Shapley Ames Catalog
(\cite{sandage81}). It was performed by GG in two steps. The first step was
to make a rough classification based on the search for three features in each
galaxy: disc and/or 
bulge and/or spiral arms. If a galaxy could not be included in any of these 
3 categories, it was assigned a ``peculiar'' morphology. Note was taken of
signs of merging or interaction when present. The second step was to define
sub-classes within ellipticals (bulges with/without discs) and
spirals (disc and/or spiral arms). This task is 
difficult (but not impossible!) because the objects are small ($9^{\prime\prime}
\la D \la 12^{\prime\prime}$). The discs are
therefore poorly visible and the contrast of the spiral arms is
weak. In Figure \ref{eye_gal}, the spiral arms are clearly visible
in object \# 1, and discs are visible in \# 17, 30, 31, etc... 
Careful visual inspection of the images of the galaxies using variable
contrast allows to discover the presence of spiral arms in many
cases. This can be done for R$_c$ $\la$ 18.0 (objects \# 1 to 17), for which
the typical apparent diameter 
of the galaxies is $\ga 12^{\prime\prime}$. For fainter 
magnitudes, visual detection of the
spiral arms is very difficult. We note that the spectral type of each galaxy
was kept unknown prior to the morphological classification, in order to avoid
a psychological bias. The morphological classification was repeated 
one month later after sorting at random the galaxies to be classified. In
general, the second 
morphological type does not differ from the first assigned type by more than
one morphological type (see Figure 17). 

\begin{figure*}
\centerline{\hbox{\psfig{figure=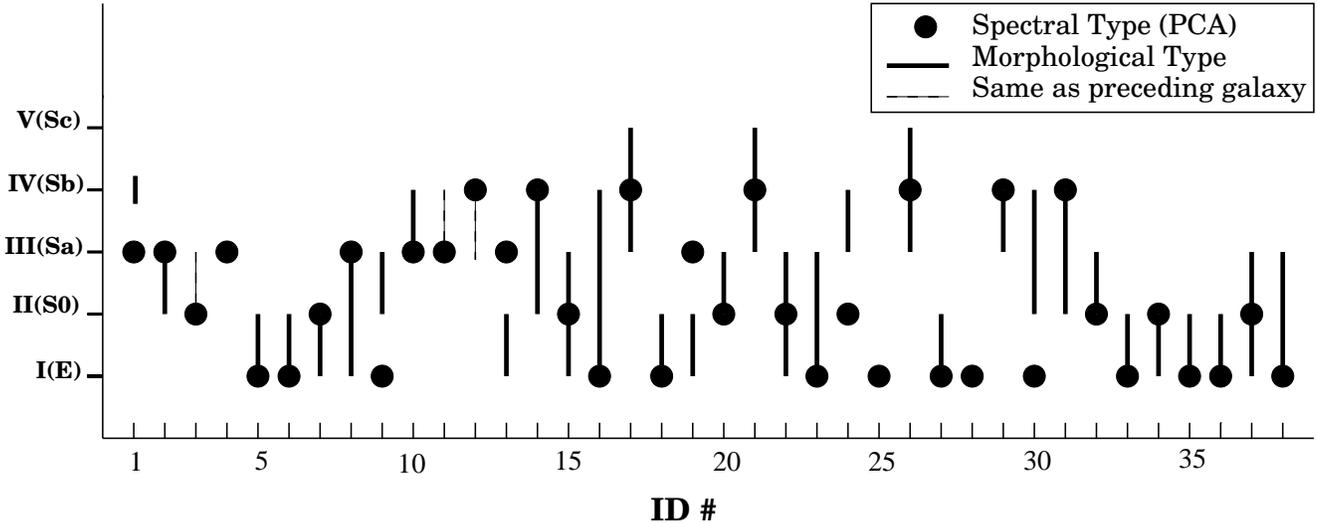,angle=-90,height=7cm}}}
\caption[]{Comparison between the morphological and spectral classification
for the 35 galaxies listed in Table \ref{eye_types} and
displayed in Figure 
\ref{eye_gal}. There are 35 galaxies and 38 spectra, some galaxies having
more than one spectroscopic measure: dashed lines for the morphological
classification indicate 
that the galaxy is the same as the preceding one, and the 2 (or 3) spectra provide 2
measures of the spectral type.}
\label{compar_class}
\end{figure*}
\renewcommand{\baselinestretch}{0.8}
\begin{table*}
\caption[]{The redshift, R magnitude, visual morphological type and PCA
spectral type for the 35 brightest galaxies (and their 38 spectra).}
\label{eye_types}
\begin{center}
\begin{tabular}{lccccl}
\hline \hline
\#   &   $z$ & R$_c$ & Morphological type   &   Spectral type (PCA) & 
Comments  \\ \hline
1   & 0.11 & 15.82 &  SBb   &  III(Sa) &    \\
2   & 0.12 & 16.62 &  S0/Sa &  III(Sa) &        \\
3   & 0.12 & 16.62 &        &  II(S0) & Same galaxy as 2 \\
4   & 0.25 & 16.91 &        &  III(Sa) & M? \\
5   & 0.11 & 17.08 &  E/S0  &  I(E)   &   \\
6   & 0.16 & 17.12 &  E/S0  &  I(E)   &   \\
7   & 0.18 & 17.35 &  E/S0  & II(S0)   &   \\
8   & 0.13 & 17.38 &  E/Sa  & III(Sa)  & D \\
9   & 0.19 & 17.56 &  S0/Sa & I(E)   & D,M \\
10   & 0.23 & 17.68 & Sa/Sb & III(Sa)  & D \\
11   & 0.23 & 17.68 &       & III(Sa)  &  Same galaxy as 10 \\ 
12   & 0.23 & 17.68 &       & IV(Sb)  &  Same galaxy as 10 \\
13   & 0.19 & 17.74 & E/S0  & III(Sa)  & G \\
14   & 0.18 & 17.80 & S0/Sb & IV(Sb)  &   \\
15   & 0.23 & 17.82 & E/Sa  & II(S0)  & VD \\
16   & 0.16 & 17.89 & E/Sb  & I(E)   & VD \\
17   & 0.23 & 17.98 & Sa/Sc & IV(Sb)  &    \\
18   & 0.22 & 18.03 & E/S0  & I(E)   & M, VD  \\
19   & 0.27 & 18.12 & E/S0  & III(Sa)  & M, VD  \\
20   & 0.18 & 18.19 & S0/Sa & II(S0)  &    \\
21   & 0.18 & 18.20 & Sa/Sc & IV(Sb)  &    \\ 
22   & 0.26 & 18.21 & E/Sa  & II(S0)  & VD   \\
23   & 0.22 & 18.22 & E/Sa  & I(E)   & M, D  \\
24   & 0.18 & 18.23 & Sa/Sb & II(S0)  & M, P  \\
25   & 0.32 & 18.25 &       & I(E)   & M?  \\
26   & 0.18 & 18.27 & Sa/Sc & IV(Sb)  &     \\
27   & 0.23 & 18.33 & E/S0  & I(E)   & D  \\
28   & 0.42 & 18.33 &       & I(E)   & P, M, VD \\
29   & 0.19 & 18.34 & Sa/Sb & IV(Sb)  &       \\
30   & 0.13 & 18.36 & S0/Sb & I(E)   &     \\   
31   & 0.17 & 18.39 & S0/Sb & IV(Sb)  &     \\
32   & 0.19 & 18.40 & S0/Sa & II(S0)  &     \\
33   & 0.26 & 18.42 & E/S0  & I(E)   & D, M, G  \\
34   & 0.23 & 18.43 & E/S0  & II(S0)   & G   \\
35   & 0.19 & 18.50 & E/S0  & I(E)   &     \\
36   & 0.28 & 18.50 & E/S0  & I(E)   &  M, VD \\
37   & 0.26 & 18.54 & E/Sa  & II(S0)  & M   \\
38   & 0.27 & 18.58 & E/Sa  & I(E)   &     \\ \hline
\end{tabular}
\smallskip 
\\
\end{center}
\footnotesize
\underline{Notes:} \\ \\
P: Peculiar. \\
M: Merger. \\
G: Group. \\
D : Classification difficult. \\
VD: Classification very difficult. \\
\end{table*}
\renewcommand{\baselinestretch}{1.0}

Figure \ref{compar_class} shows that there is good agreement between the visual 
morphological classification and the PCA spectral classification, for most of
the selected galaxies. The mean difference between spectral and morphological
type is $\pm$ 0.8 type, with an r.m.s. type dispersion of 0.5. 
We stress that the good agreement
between spectral type and morphological type for the objects in Table
\ref{eye_types} suggests that we have been successful in our visual
classification in reproducing the typical morphological criteria used in the 
Kennicutt sample. 
This also suggests that the low redshift members of the ESS survey
($z\sim0.1-0.25$) have a similar morphological-to-spectral relationship than
the nearby galaxies ($z\sim0$) in the Kennicutt sample.
Note that the morphological classification of the Kennicutt
{\em local} sample is done in the B band, and this
is consistent with our classification being performed on the R images: the
B filter ``redshifts'' to the R filter at
$z\sim0.21$ (the average redshift of the objects in Table \ref{eye_types}).

\subsection{Peculiar objects revealed by the PCA}

\begin{figure*}
\centerline{\hbox{\psfig{figure=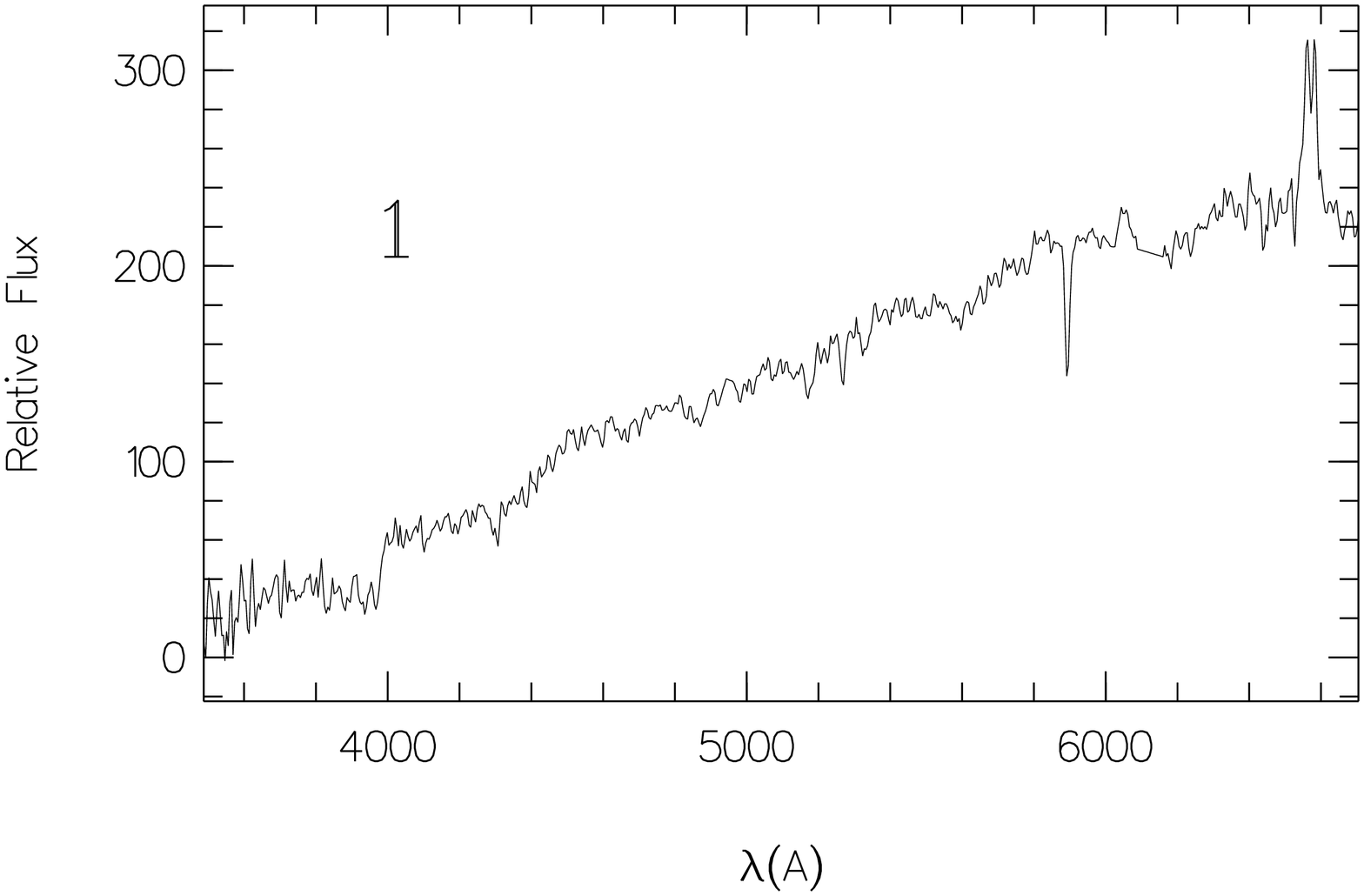,angle=-90,height=3.0cm}\psfig{figure=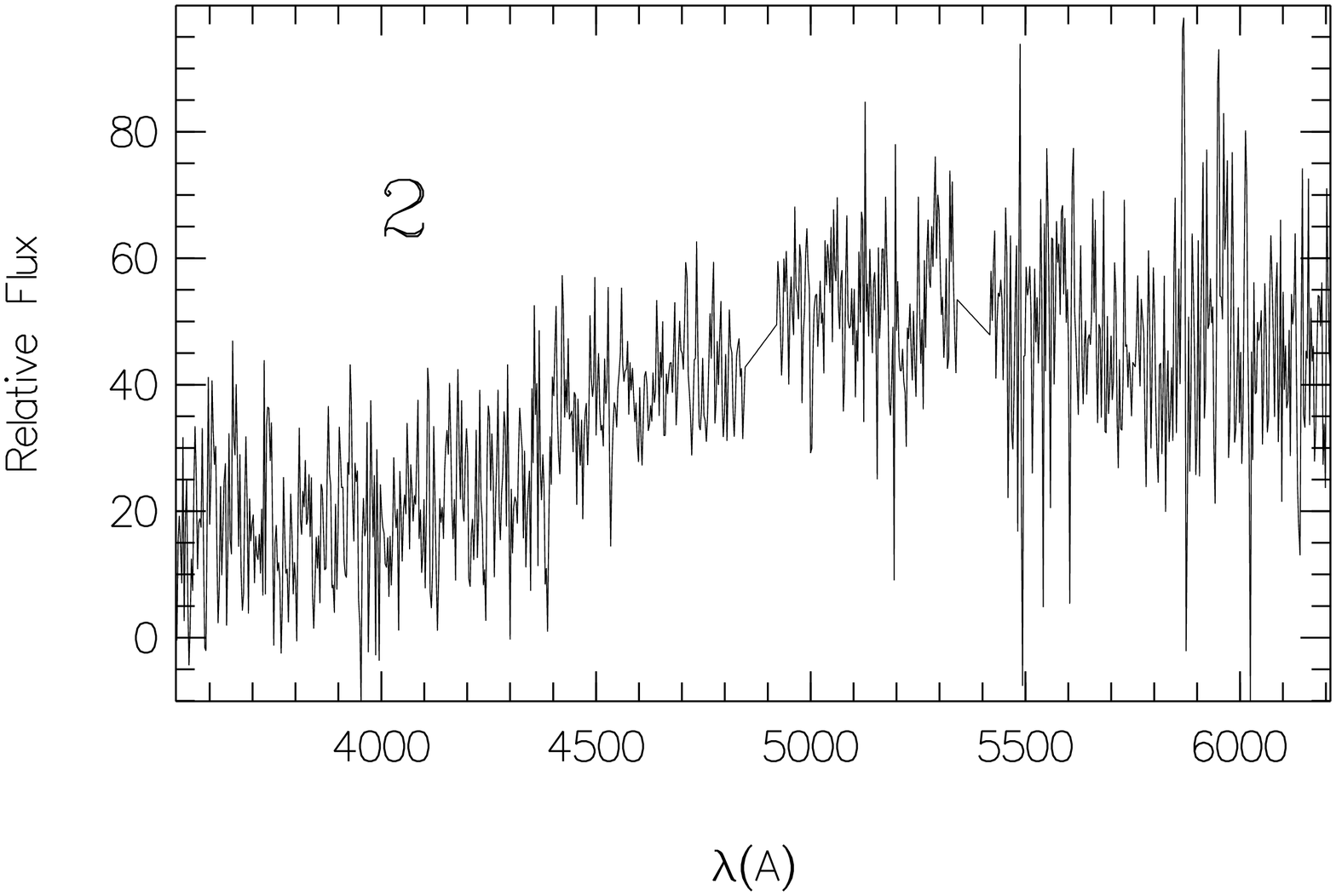,angle=-90,height=3.0cm}\psfig{figure=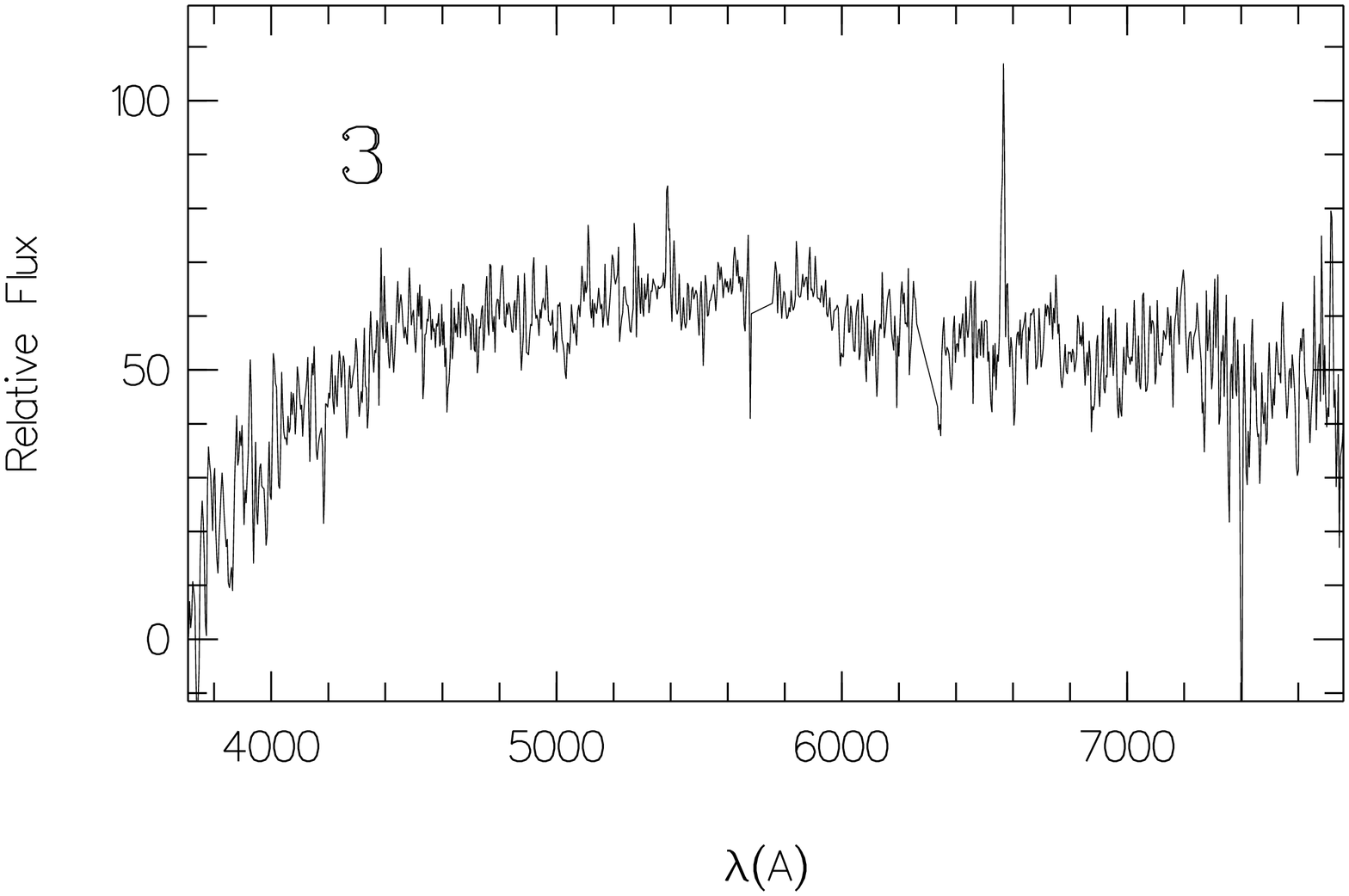,angle=-90,height=3.0cm}}}
\centerline{\hbox{\psfig{figure=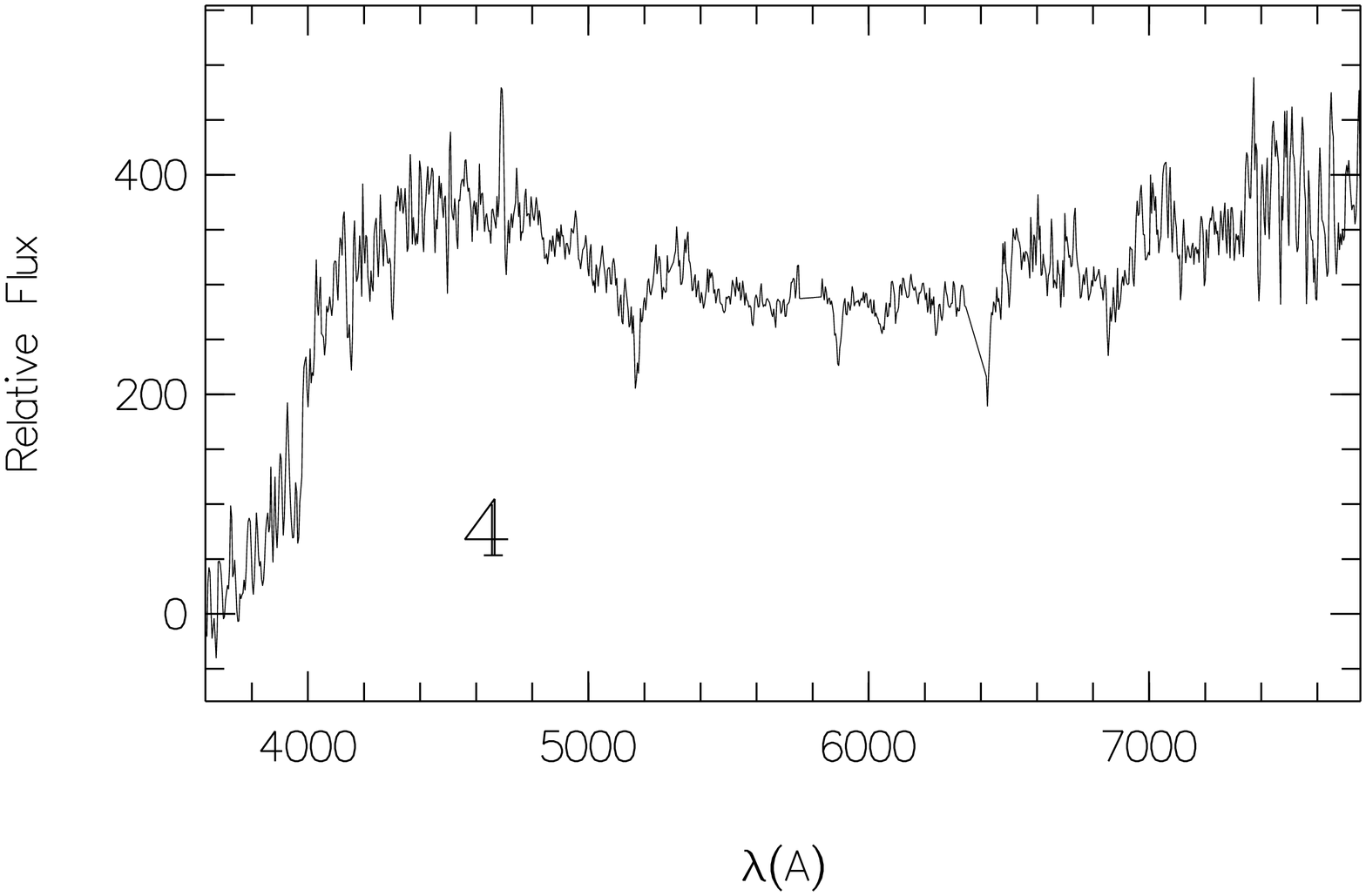,angle=-90,height=3.0cm}\psfig{figure=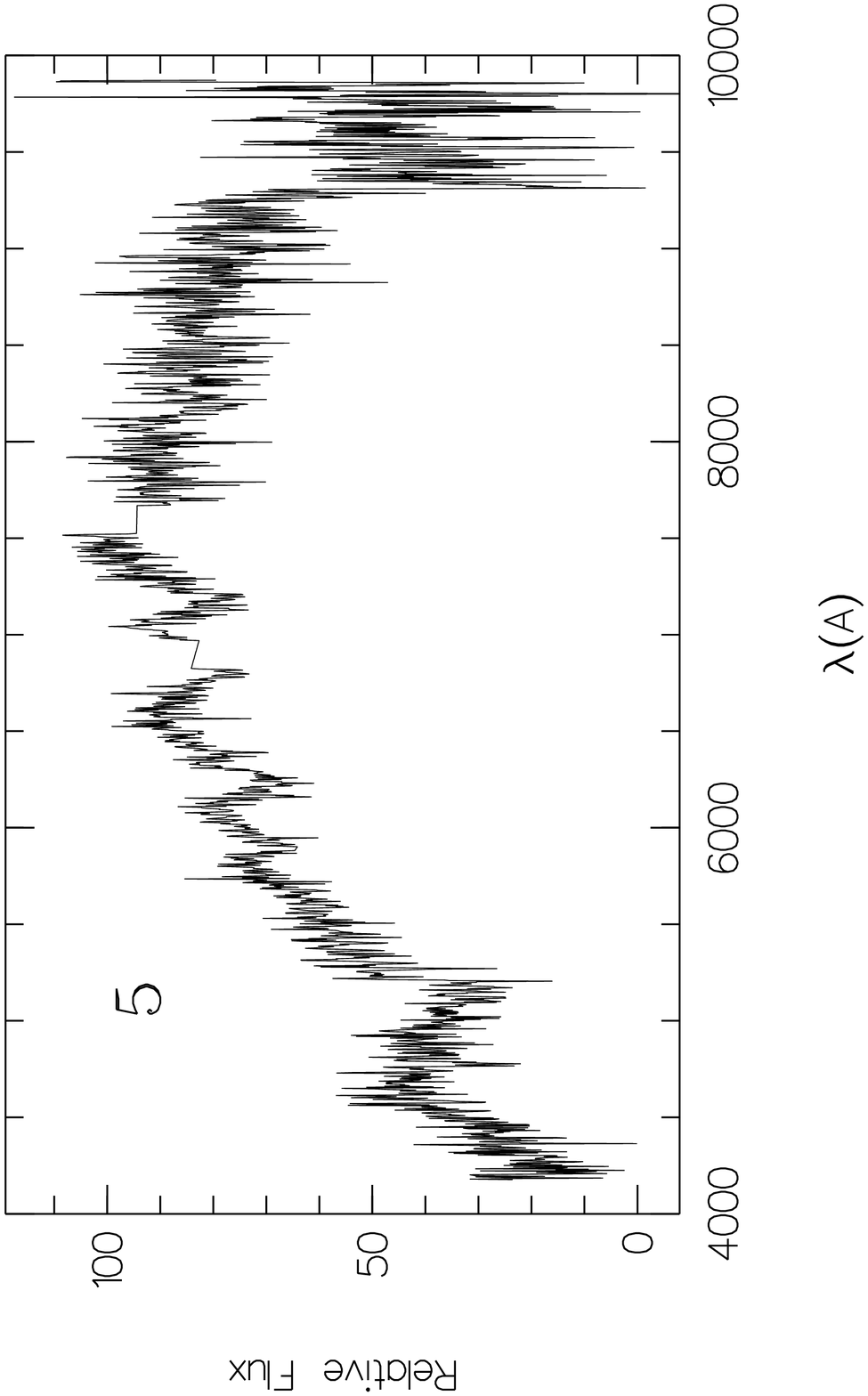,angle=-90,height=3.0cm}\psfig{figure=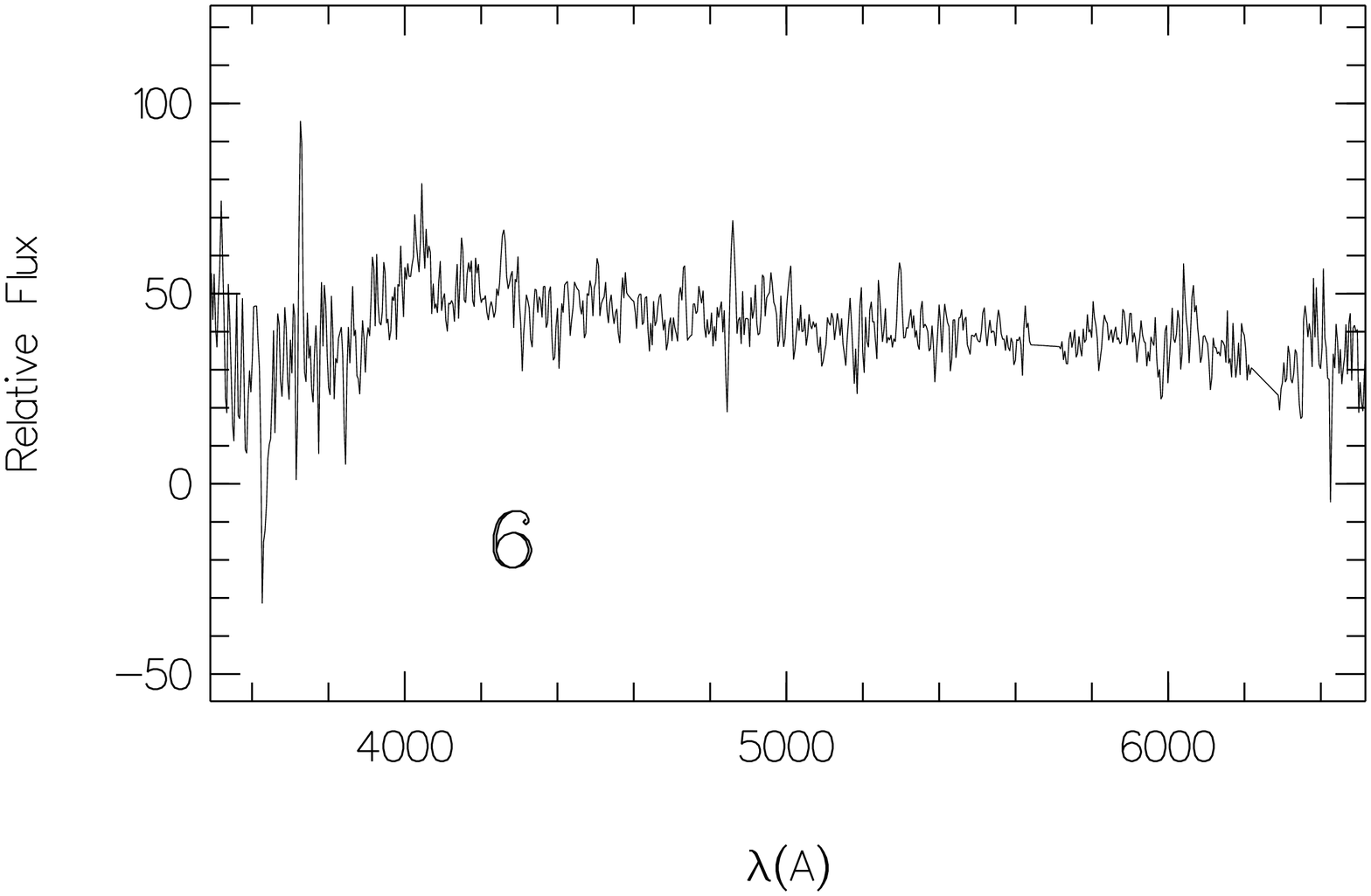,angle=-90,height=3.0cm}}}
\centerline{\hbox{\psfig{figure=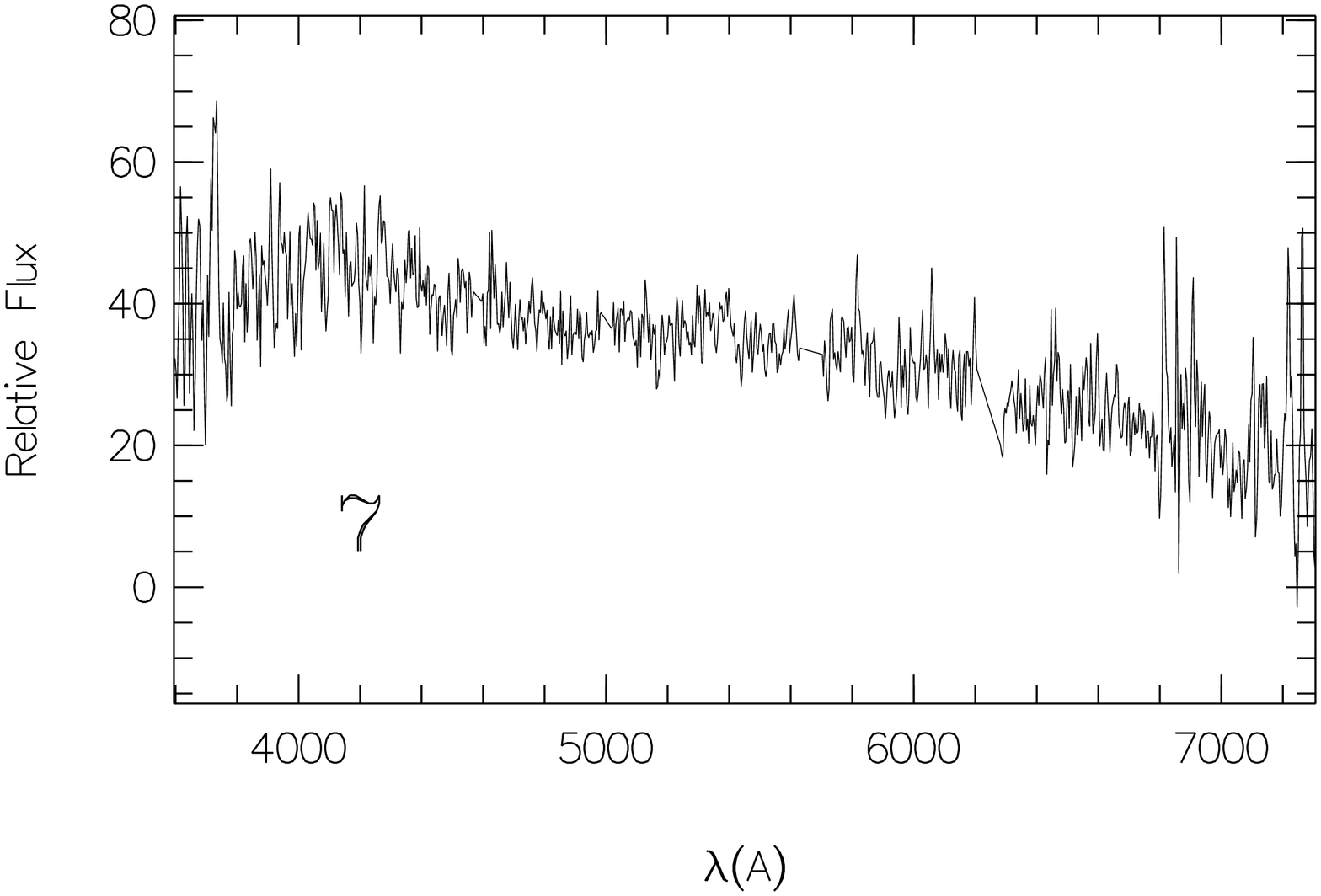,angle=-90,height=3.0cm}\psfig{figure=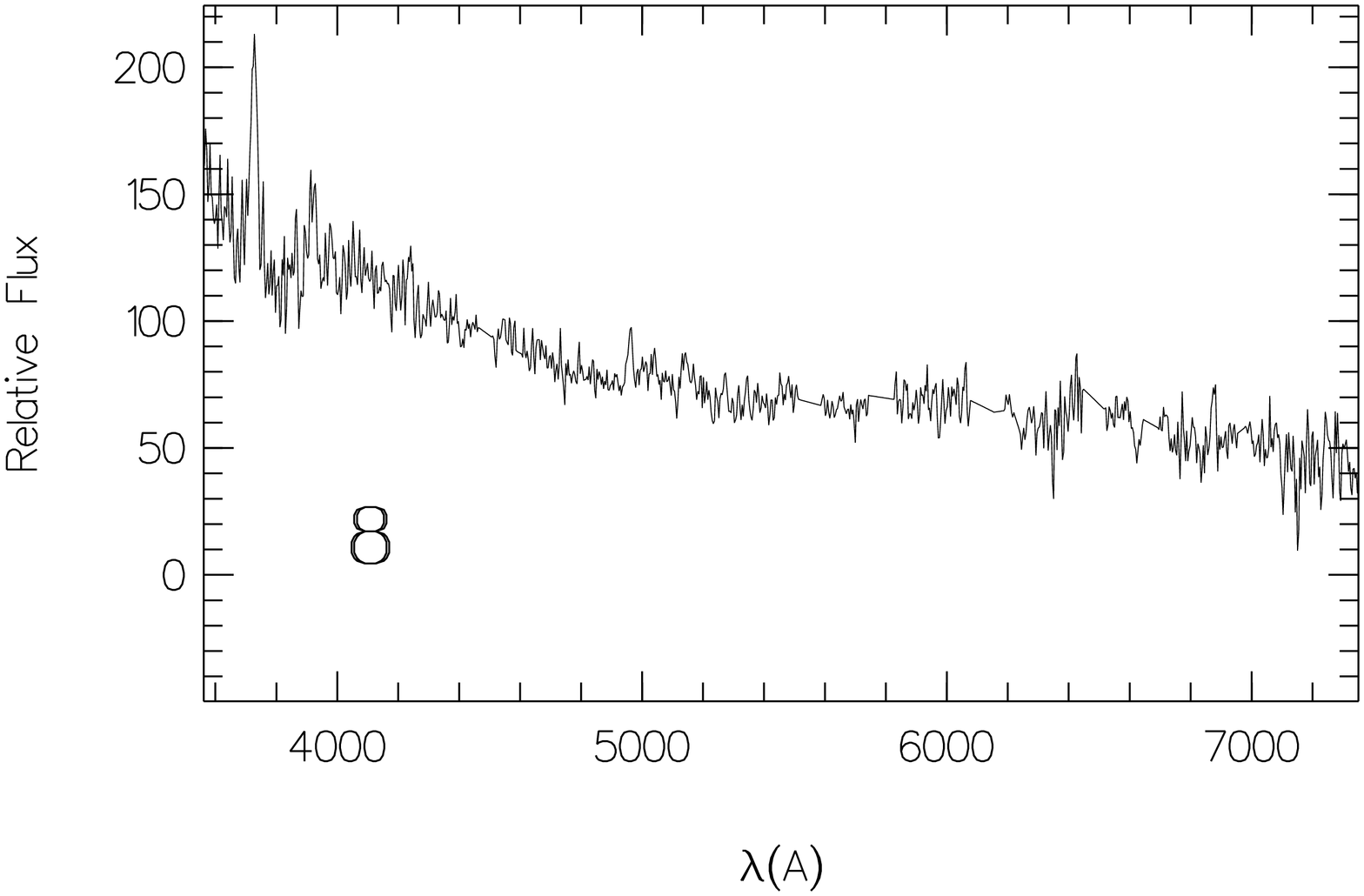,angle=-90,height=3.0cm}\psfig{figure=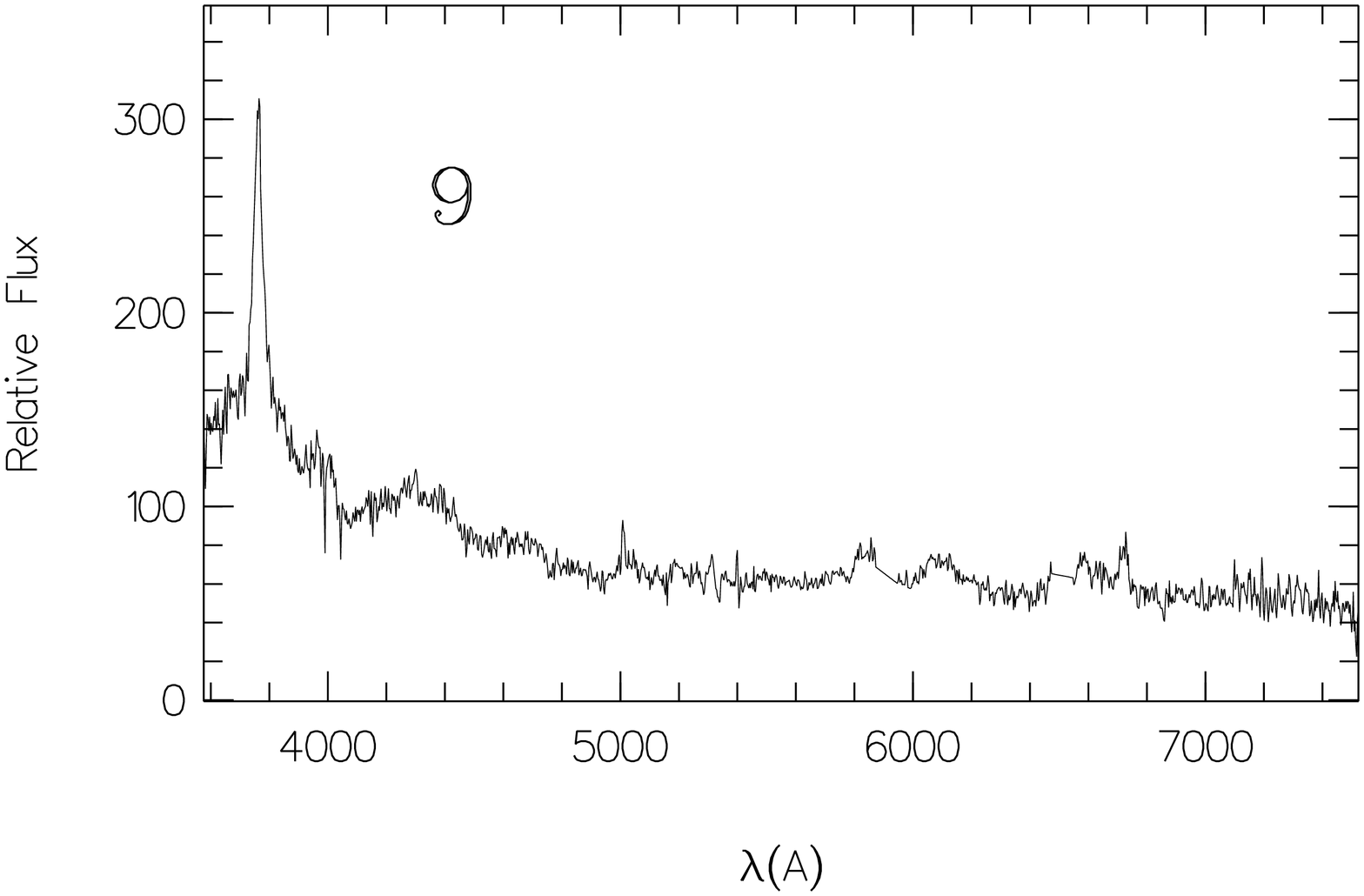,angle=-90,height=3.0cm}}}
\caption[]{Galaxies deviating from the general trends of the ESS spectral
characteristics, in rest-wavelength.} 
\label{raros_fig}
\end{figure*}

One of the useful features of the PCA is its capability to detect objectively
objects
which deviate from the general trend. In Figure \ref{reg12_Oxg}, we mark with
open circles galaxies which lie outside of the
main ($\delta$,$\theta$) sequence and outside the sequence of emission-line
galaxies (see \S 6.3) in order to show that the degree of peculiarity is 
related to the departure from the main sequence. Although galaxy \#5 has a
close neighbor in the ($\delta$,$\theta$) plane, we only
show object \# 5 because two spectra with $\delta$ values differing by less
than 5\% are indistinguishable. The spectra
of the selected objects are shown in Figure \ref{raros_fig}. 
Redshift, R$_c$ magnitudes, colors, spectral and morphological types of these
objects are listed in Table \ref{raros}. Visual examination of the R CCD
images of these objects has been performed. 
In general, the spectral and morphological type agree within the
uncertainties in determining the morphology (see \S 6.6). It is
interesting to note that the images of the six of nine galaxies show clear
signs of 
peculiarities and/or merging. One galaxy even shows a jet-like feature (\#
7). Galaxy \# 1 shows a very red continuum, steeper than in any Kennicutt
galaxy, a strong break, as well as strong Na$\lambda$5892 and H$\alpha$
emission lines. Note that the image shows a regular morphology, and therefore
the spectral classification provides additional physical information.
Galaxy \# 2 presents a very low S/N ratio which is just at the chosen limit
for sample 1 (S/N $\ga$ 5.0). The unusual continuum shape of this spectrum is
probably responsible for the deviation from the sequence but the low S/N
ration does not allow any further study. 
Galaxies \# 3, 4, 6 and 7 exhibit Markarian signatures
(galaxies with bursts of star formation). The continuum shape and absorption
bands, typical of these galaxies, are clearly seen in Figure \ref{raros_fig}
and are comparable to other known spectra (see for example
\cite{contini96}). Such objects have 
in general a spiral morphology (\cite{contini96}), which is in marked
contrast with the I(E) spectral type of objects \# 3 and 4. This shows that, for
galaxies which significantly 
deviate from the PCA spectral sequence, the spectral/morphological
relationship is no longer valid. 
Object \# 5 is a star which was willingly introduced in the sample as a test,
and as we can see, it is far from the main sequence. It provides an
additional test of the capability of the PCA to distinguish abnormal spectra.
Galaxy
\# 8 is a typical galaxy with strong star formation. Note that H$\alpha$ was
blanketed by a sky line. 
Galaxy \# 9 resembles a QSO at $z \sim 0.51$, if one identifies the broad
emission line with MgII.

Other objects generally labeled as peculiar are the AGN (Seyfert, LINER's, N 
galaxies, etc...) and the E+A galaxies (see \cite{zabludoff96}). As discussed 
in \S 6.3, the AGN galaxies are included in the ``emission-line sample''. The
PCA is an excellent tool to detect these objects and quantify their spectral
features  
via the 3$^{rd}$ PC. The E+A galaxies are however difficult to identify from
the PCA classification. Because {\em classical} E+A galaxies have strong
Balmer absorption lines, but {\em no} emission lines in the region 3500 to
7000  \AA,  
these galaxies lie inside the normal sequence of early to intermediate-type 
galaxies (see Figure \ref{delta_theta_type}). In order to identify the E+A
galaxies in the ESS sample, we have measured the equivalent
width of H$\beta$, H$\delta$ and H$\gamma$ for the galaxies of sample 1,
using a similar criterium to that used by Zabludoff \etal (1996): W$[H\delta]
\ge 5$ \AA$\;$ and no sign of [OII] emission. We found 
9 galaxies which satisfy the selection criterium. 
The PCA spectral type of these galaxies is, 
except in one case, V/Sb. These  
galaxies imply a 3\% fraction of E+A galaxies in the ESS in the redshift
range $0.1 < z < 0.5$. 
This is in marked contrast with the 21 E+A out of 11113 field galaxies found
by Zabludoff \etal in the Las Campanas Redshift Survey to $z \sim 0.2$, which
gives $\sim 0.2\%$ of E+A galaxies. Two E+A 
candidates of sample 1 have $z \le 0.20$, and the fraction does not change if
we consider the 66 galaxies of sample 1 with $z \le 0.20$. Because of the 
large redshift of these objects ($0.2 \la z \la 0.42$), the CCD images of
these galaxies subtend too 
small solid angle for a detailed morphological study. 
Although we do not see evidence of merging, four
galaxies form two different pairs, and so we cannot discard the possibility
of interaction between some of the E+A galaxies and their neighbors. The
other E+A do not show evidence of interaction.

We have verified that all of the objects which strongly deviate 
from the sequence in which lie the normal galaxies are indeed peculiar 
objects. On the other hand, the spectra of the objects which lie inside the
spectral sequence were visually inspected and do 
not show signs of any peculiarity, except the possibility that they could
have strong absorption lines (for example E+A galaxies).
The PCA is therefore 
efficient at detecting in a quantitative manner both the normal and most
frequent objects, and the rare  
and peculiar objects.

\begin{table*}
\caption[]{Peculiar objects revealed by the PCA.}
\label{raros}
\begin{tabular}{lcccccccc}
\hline \hline
\#  & $z$  & R$_c$ & B$_j-$V$_j$ & V$_j-$R$_c$  & B$_j-$R$_c$ &
Spec. Type(PCA) & PCA$_{error}^{(c)}$ & Morphology \\ \hline 
1$^{(a)}$ & 0.12 & 18.36  & 1.17 & 0.74 & 1.91	& I(E) & 0.003 & S0/Sa \\
2  & 0.41 & 20.35  & 1.24 & 1.37 & 2.61		& I(E) & 0.023 & E/Sa merger \\
3  & 0.20 & 19.78  &      & 0.69 & 		& I(E) & 0.010 & E/Sa merger \\
4$^{(b)}$ & 0.19 & 17.56 &      & 1.10 & 	& I(E) & 0.016 & S0/Sa \\
5  & -- & 19.87  & 1.28  & 0.77  & 2.05	& II(S0) & 0.014 & star \\
6  & 0.21 &        &      &      & 		& V(Sc) & 0.011 &  \\
7  & 0.21 & 20.47  & 0.64 & 0.42 & 1.06		& V(Sc) & 0.010 & Sa/Sb +
jet? \\
8  & 0.24 & 19.46  & 0.02 & 0.82 & 0.84		& VI(Sm/Im) & 0.011 & Sa/Sb
Pec + merger \\
9  & 0.51? & 19.43 & 0.27 & 0.05 & 0.32		& VI(Sm/Im) & 0.048 & Stellar
profile+merger \\
\hline
\end{tabular}
\\
\smallskip 
\underline{Notes:} \\
$^{(a)}$ Same as galaxy \# 30 of Table \ref{eye_types}. \\
$^{(b)}$ Same as galaxy \# 9 of Table \ref{eye_types}. \\
$^{(c)}$ Interval error in the PCA reconstruction, as defined in equation 7
(see \S 6.1). 
\end{table*}

\section{Discussion}

\subsection{Aperture Bias}

When performing a spectral classification, one must keep in mind 
that, in general, galaxy slit spectroscopy provides only a partial sampling  
of the objects. Here we comment on some of the biases which could be present
in the spectra, and could produce an erroneous interpretation of the
results. Two different phenomena produce what is called an ``aperture
bias''. A first bias originates from the fact that the gathered light from
galaxies at different distances, sampled with a slit 
of fixed width (in our case 1.3$^{\prime\prime}$ to 1.8$^{\prime\prime}$),
correspond to different spatial extensions (and stellar populations) in the
corresponding galaxies (this is called ``aperture bias'' hereafter). A
second bias (called ``orientation bias'' hereafter) comes from the varying
spatial  
orientation of the slit with respect to the observed galaxies due to
observational constraints (see Figure 16). In order to study the aperture
bias, Zaritsky, Zabludoff \& Willick (1995) have made some simulations 
for characterizing the influence of the size
of an optical fibre onto their spectral classification 
for galaxies with $ 0.05 \la z \la 0.2$.
Following their conclusions, our slits of 1.3$^{\prime\prime}$ to 
1.8$^{\prime\prime}$ should not cause any significant aperture bias for galaxies 
with $z \ga 0.1$. 
Moreover, the slits used for the ESS observations contain $\ga 95\%$ of the
bulge and disc emission for a typical face-on spiral of $\sim 20$ kpc in
diameter at $z \ga 0.1$ (for spirals with other orientations, the fraction of
the bulge and disk sampled is even larger). Therefore the results
presented here are weakly affected by the aperture bias.
For elliptical galaxies, the  ``orientation bias'' is null, because
the stellar constituents generally have uniform distributions along the galaxy extent.
On the other hand, for spiral galaxies, the early and late stellar populations
are not equally distributed along the galaxy profile, and slits 
oriented along the minor axis will over-sample the bulge of the galaxy
with respect to the disc. In this case, and assuming a 
standard morphological-spectral relationship, the young stellar 
populations are under-sampled and therefore the spectral type would appear 
earlier than the morphological type, or at least, earlier than the real 
integrated spectrum. This effect could have important consequences in the
fraction of types found as elliptical and spiral. Observing our CCD images,
along with the slit orientations, allows us to conclude that an orientation
bias 
could exist for spiral galaxies which have $z \la 0.1$ (where only a fraction
of the galaxy profile is inside the slit). However, 
less than 4\% of the ESS galaxies have $z \la 0.1$.
We emphasize that only a rigorous study of the data from spectro-imaging
surveys can quantify both the aperture and orientation bias present in
existing and future redshift surveys (see \cite{hickson94}). 

\subsection{Comparison with other surveys}

The fraction of different galaxy types found in the ESS sample may be
compared to those found in other redshift surveys. 
Table \ref{type_dif} shows the
fractions of the different morphological types found in the 6 other redshift
surveys which provide the adequate information for comparison with the ESS
data. For the ESS sample, we assimilate the
spectral type with the corresponding morphological type from the Kennicutt
average templates used for the classification.
When examining Table \ref{type_dif},
the reader should be aware of the different classification criteria used in
the various surveys. Moreover, some morphological
differences are subtle and can 
only be distinguished from high-quality imaging. The uncertainties
in a given morphological type are in general not provided. However, some
studies (\cite{naim95}, and references therein) have shown that, when
comparing the morphological classification of nearby galaxies by 6 different
experienced astronomers, the r.m.s. scatter between two observers is
typically 2-T units (\cite{devaucouleurs59}). In the case of deep surveys,
the errors in the 
classification are complicated by the appearance of new morphological types
as 
in the case of HST images (see \cite{vandenbergh96}), and/or by the image
quality (\cite{dalcanton96}). 
For the ESS, we are able to check and quantify the different error
sources. The major error originates from the flux calibration, whose {\em
measured} uncertainty leads to an absolute error 
in the type fractions of $\sim 5\%$ (see \S 6.1). 

We first examine the non-uniform binning. Given our 
classification errors for elliptical and spirals
($\sim$ 5\% of the fraction in each class), the fraction 
of E/S0 is comparable to that given
by RSAC and by the faint DWG (references in the Table
\ref{type_dif}). Clearly, we found more 
ellipticals than Griffiths \etal and than in the HDF, as expected from the
fainter limiting magnitude of the 2 surveys. In these cases the
large number of spirals is due to the combination of evolution and
``morphological K-corrections'': in the case of the HDF, the survey is
sampling the UV
spectral bands, which trace the peculiar morphologies related to the
star-forming regions within the galaxies, (see \cite{vandenbergh96}). 
On the other hand, the fraction of
E/S0 in the ESS is smaller than the fraction found in the CfA1 and CfA2
surveys. For the CfA2 survey, the fraction of ellipticals is larger probably
due to the fact that the eyeball classification is performed from photographic
plates, and some proportion of spirals is likely classified as
ellipticals due to the disappearance of the spiral arms when the bulges are
saturated (\cite{huchra95}). The fraction of spirals (Sa/Sb/Sc) in the ESS
sample is quite similar to that for the RSAC but is significantly larger that
in all other surveys. The number of Sa/Sb/Sc for the ESS sample probably
includes some fraction of Sd galaxies, because this morphological
type is absent from our classification (see Table \ref{results_pca} and \S
6.1).  

It is interesting to note that if we consider the
classification provided by the uniform binning in the $\delta$ parameter (see
\S 6.1 and last row of Table \ref{type_dif}), our fractions of E/S0 and
Sa/Sb/Sc agree quite well with the respective fractions for the CfA1
survey. However, we emphasize that such a uniform binning  system is not
realistic, in the sense that we obtain a poor correlation between the
spectral properties and the morphological type. This leads to the important
issue of the discreteness of type definition. It has been shown
(\cite{morgan57}, \cite{aaronson78}, \cite{abraham94}), and we confirm in
this paper, that the spectral {\em and} morphological properties span non
uniformly but continuously over most sets of parameters used for classifying
galaxies. The assignment of a type in the Hubble system suffers from an
artificial discretization which can lead to significant differences depending
on the classification procedure which is adopted. This is the case for the
E/S0/Sa types, for which the spectral properties show small variations due to
the small changes in the stellar populations among these
types (this is confirmed by the small range in $\delta$ describing these
types; see Figure \ref{reg123_bin}). 

The ESS tends to have a similar type distribution 
to that observed in local or intermediate redshift surveys, as the CfA1, the 
RSAC, and the DWG survey. This seems to
indicate that the galaxy distribution as a 
function of type is rather stable up to $z\sim0.5$. Other surveys
indeed indicate that up to this redshift, only a marginal galaxy evolution is
present (Hammer \etal 1995, Hammer \etal 1997, \cite{lilly96}): the [OII]
equivalent width does 
not increase significantly and galaxy types are roughly uniform.  
The comparison of the ESS spectral classification with other
surveys constitutes not only a test of our classification procedure, but also
provides a quantitative insight into the origin of the observed differences,
which can be 
related to galaxy evolution and the well-known morphology-density relation 
(\cite{postman84}). These effects will be further investigated in forthcoming
papers.  

\begin{table*}
\caption[]{Different morphological mix obtained by other surveys. The ESS
classifications are summarized in the last 2 rows. All the type fractions are
percentages.} 
\label{type_dif}
\begin{center}
\begin{tabular}{lllllll}
\hline \hline
Source	   &	Magnitude$^{(a)}$  	&  E/S0  &  Sabc  &  Sd/Irr  &
Sp/Irr$^{(b)}$ & 
Unclassified \\ \hline
CfA1$^{(c)}$   &	m$_z \leq 14.5$ &  35	 &  54    & 10 &  65  &  1 \\
CfA2$^{(c)}$   &	m$_z \leq 15.5$ &  42	 &  48    & 8  &  56  & 2  \\
RSAC$^{(d)}$	   &    local	  	&  29	 &         &   & 71$^{(b)}$  &   \\   
Shanks$^{(e)}$  &	m$_{bj} \leq 16.0$  & 43	 &  45    & 12 &  57  & 0  \\
Griffiths$^{(f)}$ &	m$_I \leq 22.25$  & 19   &  44    & 13 &  57  & 25 \\
DWG faint$^{(g)}$  &	m$_I \leq 21.75$  & 28   &  50    & 14 &  66  & 6  \\
HDF$^{(h)}$      &	m$_I \leq 24.5$	  & 16	 &  37    & 47 &  84  & 0  \\
ESS$^{(i)}$     &	m$_R \leq 20.5$   & 26 & 71 &
3$^{(k)}$ & 74 & 0 \\
ESS$^{(j)}$     &    		  & 38 & 55 &
7$^{(k)}$ & 62 & 0 \\ \hline 
\end{tabular}
\smallskip 
\\ 
\end{center}
\footnotesize
$^{(a)}$ Limiting magnitude for the classification. \\
$^{(b)}$ Sa/Sb/Sc/Sd/Irr. \\
$^{(c)}$ \cite{marzke94}. \\
$^{(d)}$ \cite{sandage81}. \\
$^{(e)}$ \cite{shanks84}. \\
$^{(f)}$ \cite{griffiths94}.   \\
$^{(g)}$ \cite{driver95a}. \\
$^{(h)}$ Hubble Deep Field. \cite{driver95b}.   \\
$^{(i)}$ ESO-Sculptor Survey, using the non-uniform rebinning 
of column 6 of Table \ref{results_pca}. \\
$^{(j)}$ ESO-Sculptor Survey, using the uniform rebinning 
of column 2 of Table \ref{results_pca}. \\
$^{(k)}$ Sm/Irr types.
\end{table*}

\section{Conclusions and prospects}

In this paper we show that we can classify the galaxies 
of the ESO-Sculptor survey (ESS) from their flux calibrated spectra
using the Principal Component Analysis (PCA) technique. The PCA allows to
define a {\em continuous} spectral sequence highly correlated with the Hubble
morphological type. This sequence can be written as a linear 
combination of a reduced set of parameters and vectors (3) which account for
$\sim$ 98\% of the total flux of each spectrum. The parameters are also
sensitive to the strength of emission lines. Our main results can
be summarized as follows: 

\begin{enumerate}
\item[(1)] Using Kennicutt spectra for galaxies of known Hubble types, 
we establish the 
strong correlation between the spectral galaxy type and the underlying old
(red) and young (blue) stellar population within the galaxy. These
populations can be  
quantitatively separated in the PCA approach using a sequence which arises
mainly from the changes in the shape of the continuum and the relative
strength of the absorption features. 
\item[(2)] By application to the ESS data, we show that the PCA is a 
flexible and powerful tool to
classify galaxies using the spectral information. Galaxies can be classified
using one continuous parameter ($\delta$). We also find that the 
presence and strength of the emission lines are correlated with the spectral type 
(late galaxies tend to have strong emission lines), and can be quantified by
a second continuous parameter ($\theta$). The continuous
nature of the classifying parameters $\delta$ and $\theta$ 
provides a powerful tool for an objective study of the systematic
and non-systematic properties ({\it i.e.\/,} peculiar objects) of spectral
data. Moreover, it allows us to 
use one (or two) fundamental parameter(s) to construct an analytical relation
between the classification parameter(s) and other quantitative properties of
the galaxies (for example K-corrections, local galaxy density, etc...).
\item[(3)] We illustrate using the ESS data how the PCA acts as a powerful
filter of noisy spectra, inherent to deep redshift surveys. Reconstruction
of the ESS spectra with 3 principal components increases the S/N from the
range 8-20 to the range 35-80.  
\item[(4)] The spectral sequence given by the PCA is not uniformly populated:
the early types are more concentrated in the classification plane than the
late types. This non-uniform distribution of the different {\em spectral}
types is closely related to the fact that systematic differences between two
consecutive {\em morphological} types are larger among spiral galaxies than
among elliptical or lenticular galaxies. This leads to construct a variable
binning when comparing the observed spectral sequence with morphological
classifications from other surveys.
\item[(5)] When making the analogy between spectral type and morphology via
the Kennicutt spectra, we find that the ESS sample contains 26$\pm$7\% of
E/S0, 71$\pm$9\% of Sa/Sb/Sc and 3$\pm$7\% of extreme late spirals (Sm/Im). 
The type fractions for the ESS show no significant changes in the redshift
interval $z \sim 0.1-0.5$, and 
are comparable to those found in other galaxy surveys at intermediate
redshift. For the $\sim 277$ galaxies in the ESS analyzed here, the 
dominant type is Sb, followed by Sc, Sa, and early types.
We do not detect any strong evolution in the ESS data as a function of
redshift, up to the depth of the ESS spectroscopic
catalogue ($z \sim 0.5$). Other surveys have  
detected only a marginal evolution at $z \la 0.5$, like the CFRS
(\cite{lilly96}) and the Autofib survey (\cite{heyl96}), and
significant evolution appears to occur at $z \ga 0.4-0.6$. 
In the ESS sample, we note a significant excess of early types at $z \sim
0.4-0.5$. The nature of this excess will be further investigated in a future
paper using the complete redshift sample.
\end{enumerate}

Application of the PCA method to the ESS shows that it may be applied
to any set of flux-calibrated spectra, and that it is a promising technique for
on-going and future massive galaxy surveys. 
The major interest of the PCA technique is that the spectral trends followed
by the sample used are independent  
from any set of templates (the classification space is continuously 
populated). The PCA technique therefore offers clear
advantages over other 
discrete methods like the $\chi^2$ (see \cite{zaritsky95}) or the 
cross-correlation method (see \cite{heyl96}), which 
are fully dependent on the set of templates used: in these approaches, it is
difficult to discriminate differences in the results from differences 
in the input templates. Also, such classification procedures are 
sensitive to fluctuations due to the noise of each target spectrum, an 
undesirable phenomenon. In contrast, the PCA offers an unsupervised
classification system (\cite{naim95}), in which one does not make any
assumption on the general trends followed by the sample.
Moreover, the PCA classification shows that the spectral sequence is essentially
determined by the variations in the shape of the continuum. Any
spectral classification method dependent only on the strength of the
absorption lines (\cite{zaritsky95},
\cite{heyl96}) is therefore very sensitive
to instrumental effects and/or physical phenomena not necessarily correlated
with spectral type, and must be interpreted with caution. 

The spectral classification for the ESS sample will be used to derive
precise K-corrections, which are fundamental for
deriving absolute magnitudes. Those in turn will allow to calculate the
luminosity functions as a function of spectral type
(\cite{galaz97}; \cite{galaz97b}). With the specific galaxy
luminosity function, we can 
investigate in detail the morphology-density relation in the field (see
\cite{marzke94}) and more generally the variations in galaxy properties with
local environment and location within the large-scale structure. These
various analyses will be reported in subsequent papers. 

\begin{acknowledgements}
We would like to thank Fionn Murtagh, Eric Slezak, Albert
Bijaoui, Florence Durret, Gary Mamon, Ren\'e M\'endez, Patrick Petitjean and
Catarina Lobo for useful discussions. GG is fully supported by a fellowship
from the French Ministry of Foreign Affairs.
\end{acknowledgements}

\end{document}